\documentclass[useAMS,usenatbib]{mn2e}
\usepackage{graphicx,url,multirow,mathrsfs,amssymb,amsmath}
\def\n4{NGC\,4051}
\def\apj{ApJ}%
\def\apjl{ApJ}%
\def\apjs{ApJS}%
\def\aap{A\&A}%
\def\cjaa{Chinese J. Astron. Astrophys.}%
\def\mnras{MNRAS}%
\def\pre{Phys.~Rev.~E}%
\def\prl{Phys.~Rev.~Lett.}%
\def\pasp{PASP}%
\def\solphys{Sol.~Phys.}%
\def\nat{Nature}%
\def\physrep{Phys.~Rep.}%
\newcommand{\eqb}{\begin{eqnarray}}
\newcommand{\eqe}{\end{eqnarray}}
\DeclareMathOperator{\cov}{cov}
\title[Generating artificial light curves]{Generating artificial light curves: Revisited and updated}
\author[D.~Emmanoulopoulos]{D.~Emmanoulopoulos,$^{1}$\thanks{E-mail: D.Emmanoulopoulos@soton.ac.uk} I.~M.~M\textsuperscript{c}Hardy$^{1}$ and I.~E.~Papadakis$^{2,3}$ \\ 
$^{1}$Physics and Astronomy, University of Southampton, SO17 1BJ Southampton, United Kingdom\\
$^{2}$Physics Department, University of Crete, PO Box 2208, 71003 Heraklion, Greece\\
$^{3}$IESL, Foundation for Research and Technology, 71110 Heraklion, Greece}
\begin{document}

\date{Accepted 2013 April 30. Received 2013 April 29; in original form 2013 April 3}
\pagerange{\pageref{firstpage}--\pageref{lastpage}} \pubyear{2002}
\maketitle

\label{firstpage}
\begin{abstract}
The production of artificial light curves with known statistical and variability properties is of great importance in astrophysics. Consolidating the confidence levels during cross-correlation studies, understanding the artefacts induced by sampling irregularities, establishing detection limits for future observatories are just some of the applications of simulated data sets. Currently, the widely used methodology of amplitude and phase randomisation is able to produce artificial light curves which have a given underlying power spectral density (PSD) but which are strictly Gaussian distributed. This restriction is a significant limitation, since the majority of the light curves e.g.\ active galactic nuclei, X-ray binaries, gamma-ray bursts show strong deviations from Gaussianity exhibiting `burst-like' events in their light curves yielding \textit{long-tailed} probability distribution functions (PDFs). In this study we propose a simple method which is able to precisely reproduce light curves which match both 
the PSD and the PDF of either an observed light curve or a theoretical model. The PDF can be representative of either the parent distribution or the actual distribution of the observed data, depending on the study to be conducted for a given source. The final artificial light curves contain all of the statistical and variability properties of the observed source or theoretical model i.e.\ same PDF and PSD, respectively. Within the framework of \textit{Reproducible Research}, the code, together with the illustrative example used in this manuscript, are both made publicly available in the form of an interactive {\sc mathematica} notebook.
\end{abstract}


\section{INTRODUCTION}
\label{sect:intro}
Currently in astrophysics, artificial light curves are usually constructed using the procedure of \citet{timmer95} (hearafter TK95). This method is able to produce ensembles of non-deterministic, normally distributed time series from a given underlying power spectral density (PSD) model, $\mathscr{P}(f)$, which represents the variability power as a function of temporal frequency, $f$. Resembling the method proposed by \citet{davies87}, it randomises correctly both the phase and the amplitude of the Fourier components, thus advancing on the previous method of \citet{done92}, which randomises only the phase, assuming a deterministic amplitude which causes a long-term trend in the resulting simulated data sets.\par
There are numerous applications of the TK95 procedure in a plethora of astrophysical fields, including: 
\begin{itemize}
\item Its use in the establishment of statistical confidence intervals during cross-correlation studies \citep[e.g.][]{agudo11,bartlett13}.
\item Its application during microlensing studies \citep[e.g][]{ofek03,koptelova10,tewes12}.
\item The detection of variability in large catalogues and surveys \citep{bauer09,villforth10,macLeod10,primini11}, as well as for the detection of the smallest variability time-scales embedded in a data set \citep[e.g.][]{aharonian07_pks2155_304}.
\item Determination of the detection limits of astronomical instruments for a given type of astrophysical source (e.g.\ AGN, GRBs)  \citep[e.g.][]{greene10,primini11,doro13,khabibullin12}.
\item Its use in the derivation of confidence intervals during the study of quasi-periodic oscillations of galactic and extra-galactic objects \citep[e.g.][]{benlloch01,gierlinski08,do09}, as well as for the statistical characterisation of periodic and pulsed patterns in stellar photometric data \citep{stanishev02,grosso10,blomme11}.
\item Its central role in the estimation of the underlying PSD of irregularly sampled AGN light curves within the procedure proposed by \citet{uttley02}.
\item Its vital use in the study of both the powers and limitations of a given statistical method \citep[e.g.][]{zhang04,vaughan05,emmanoulopoulos10,gora11}.
\item Its use in the fields of Solar astrophysics \citep{rajaguru04} and geophysics \citep{venema06}.
\end{itemize}
At this point, it is very important to note that the abovementioned method of TK95 is appropriate for the production of Gaussian artificial light curves {\it only}. This means that the resultant surrogate data sets\footnote{The term `surrogate' data set \citep[after][]{theiler92} will be used throughout the manuscript in exactly the same way as the terms `artificial', `simulated' and `synthetic' data set.} preserve only the first two statistical moments of the original data set, i.e.\ the mean value, $\mu$, and the variance, $\sigma^2$, ignoring potential higher order statistical moments, such as skewness, kurtosis, or multi-modes found in the normalised flux distribution of the data, i.e.\ probability density function (PDF), corresponding either to the parent or observed distribution. Thus, Gaussian light curves show on average the same amplitude variations above and below the mean, resulting in a zero skewness distribution of data points. Another characteristic is that since the surrogates are normally 
distributed, there is always a finite probability for the artificial data points to become negative\footnote{For a given point this probability is equal to $0.5\,\text{erfc}[{\mu/(\sigma\sqrt{2})}]$ where $\text{erfc}$ denotes the complementary error function.}. However, the light curve of any astronomical source must by default remain positive and so must the resulting PDF.\par
In this framework, TK95 methodology can be used to simulate observed data sets which are Gaussian distributed in the broad sense i.e.\ having a negligible skewness and/or kurtosis. On a large number of occasions however, light curves exhibit a `burst'-like behaviour e.g.\ in X-rays with \textit{RXTE} and \textit{XMM-Newton} \citep[e.g.][]{chitnis09,vaughan11}, in $\gamma$-rays with the \textit{Large Area Telescope} (\textit{LAT}) on-board \textit{Fermi} \citep[e.g.][]{chatterjee09,agudo11}, in which the events are distributed following right \textit{heavy-tailed} distributions. This implies occurrence probabilities of high flux values larger than those expected from Gaussian distributions and thus the Gaussian TK95 products can not be used for the establishment of confidence intervals e.g.\ in cross-correlation studies.\par
Furthermore, the rms-flux relation, i.e.\ the linear scaling of the fractional root mean square (rms) variability amplitude with the flux, observed in both AGN and X-ray binaries (XRBs) \citep{uttley01,uttley05a,gandhi09,mchardy10}, cannot be reproduced by the TK95 algorithm. \citet{uttley05a} show that such behaviour can arise from a non-linear multiplicative variability process in which the parent distribution follows a log-normal distribution. The authors therefore suggest a modification to the TK95 products involving exponentiation (in base $e$) of the normally distributed artificial data sets, yielding light curves which both possess a log-normal distribution and exhibit the rms-flux relation. Although the normalisation of the input PSD, $\mathscr{P_{\rm rescale}}(f)$, is selected in such a way that the variance of the final products matches that of the observed light curve, the actual shape of the PSD is distorted from the original one, $\mathscr{P}(f)\delta f$, in such a way that the actual 
variability power within a given frequency range, $\mathscr{P_{\rm rescale}}(f)\delta f$, differs from the genuine one, $\mathscr{P}(f)\delta f$.\par
Apart from this PSD distortion, the exponentiation transformation cannot be generalised to arbitrary parent or observed distributions (depending on the type of study). The specification of the parent distribution requires either very large data set \citep[e.g.\ for the case of Cyg\,X-1 253144 data points are required to form the parent log-normal distribution][]{uttley05a} or a theoretical model \citep[e.g.][]{kelly11}. Employment of the parent distribution is of vital importance in comparing variability properties of data sets obtained over a long period of time which map the complete variability behaviour of the source. Nevertheless, it is sometimes crucial to establish the detection significance of a given result coming from a single observed data set. This approach has been used several times in the field of reverberation studies, in the form of flux redistribution or random subset selection \citep[e.g.][]{peterson98}, or the detection of time lags in very high energy (VHE) Cherenkov astronomy in the 
framework of Quantum Gravity \citep{aharonian08_QG}. For the case of transient phenomena in particular, e.g.\ GRBs, in which the concept of a parent distribution is not applicable, only a single realisation is available for each observation and thus this should be used as the PDF.\par
In this paper we put forward a simple method which combines the routine of TK95 and the iterative amplitude adjusted Fourier transform algorithm of \citet{schreiber96} (hearafter SS96), which produces artificial light curves which possess exactly the same PSD and PDF as the originally observed light curve (or a theoretical model). Thus, the surrogates will have exactly the same variability and statistical properties as the observed light curve. Initially, in Sect.~\ref{sect:methodology} we describe in detail the method. For illustrative purposes, in Sect.~\ref{sect:appli_ngc4051} we then apply it to the case of the well-studied type-I Seyfert AGN \n4, using \textit{XMM-Newton} observations. Following that, in Sect.~\ref{sect:compl_applic}, we produce artificial light curves for the $\gamma$-ray blazar 3C\,454.3, using \textit{Fermi-LAT} observations, and the XRB Cyg\,X-1 using observation obtained by the \textit{All Sky Monitor} (\textit{ASM}), on-board \textit{RXTE}. In Sect.~\ref{sect:ccf_analysis} we 
present an application in cross correlation analysis, and in Sect.~\ref{sect:rms_flux} we reproduce the rms-flux relation for the case of log-normally distributed light curves. In Sect.~\ref{sect:invariants} we discuss which properties of the light curves are preserved during our simulation process and finally, a discussion together with a summary of our results, can be found in Sect.~\ref{sect:summar_discu}. In the Appendix~\ref{app:def_nom} we give the basic definitions and properties of the various quantities i.e.\ periodogram, PSD and PDF, as well as the various fitting procedures that will be used throughout this manuscript. In Appendix~\ref{app:cumul_statiDependnce} we elucidate the differences between statistical moments and cumulants, which are commonly confused in the astronomical literature.\par
Throughout the manuscript the error estimates for the various best-fitting model parameters correspond to the 90 per cent confidence intervals unless otherwise stated. The error bars of the plot points in all the figures indicate the 68.3 per cent confidence intervals.

\section{METHODOLOGY}
\label{sect:methodology}
\subsection{The algorithm}
\label{ssect:algorithm}
This method is a combination of TK95 and the SS96, with some significant alterations and modifications which join the two together.\par
Consider an observed light curve $x_{\rm obs}(t)$ consisting of $N$ uniformly sampled observations (sampling rate $\Delta t$), $\{t_i, x_{\rm obs}(t_i)\}$ for $i=1,2,\ldots,N$. The light curve has an underlying PSD, $\mathscr{P}(f)$, and an observed (or `parent', depending on the purpose of the statistical study) PDF, ${\rm PDF}\left[0\leq x_{\rm obs}(t)<\infty\right]$. Note that both/either PSD and/or PDF can also originate from a theoretical model which we want to check the statistical properties of its products i.e.\ time series. Note that if one wishes to take into account the various spectral distortion effects (Sect.~\ref{ssect:spec_alter}) then one should adjust both the simulation length $N$ and the time resolution accordingly as described in Sect.~\ref{ssect:spec_alter}.
\begin{enumerate} 
\item Using the TK95 procedure, a normally distributed time series\footnote{Actually this is an asymptomatically normally distributed time series. Despite the fact that TK95 corresponds to a realisation of a Gaussian process, individual artificial data sets products (of finite length) may not be necessarily normally distributed since the Gaussianity of the process limits the asymptotic distribution only, $N\rightarrow\infty$.} is produced, $x_{\rm norm}(t)$, consisting of $N$ values and an underlying PSD identical to $\mathscr{P}(f)$. Then, for each Fourier frequency, $f_j$, the discrete Fourier transform (DFT), $DFT_{\rm norm}(j)$, is estimated and from this the corresponding amplitudes, $\mathscr{A}_{\rm norm}(j)$, phases, $\phi_{\rm norm}(j)$, and periodogram, $P_{\rm norm}(f_j)$ (equations \ref{eq:amplitude}, \ref{eq:phase} and \ref{eq:periodogram}, respectively). Note that since the iteration algorithm aims to produce artificial source light curves, the input PSD should not contain the Poisson 
noise component.
\item From the ${\rm PDF}\left[0\leq x_{\rm obs}(t)<\infty\right]$ a series of $N$ pseudo-random numbers is produced which forms a white noise data set, $x_{\rm sim,1}(t)$. Then, at each Fourier frequency, the discrete Fourier transform of $x_{\rm sim,1}(t)$ is estimated, $DFT_{\rm sim,1}(j)$, and from that the corresponding amplitudes, $\mathscr{A}_{\rm sim,1}(j)$, phases, $\phi_{\rm sim,1}(j)$, $f_j$ and periodogram, $P_{\rm sim,1}(f_j)$\footnote{\label{ftnt:flatPSD}For $x_{\rm sim,1}(t)$ the periodogram, $P_{\rm sim,1}(f)$, corresponds by default to an underlying PSD with a slope of $\alpha=0$ (since it represents a white noise process).}.
\item \underline{Spectral adjustment}: For each frequency, $f_j$, the amplitudes $\mathscr{A}_{\rm sim,1}(j)$ are replaced with the amplitudes $\mathscr{A}_{\rm norm}(j)$, whilst keeping the phases $\phi_{\rm sim,1}(j)$ unaltered. This yields the adjusted DFT of $x_{\rm sim,1}(t)$, $DFT_{\rm sim.adjust,1}(j)$, on which we then perform an inverse discrete Fourier transform (IDFT), yielding the time series, $x_{\rm sim.adjust,1}(t)$. This time series has an identical underlying PSD to the desired one, $\mathscr{P}(f)$, but with a distribution of measurements which has been altered from that of ${\rm PDF}\left[0\leq x_{\rm obs}(t)<\infty\right]$.  
\item \underline{Amplitude adjustment}: A new time series is created from the values of $x_{\rm sim,1}(t)$ ordered based on the ranking of $x_{\rm sim.adjust,1}(t)$. This means that the the highest value of $x_{\rm sim.adjust,1}(t)$ is replaced by the highest value of $x_{\rm sim,1}(t)$, the second highest value of $x_{\rm sim.adjust,1}(t)$ is replaced by the second highest value of $x_{\rm sim,1}(t)$ and so on. The resulting data train, $x_{\rm sim,2}(t)$, is distributed exactly as ${\rm PDF}\left[0\leq x_{\rm obs}(t)<\infty\right]$ but its PSD differs from the target one, $\mathscr{P}(f)$.
\item The same process is repeated in an iterative fashion $\kappa$-times, starting from step (ii), until the resulting products remain the same i.e.\ $x_{\rm sim,k+1}(t)\equiv x_{\rm sim,k}(t)$ (convergence):
\begin{itemize}
\item 2\textsuperscript{nd} iteration: $x_{\rm sim,1}(t)$ is replaced by $x_{\rm sim,2}(t)$
\item 3\textsuperscript{rd} iteration: $x_{\rm sim,2}(t)$ is replaced by $x_{\rm sim,3}(t)$
\item $\kappa$\textsuperscript{th} iteration: $x_{\rm sim,\kappa-1}(t)$ is replaced by $x_{\rm sim,\kappa}(t)$
\end{itemize}
\end{enumerate}

\begin{figure}
\hspace{-2em}\includegraphics[width=4.3in]{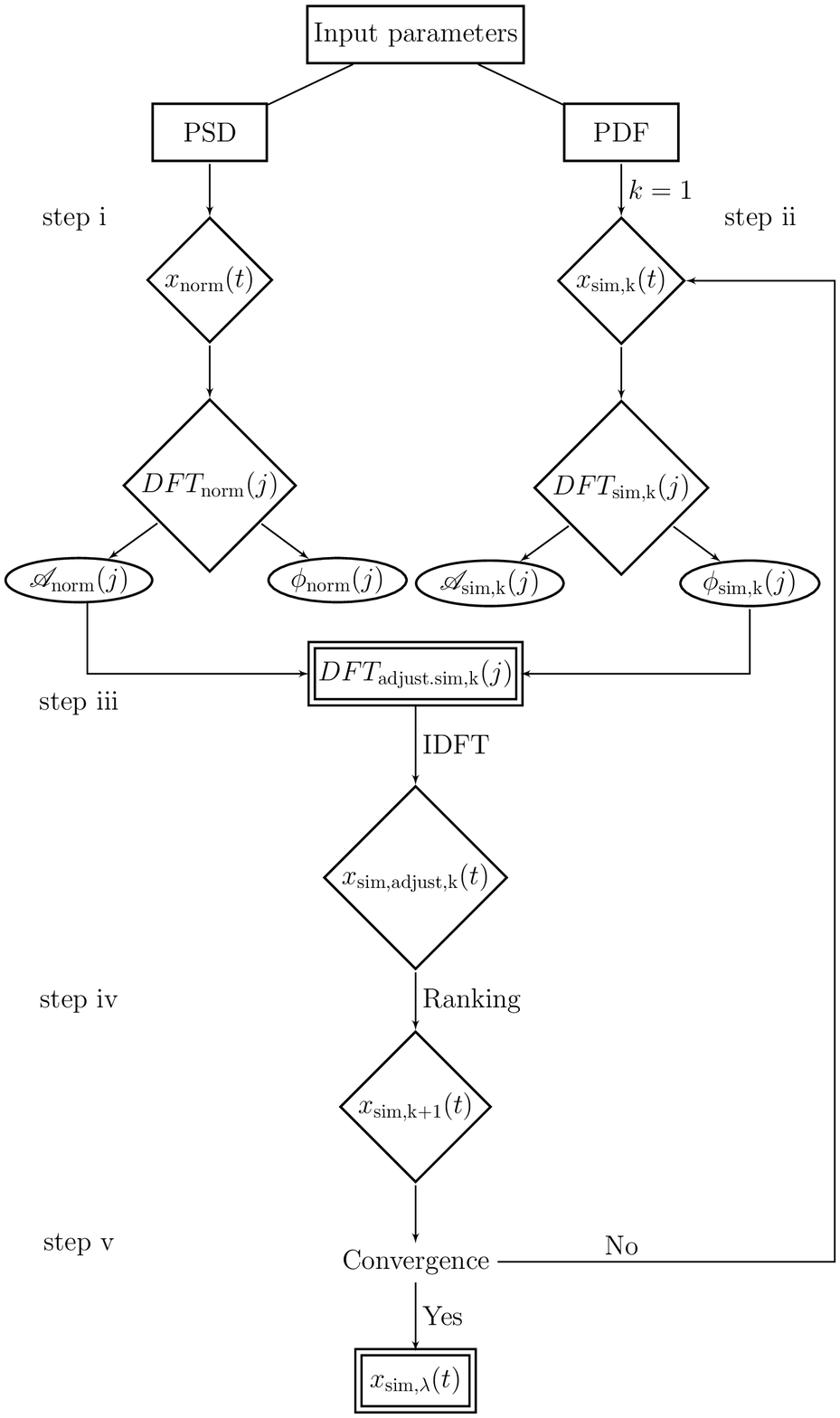}
\caption{The flow chart diagram showing the various steps of the method.}
\label{fig:methodFlowChart}
\end{figure}

After a given number of iterations, e.g.\ $\lambda=\kappa+1$, the synthetic light curve products do not change (i.e.\ convergence) and thus the $x_{\rm sim,\lambda}(t)$ iterated product comprises the final artificial light curve product. The exact number of iterations depends on the length of the original data set, the underlying input PSD and the input PDF. More about the convergence can be found in Sect.~\ref{sssect:convergence_singleDS} and Sect.~\ref{sssect:convergence_ensembleDS} using Monte Carlo simulations. Note that for the case of a Gaussian PDF the iteration process gives exactly equivalent results with the TK95 products since (step i) yields products which are already Gaussian distributed. The flowchart of the abovementioned method is given in Fig.~\ref{fig:methodFlowChart}.\par

\subsection{Appropriate treatment of the Poisson noise}
\label{ssect:poisson}
The final simulated product, $x_{\rm sim,\lambda}(t)$, has the desired distribution and PSD corresponding to the source light curve produced. Since the observed light curve is a product of a counting detector-process the observations are affected by Poisson noise, which is imprinted in the corresponding PSD as a constant component. In order to mimic this effect, each light curve point $x_{\rm sim,\lambda}(t)=\{x_{\rm sim,\lambda}(t_1),x_{\rm sim,\lambda}(t_2),\ldots,x_{\rm sim,\lambda}(t_N)\}$ is replaced by an appropriate Poisson random variate
\eqb
x_{\rm sim,Pois,\lambda}(t_i)\sim\frac{Pois[\mu=x_{\rm sim,\lambda}(t_i)\Delta t]}{\Delta t}\;\;{\rm for}\;i=1,\ldots,N
\label{eq:poisson_variate}
\eqe
where $Pois[x_{\rm sim,\lambda}(t_i)\Delta t]$ dictates the probability mass function of the Poisson distribution with a mean value of $x_{\rm sim,\lambda}(t_i)\Delta t$.

\subsection{Spectral distortions: Red noise leak and aliasing.}
\label{ssect:spec_alter}
In the case of a non-white noise PSD (as in the case of AGN light curves) the periodogram estimates tend to be biased due to `red noise leak' \citep[the transfer of variability power from the low to high frequencies due to the finite length of observations;][]{deeter82,deeter84} and aliasing effects \citep[fold-back of variability power from high frequencies to lower frequencies due to the finite time resolution;][]{kirchner05}.\par These are two very well understood spectral distortions induce by the sampling properties of the data set and they can be taken into account in the usual manner \citep[e.g.][]{uttley02}. In order to take into account the `red noise leak' effect we produce surrogate data sets which are much longer than the observed data set (e.g.\ 100 times) and then we randomly select a subset having the desired length.\par
With respect to the aliasing, since we are dealing with data averaged over time intervals, $\Delta t_{\rm sample}$, rather than simply sampled data its effect is very much reduced \citep{vanderKlis88}. Nevertheless, if one also wishes to include the aliasing effect in the simulations, for the case of unbinned data (e.g.\ count data), then one should increase the time resolution of the simulations e.g.\ to 10 per cent of $\Delta t_{\rm sample}$.\par
With these two approaches, the dependencies in both the Fourier amplitudes (equation \ref{eq:amplitude}) and the Fourier phases (equation \ref{eq:phase}) are taken into consideration. However, if one wishes to carry out statistical studies that deal only with the Fourier amplitudes (as in all the examples shown in this manuscript) then the length and binning adjustments can be applied during the first step of the abovementioned method i.e.\ during the application of the TK95 procedure. In this way the TK95 artificial light curves, $x_{\rm norm}(t)$, carry all the spectral distortion effects that involve only the Fourier amplitudes, which will then be passed to the final surrogates during the spectral adjustment stage (step iii). An major advantage to this is that the whole process is much faster since the iteration process involves data sets which have lengths equal to that of the observed data set.\par
Note once again that for statistical studies that involve Fourier phases, e.g.\ phase-lag spectra studies, one should follow the initial recipe i.e.\ carry out the whole simulation for longer and more finely sampled surrogate data sets, and select a subset which has the desired length from the final converged iteration product. The effect of red-noise leak on the phases is rarely discussed in the literature despite the fact that it adds significant dependencies to the phases.\par
\begin{figure*}
\hspace*{-0.5em}\includegraphics[width=3in]{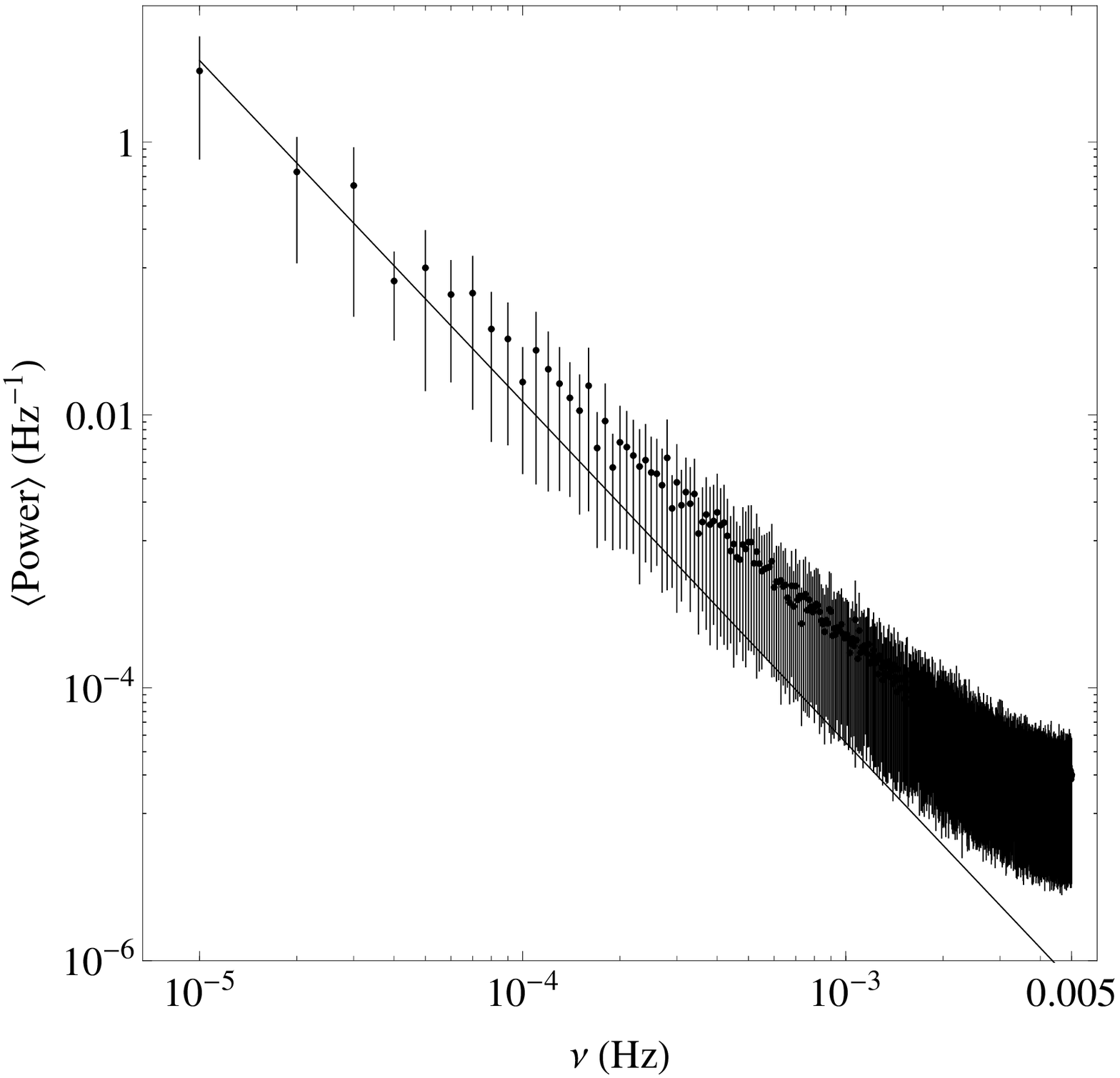}\hspace{2em}
\includegraphics[width=3in]{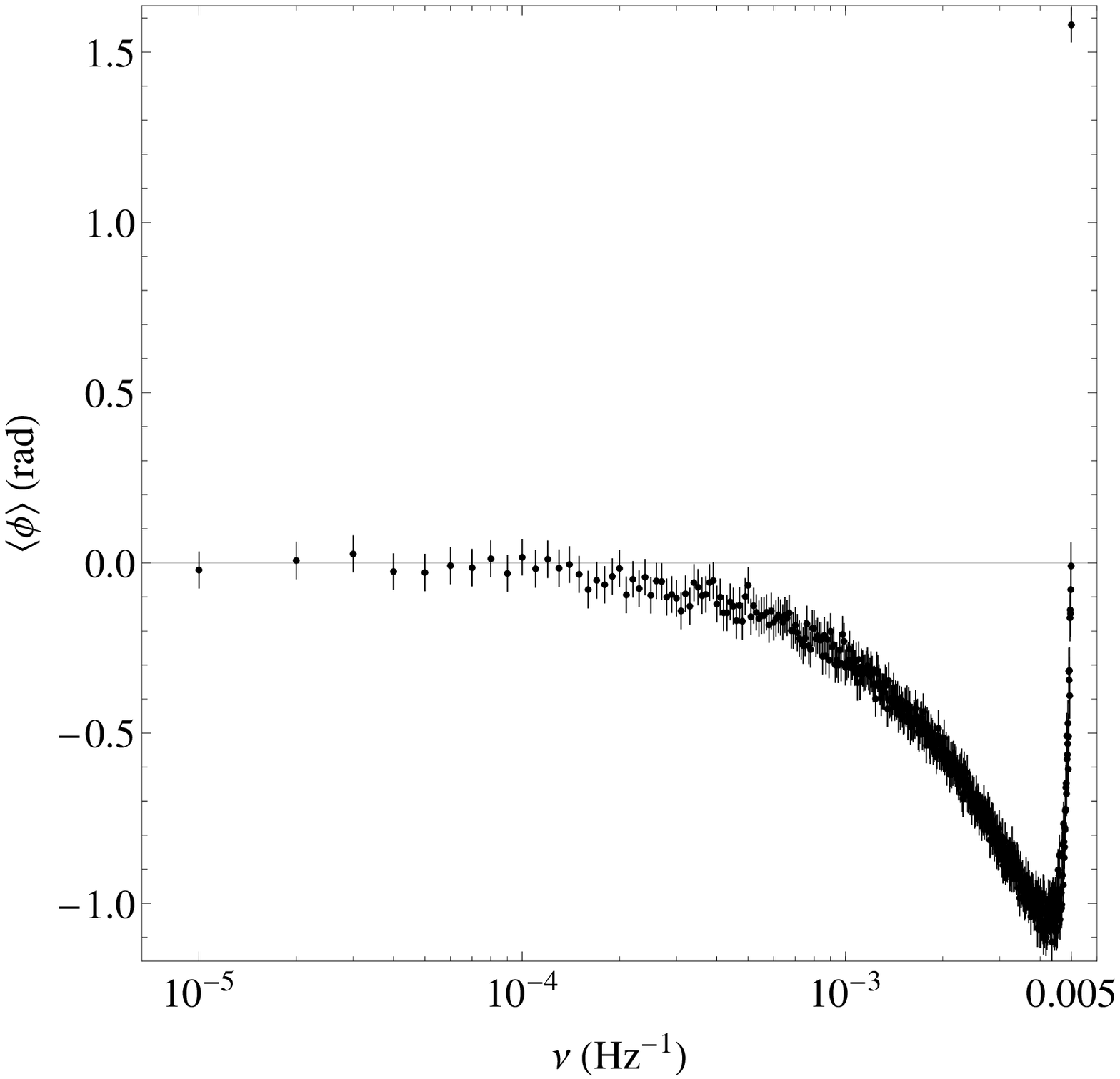}
\hspace*{-1.4em}\includegraphics[width=3.18in]{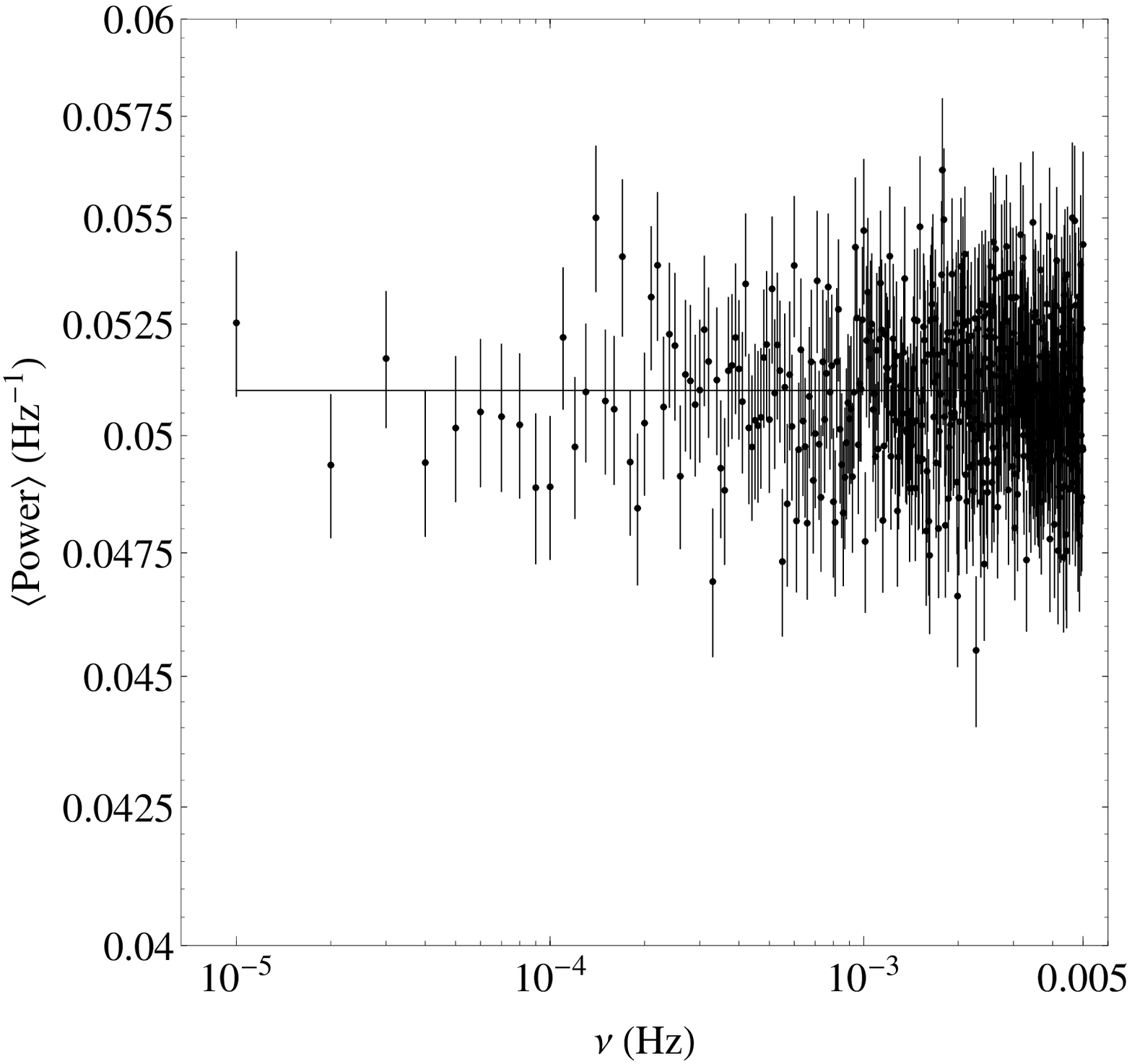}\hspace{2em}
\includegraphics[width=2.96in]{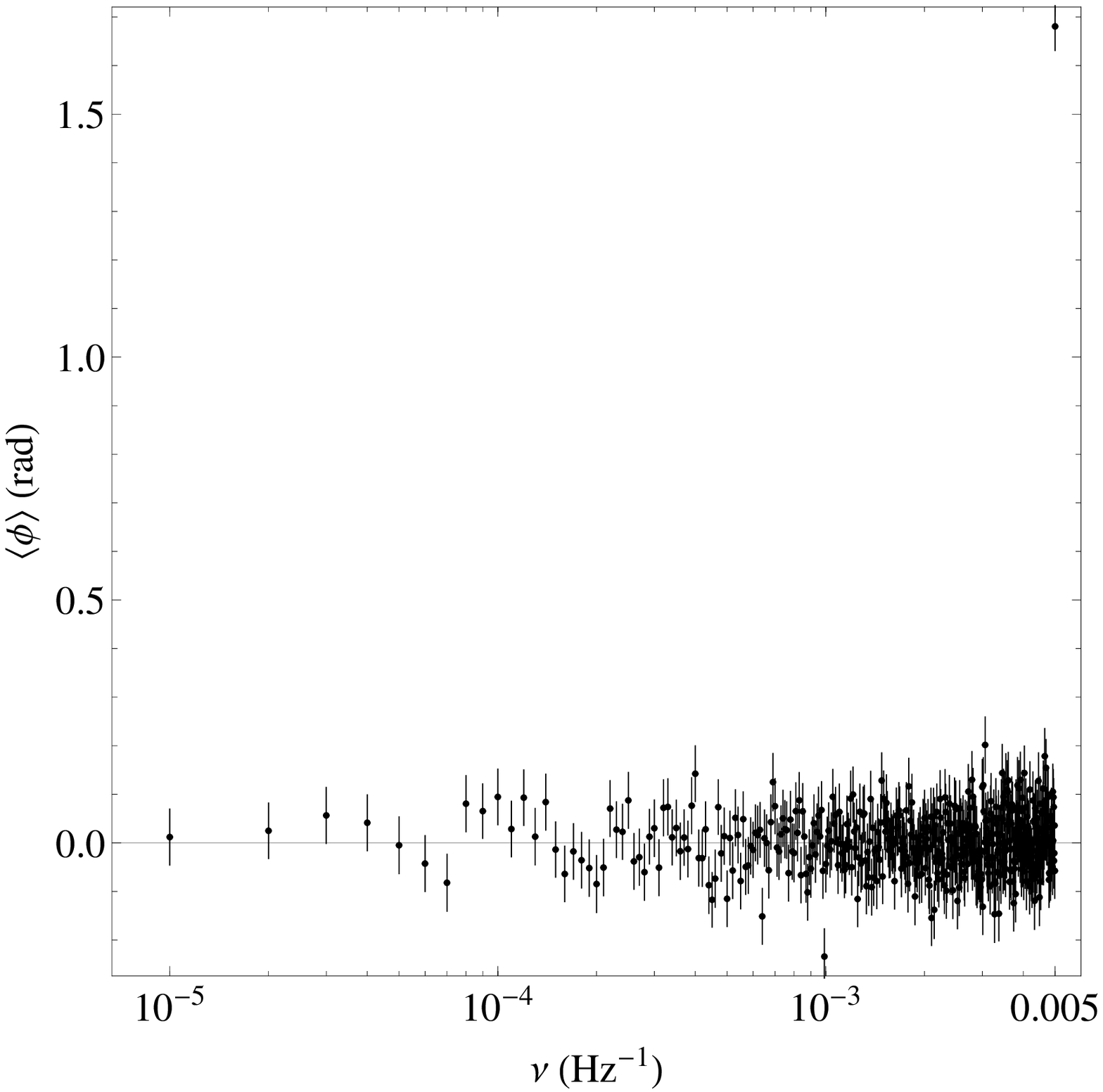}
\caption{The effect of red noise leak in the Fourier amplitudes and phases. [Top-left panel] The averaged periodogram estimates for an input PSD with a power-law shape of a slope -2.5. [Top-right panel] The averaged Fourier phases for an input PSD with a power-law shape of a slope -2.5. [Bottom-left panel] The averaged periodogram estimates for an input PSD with a power-law shape of a slope 0. [Bottom-right panel] The averaged Fourier phases for an input PSD with a power-law shape of a slope 0.}
\label{fig:redNoiseEffects}
\end{figure*}

For demonstrative purposes, we create an artificial light curve, using the TK95 procedure, consisting of $10^6$ data points, binned in 100 s, with an input PSD which has a power-law model of slope -2.5. We then chop the light curve into 1000 segments each one consisting of 1000 consecutive points. For each data set and for each Fourier frequency, $f_j$ we estimate the DFT (equation \ref{eq:dft}) and from this its amplitude (via the periodogram, equation \ref{eq:periodogram}) and its phase (equation \ref{eq:phase}). Finally, for each $f_j$, we average the the periodogram estimates and phases; the results are shown in the top panels of Fig.~\ref{fig:redNoiseEffects}.\par
The top-left panel of Fig.~\ref{fig:redNoiseEffects} shows clearly the effect of red noise leak for the amplitudes, something that has been extensively discussed in the literature. The top-right panel of Fig.~\ref{fig:redNoiseEffects} shows vividly that the red noise leak also affects the phases in a very distinctive way. The onset of the effect is around $10^{-4}$ Hz (the same as it is for the amplitudes) and from then on all the phases follow an arched trend towards negative values. The last point in this plot, corresponds to the phase estimate at the Nyquist frequency, $f_{N/2}=f_{\rm Nyq}$, for which the DFT is a always a real number (positive or negative). This means that, for $f_{\rm Nyq}$, we average only the phases corresponding to the negative values, which are always $\pi$, over the total number of points, since for the positive values the $\operatorname{arg}$ is not formally defined (in this context one could consider it equal to 0). On average one should get values around $\left<\phi_{N/2}
\right>\simeq\pi/2\simeq1.57$ (i.e.\ roughly equal numbers of positive and negative values).\par
We repeat the same process but this time with a constant underlying PSD i.e.\ a power-law with slope 0. As we can see from the bottom panels of Fig.~\ref{fig:redNoiseEffects} both the amplitudes (bottom-left panel) and the phases (bottom-right panel) are not affected by the red noise leak effect. It is clear that, as in the case of the Fourier amplitudes, the effect of red noise leak for the Fourier phases depends on the shape of the input PSD, e.g.\ the softer the power-law of the underlying PSD, the greater the effect of the red noise leak. As we discussed previously, our methodology correctly takes this effect into account by extending the total length of the surrogate data set and then chopping the converged final synthetic light curve to the desired length.

\begin{figure*}
\includegraphics[width=2.90in]{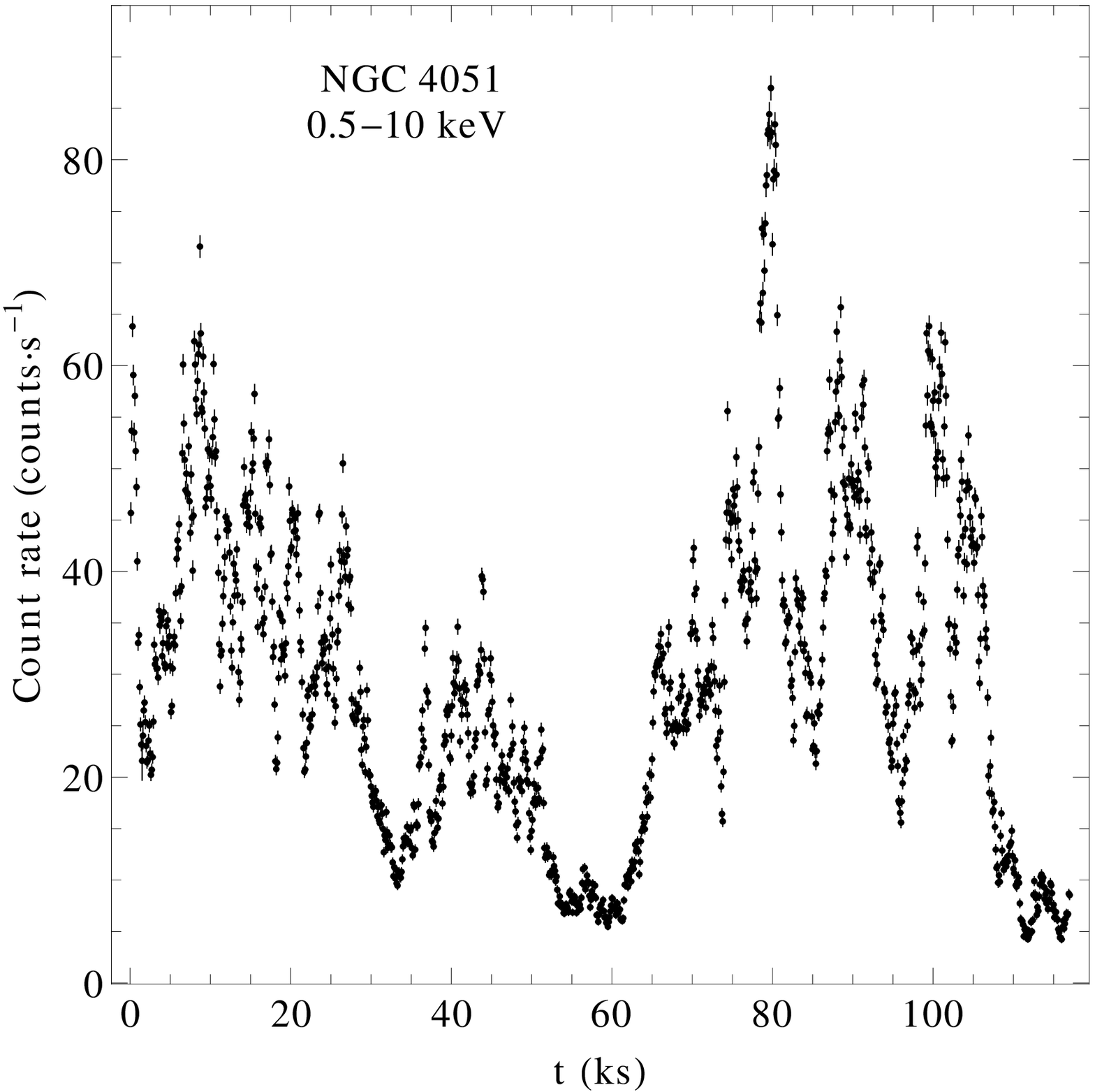}
\includegraphics[width=3.1in]{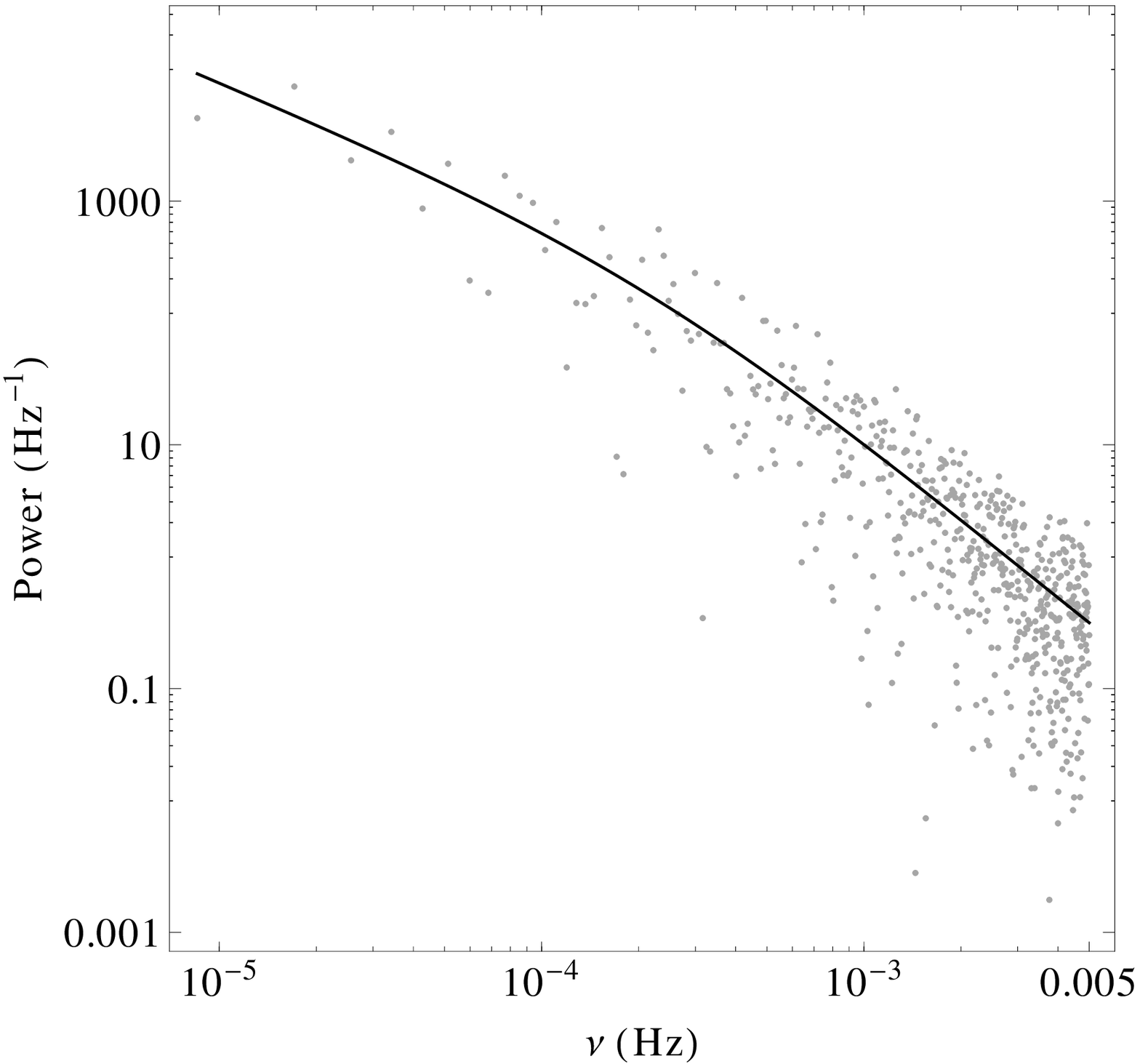}
\caption{The \textit{XMM-Newton} data set of \n4. [Left panel] The EPIC pn and MOS combined light curve in the 0.5--10 keV energy-band in bins of 100 s (obs ID: 0109141401, revolution: 0263). [Right panel] The corresponding periodogram estimates, $P_{\rm obs}(f)$, (grey points) and the underlying best-fit PSD model, $\mathscr{P}(f;\vec{\gamma}_{\rm bf},c_{\rm bf})$, (black line).}
\label{fig:ngc4051_lc_psd}
\end{figure*}

\subsection{Basic differences and advantages from previous works}
\label{ssect:diff_advant}
Our method is essentially a marriage of TK95 and SS96 algorithms. The former remains exactly the same during the application of the method but the latter (i.e.\ the iterative amplitude adjusted Fourier transform) contains several key differences, from the SS96, which makes it suitable particularly for the needs of astronomical data sets.
\begin{enumerate}
 \item We use a pseudo-random data set following the estimated distribution, rather than a shuffled version of the original observed data set.
 \item We replace the Fourier phases of the TK95 products rather than those of the original data set.
 \item All the spectral distortion effects due to the finite length and sampling rate of the observed data set (i.e.\ red noise leak and aliasing) are taken into account.
 \end{enumerate}
The coupling of the two methods allow us to study not only observed light curves but also theoretical models that give predictions about the PSD and the PDF of a given astrophysical object. The TK95 is carrying the spectral information (PSD) and the SS96 is distributing the various measurements (PDF) accordingly. In this way, we can produce, based on a theoretical model, realistic and positively defined non-Gaussian synthetic light curves as opposed to only Gaussian light curves coming from TK95.\par
The problem of generating stochastic sequences of numbers with specified properties is extensively analysed in the literature since the early 70's \citep[for a complete reference guide see][]{sowey86}. In particular, \citet{liu82} proposed a white Gaussian noise input to a linear digital filter followed by a zero-memory non linearity (ZMNL). The ZMNL is chosen so that the desired distribution is exactly realised and the digital filter is designed so that the desired autocovariance is closely approximated. \citet{hunter83} proposed a method for the generation of random number sequences with an arbitrarily specified first-order probability distribution function (PDF) and an arbitrarily specified first-order autocorrelation function (ACF). The procedure involves a stochastic optimisation algorithm which minimizes the squared sum between the desired (output) and the actual (observed) ACF estimates.\par
An iterative method was developed by \citet{yamazaki88} which generates Gaussian distributed samples with a given periodogram which are then mapped into non-Gaussian distributed numbers. This is achieved by employing the invert expression of the target PDF (distribution distortion method) and the iteration process aims to correct the altered periodogram estimates (as they come out from the mapping process) to match the \textit{desideratum} periodogram. The correlation distortion method was used by \citet{johnson94}, and consists of a nonlinear transformation which is applied to construct non-Gaussian correlated features from correlated Gaussian random draws. Finally, \citet{gurley96} presented a series of mathematical approaches using Volterra series and analytical kernels to achieve bispectral matching. In the same work a neural network system identification model is employed for simulation also demonstrating the ability to match higher order spectral characteristics.
\par
Besides the abovementioned differences, our method differs fundamentally from all the previous methodologies with respect to the matching process of the PSD. We are not interested in matching the individual periodogram estimates (derived from the observed data set), but instead in the underlying PSD. In this way, at a given Fourier frequency $f_j$, the various periodogram estimates, $P(f_j)$, are distributed asymptotically around $\mathscr{P}(f_j)$ as a gamma distribution, $\Gamma\left[\nu/2,\mathscr{P}(f_j)\right]$ (equation \ref{eqe:prdgrProb})\footnote{In the literature this is usually referred to as `scaled $\chi^2$ distribution' with 2 d.o.f. (equation \ref{eqe:prdgrChiSquare}).} with $\nu$ degrees of freedom (d.o.f.) corresponding to $\nu=1$ for the the Nyquist frequency and $\nu=2$ for all other frequencies.

\subsection{A publicly available code in the form of an active document}
\label{ssect:avail_code}
In the spirit of \textit{Reproducible Results} and \textit{Active Documents} \citep{claerbout90}, we provide an interactive {\sc mathematica} notebook (created with the version: 9.0.1.0) which contains the complete numerical code together with the example presented in Sect.~\ref{sect:appli_ngc4051}. In detail the notebook contains:
\begin{itemize}
 \item The \textit{XMM-Newton} data set of the AGN \n4 which is used in Sect.~\ref{sect:appli_ngc4051}.
 \item A version of TK95 code taking into consideration (if needed) the spectral distortions described in Sect.~\ref{ssect:spec_alter}.
 \item The iteration algorithm (SS96).
 \item The addition of the Poisson noise as described in Sect.~\ref{ssect:poisson}.
 \item An animation of the simulated products at the various iteration steps.
\end{itemize}
It can be found on the web\footnote{You can also request the {\sc mathematica} notebook via e-mail to \url{D.Emmanoulopoulos@soton.ac.uk}.} as part of this paper (Online Material) or at \url{http://www.astro.soton.ac.uk/~de1e08/ArtificialLightCurves/}. By changing the two random seeds (used for step i and step ii), the whole document is automatically updated and a new artificial light curve is produced. The simple numerical code, provided in the {\sc mathematica} notebook, can be written in a much more compact form (i.e.\ much more computationally efficient), but for clarity purposes we have split it up in various programming lines. Due to its simple nature, the code can be implemented in any programming language.\par

\section{APPLICATION: THE CASE OF \n4}
\label{sect:appli_ngc4051}
\subsection{Step-by-step procedure for a single realisation}
\label{ssect:single_ngc4051}
We will now apply the method to a single X-ray data set of the type I Seyfert galaxy \n4 ($z=0.02336$) obtained by the European Photon Imaging camera aboard \textit{XMM-Newton} observatory (obs ID: 0109141401, revolution: 0263). The observed 0.5--10 keV light curve of \n4 (sample light curve) is shown in the left panel of Fig.~\ref{fig:ngc4051_lc_psd} consisting of $N=1170$ data points  in bins of 100 s.\par
The following process is carried out for illustrative purposes only and aims to simulate a single artificial light curve with the same underlying PSD as the sample light curve, and an identical \textit{observed} PDF (as opposed to the parent PDF). Nevertheless, depending on the purpose of the study one can select an appropriate PDF depicting either the underlying or the observed statistical properties. The method, of course can also produce artificial light curves coming directly from a theoretical model, which specifies an underlying PDF and PSD, without the requirement of having an actual observed light curve. Since, the PSD and the PDF of the observed data set are the two input parameters of the method, we first estimate these. Note that since we will not be performing any studies involving Fourier phases, the effect of `red noise leak' has been taken into account during (step i) i.e.\ by producing a TK95 artificial data set 1000 times longer than the original \n4 data set. For this 
particular data set (something which is generally true for \textit{XMM-Newton} data sets), the aliasing effect is insignificant since we are dealing with averaged consecutive measurements.
\par
Initially we derive the periodogram of the sample light curve, $P_{\rm obs}(f)$, (Fig.~\ref{fig:ngc4051_lc_psd}, right panel, grey points) and then we estimate the underlying PSD by fitting the smoothly bending power-law model plus a constant, $c$, representing the Poisson noise level:
\eqb
\mathscr{P}(f;\vec{\gamma},c)=\frac{A f^{-\alpha_{\rm low}}}{1+(f/f_{\rm bend})^{\alpha_{\rm high}-\alpha_{\rm low}}}+c
\label{eqe:psd_bendMod}
\eqe
in which $\vec{\gamma}=\{A,f_{\rm bend}, \alpha_{\rm low},\alpha_{\rm high}\},$ with components the source's PSD model parameters i.e.\ normalisation, bend frequency, low and high frequency slopes, respectively. During the fit we fix the $\alpha_{\rm low}$ to 1.1 \citep[as derived from long-term \textit{RXTE} data;][]{mchardy04} and the best-fitting model parameters are $\vec{\gamma}_{\rm bf}=\{0.030\pm0.004$ Hz$^{-1}$, $2.3^{+1.2}_{-0.9}\times10^{-4}$ Hz, 1.1, $2.20^{+0.07}_{-0.04}\}$ and $c_{\rm bf}=9.2^{+0.7}_{-0.8}\times10^{-3}$ Hz$^{-1}$ (Fig.~\ref{fig:ngc4051_lc_psd}, right panel, black line). Note that the derived best-fitting values agree entirely with the ones given by \citet{vaughan11} but that the best-fit $f_{\rm bend}$ differs from the value $8^{+4}_{-3}\times10^{-4}$ Hz estimated by \citet{mchardy04} (note that the error estimates correspond to the 90 per cent confidence intervals). This best-fitting model will be the target PSD which should be matched by the surrogate data sets.\par
\begin{figure*}
\hspace{-0.8em}\includegraphics[width=3.1in]{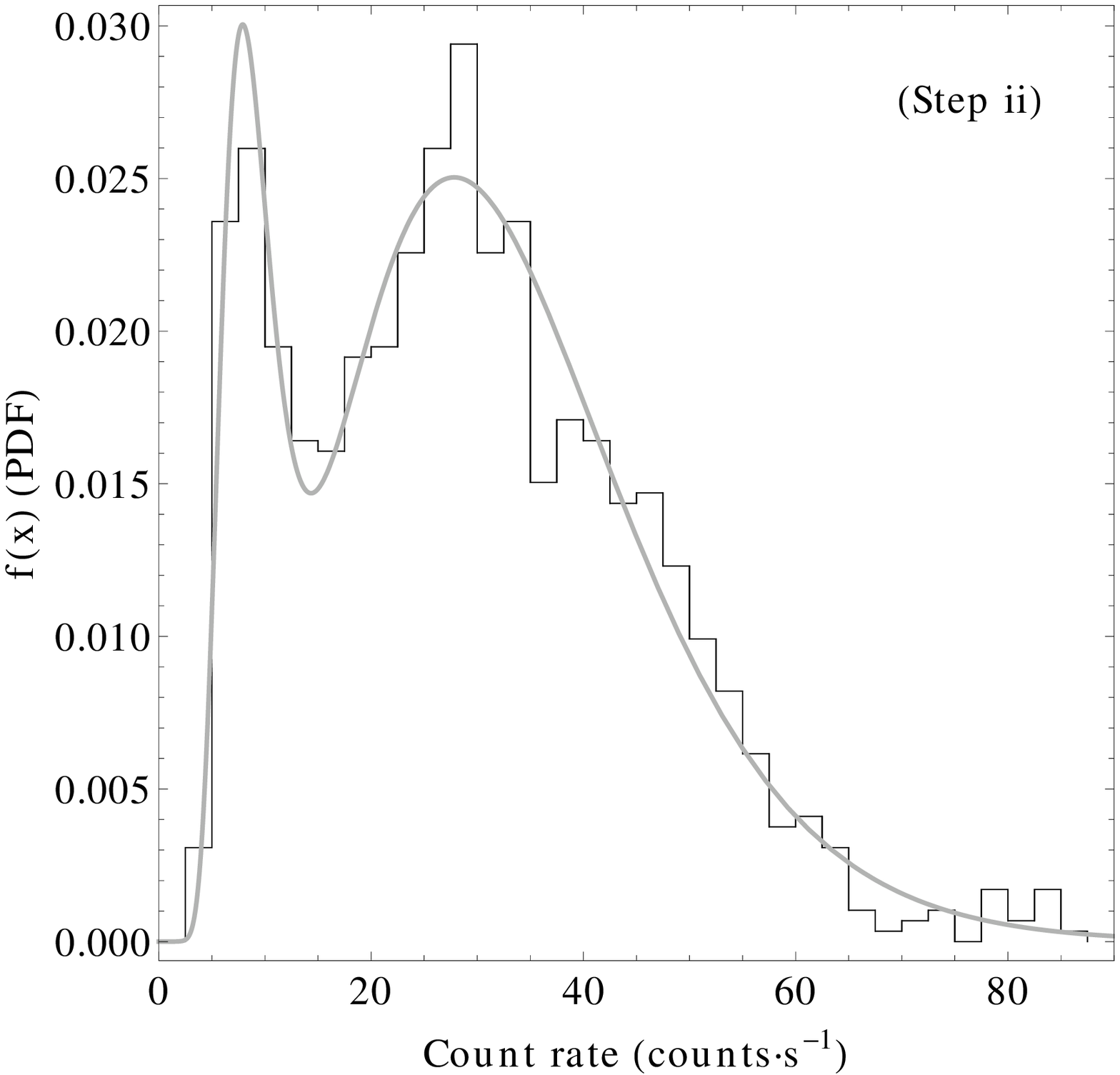}\hspace{1.3em}
\includegraphics[width=2.98in]{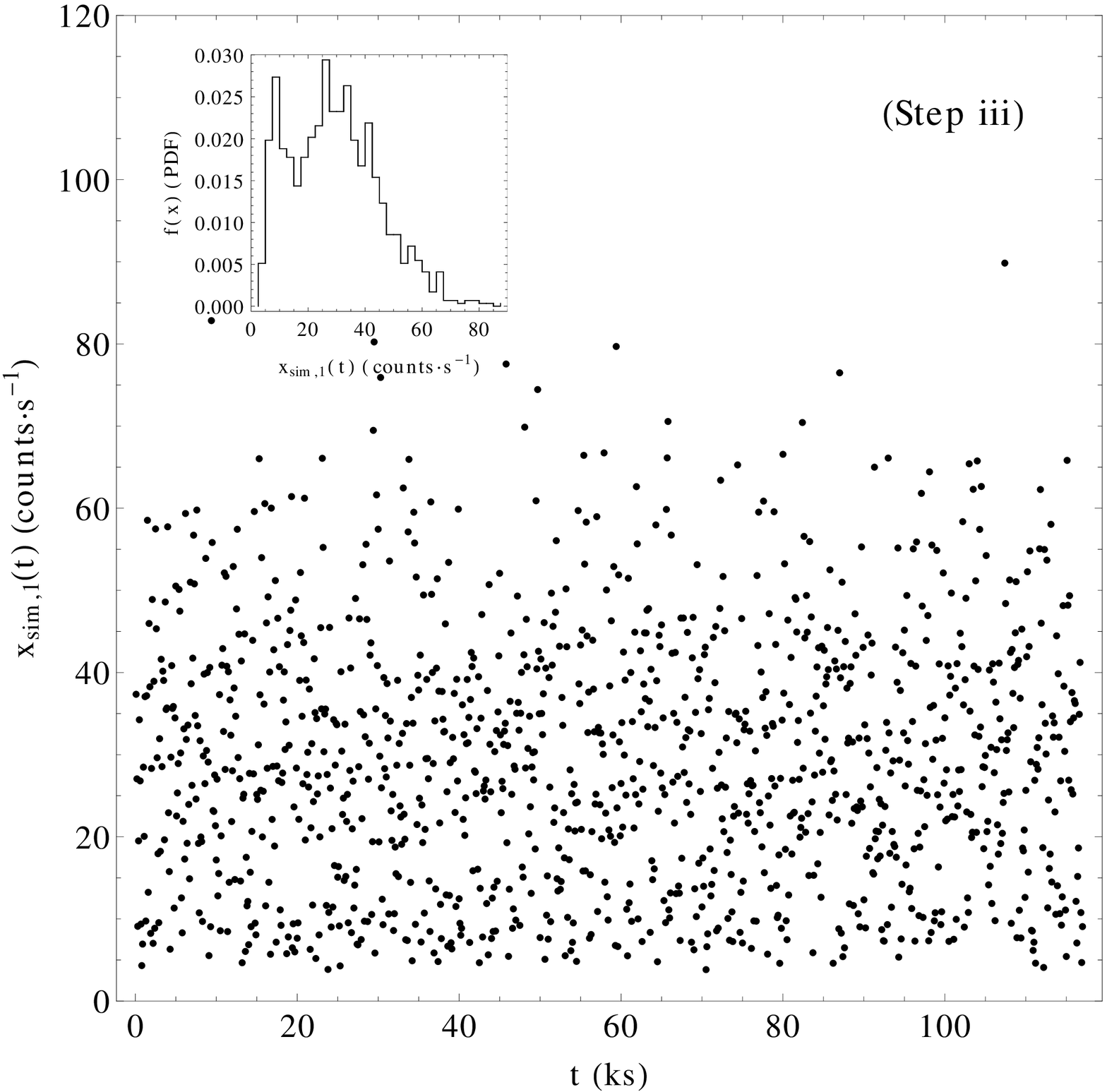}
\caption{Step ii. [Left panel] The PDF histogram of the observed data (black line) together with the best fit mixture distribution model, $\mathfrak{f}_{\rm mix}(x;\vec{\eta}_{\rm bf})$ (gray line, equation \ref{eq:pdf}). [Right panel] An ensemble of $N$ pseudo-random variates produced from equation (\ref{eq:pdf}) and the inlay shows its corresponding PDF histogram.}
\label{fig:ngc4051_simul_hist}
\end{figure*}

In order to assess the probability distribution of the sample data set we then form its probability density function histogram (Fig.~\ref{fig:ngc4051_simul_hist}, left panel, black line). The latter exhibits two clear modes: the first narrow mode corresponds to the low count rate regimes (e.g.\ the regions around 35, 55 and 110 ks in the left panel of Fig.~\ref{fig:ngc4051_lc_psd}), and the second, broader mode, to the high source states. We parametrise the observed distribution of the sample data set by fitting a probability model. For this particular data set of \n4, we select a mixture distribution model consisting of a gamma distribution, $\Gamma(\kappa,\theta)$ with $\kappa$ and $\theta$ being the shape and the scale parameters, and a log-normal distribution, $\ln\mathcal{N}(\mu,\sigma^2)$, with $\mu$ and $\sigma^2$ being the mean and the variance of the count rate's (variable $x$) natural logarithm (equation \ref{eq:pdf}). Finally, each of these component distributions contribute to the overall PDF of 
the mixture distribution, $\mathfrak{f}_{\rm mix}(x;\vec{\eta})$, with a weight of $w_\Gamma$ and $w_{\ln\mathcal{N}}=1-w_\Gamma$ respectively, with an $\vec{\eta}$ being a vector consisting of the model parameters $\vec{\eta}=\{\kappa,\theta,\mu,\sigma,w_\Gamma\}$:
\eqb
&&\mathfrak{f}_{\rm mix}(x;\vec{\eta})=\nonumber\\[1em]
&&w_\Gamma\frac{\theta ^{-\kappa } e^{-x/\theta} x^{\kappa-1}}{\Gamma(\kappa)}+w_{\ln\mathcal{N}}\frac{e^{-(\ln x-\mu )^2/(2 \sigma^2)}}{\sqrt{2\pi}x\sigma}
\label{eq:pdf}
\eqe
The best-fitting PDF model is shown superimposed on the data sample histogram in the left panel of Fig.~\ref{fig:ngc4051_simul_hist} (grey line) with best-fitting model parameters of $\vec{\eta}_{\rm bf}=\{5.67^{+0.04}_{-0.03}, 5.96^{+0.06}_{-0.04}, 2.14\pm0.06, 0.31^{+0.05}_{-0.04}, 0.82^{+0.05}_{-0.04}\}$. Having as null hypothesis, $H_0$, that the data set is drawn from the derived best-fit model distribution and alternative hypothesis, $H_{\rm a}$, that it was not drawn from that distribution, the Anderson-Darling test \citep{anderson52} yields a statistic value of 0.34 corresponding to an $H_0$ probability of 0.89 which depicts the good representation of the data by the given model.\par

\begin{figure*}
\hspace{-0.8em}\includegraphics[width=2.96in]{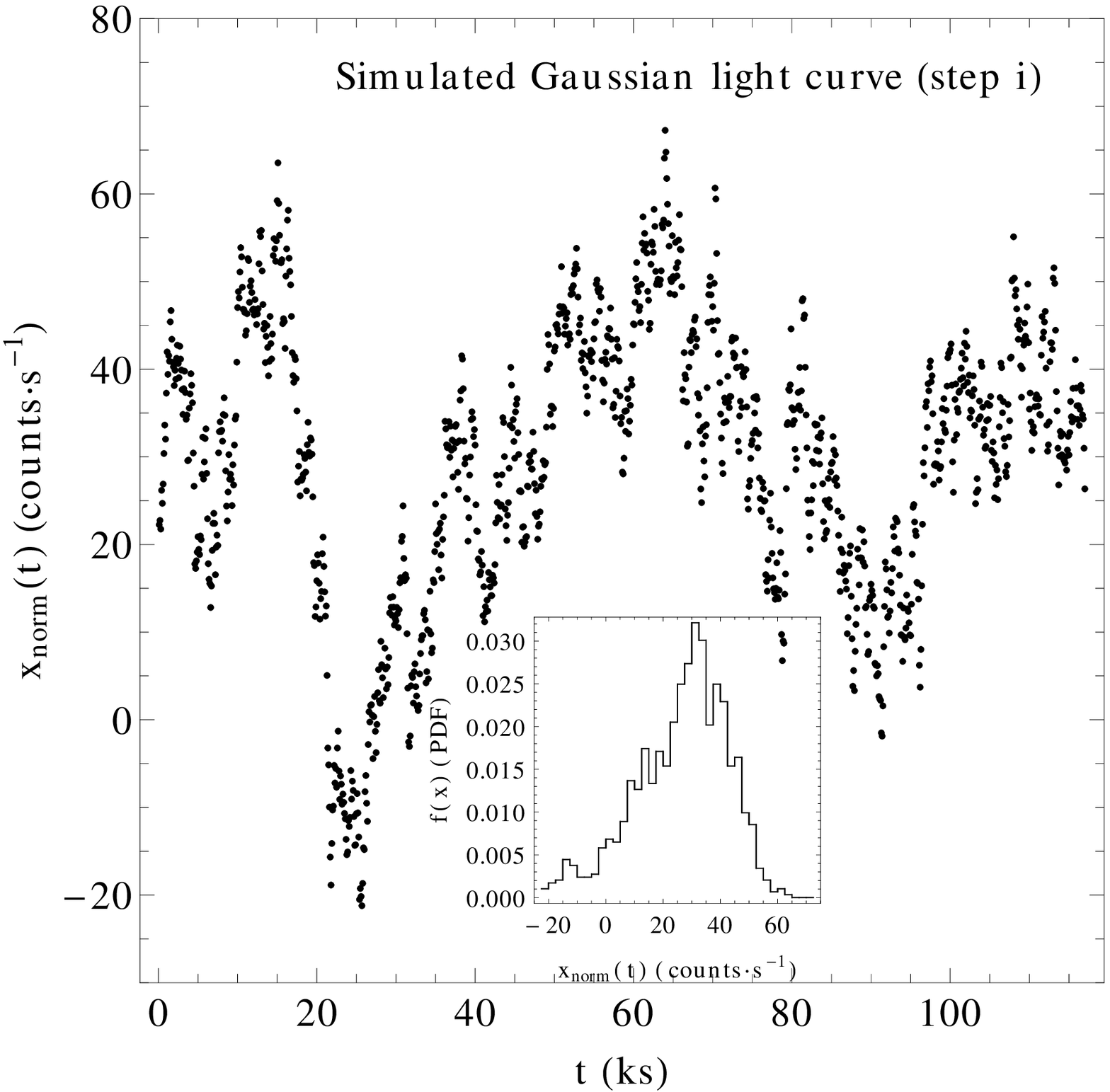}
\includegraphics[width=3.1in]{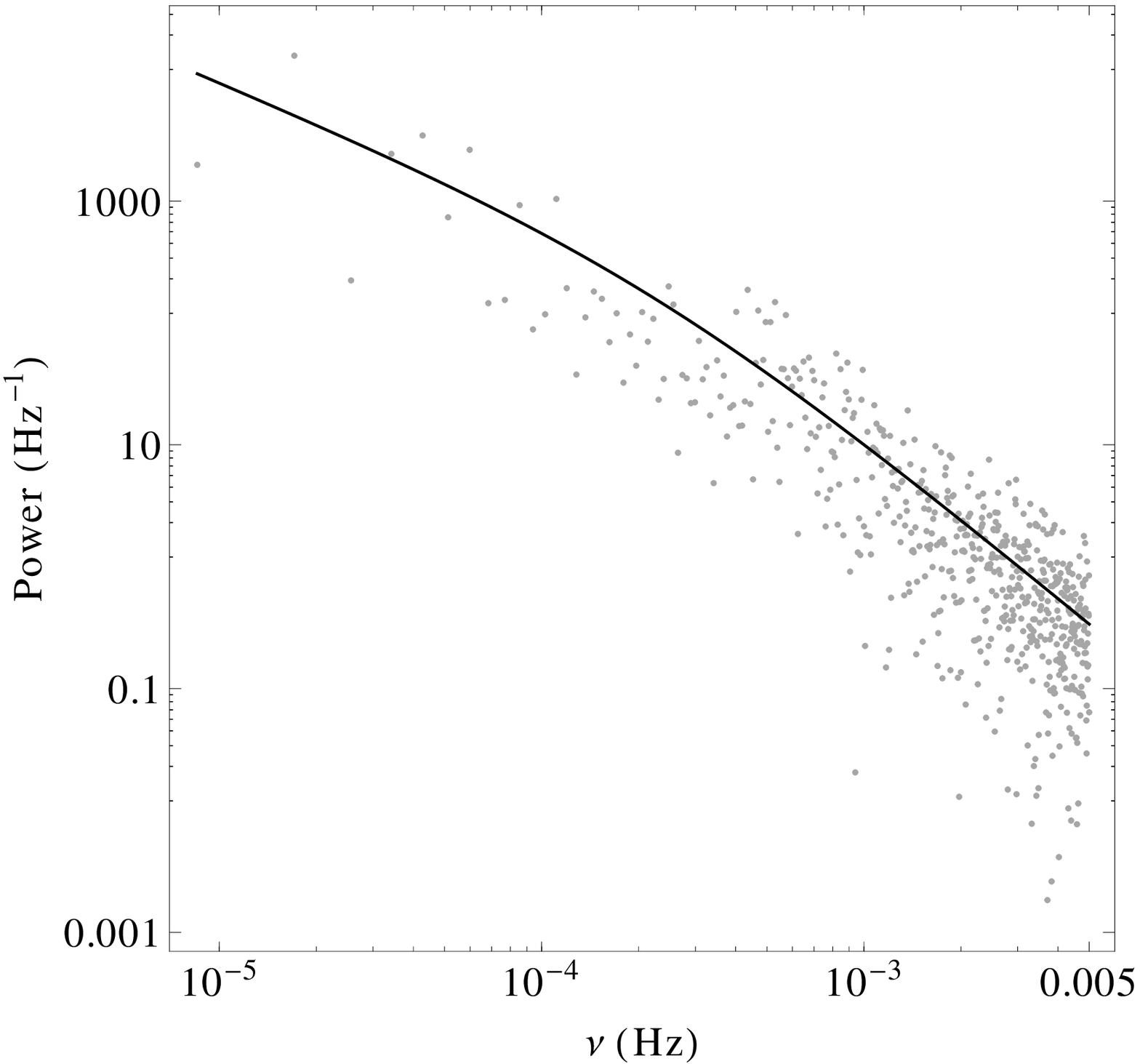}
\caption{Step i. [Left panel] A normally distributed simulated light curve created using the original data \n4's best-fitting PSD $(c=0)$ and its PDF histogram (inset). [Right panel] The corresponding periodogram estimates (grey points) and the underlying target PSD model, $\mathscr{P}(f;\vec{\gamma}_{\rm bf},0)$ (black line).}
\label{fig:artif_lc_prdgr}
\end{figure*}

\begin{figure*}
\hspace*{-4em}\includegraphics[width=2.96in]{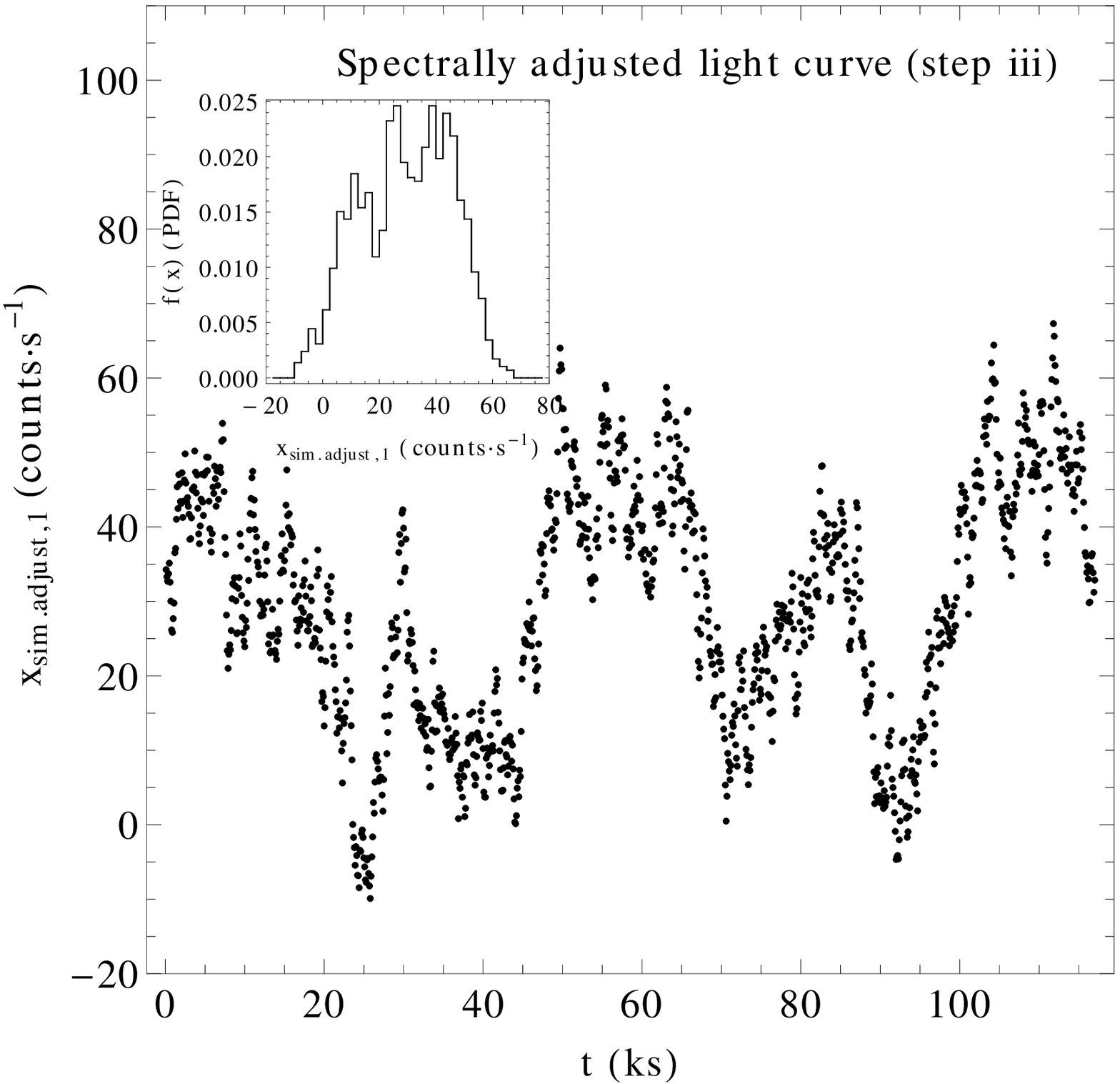}\hspace{4em}
\includegraphics[width=3.1in]{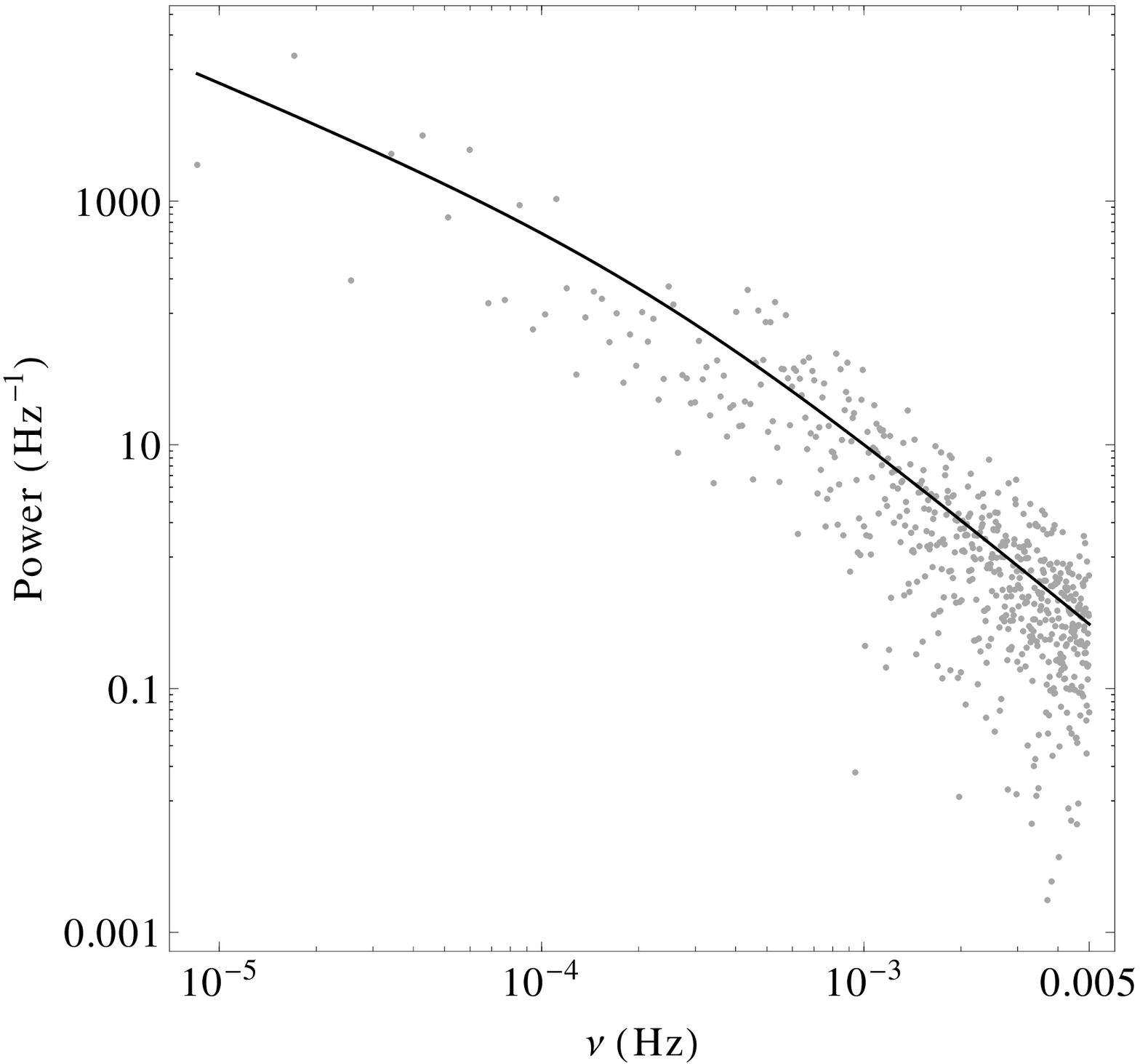}
\caption{Step iii. [Left panel] The spectrally adjusted light curve together with its PDF histogram (inset). [Right panel] The corresponding periodogram estimates (grey points), are by construction identical to those shown in the right panel of Fig.~\ref{fig:artif_lc_prdgr}, and the underlying target PSD, $\mathscr{P}(f;\vec{\gamma}_{\rm bf},0)$ (black line).}
\label{fig:sim1_lc_prdgr}
\end{figure*}
\begin{figure*}
\parbox{1\linewidth}{\includegraphics[width=2.96in]{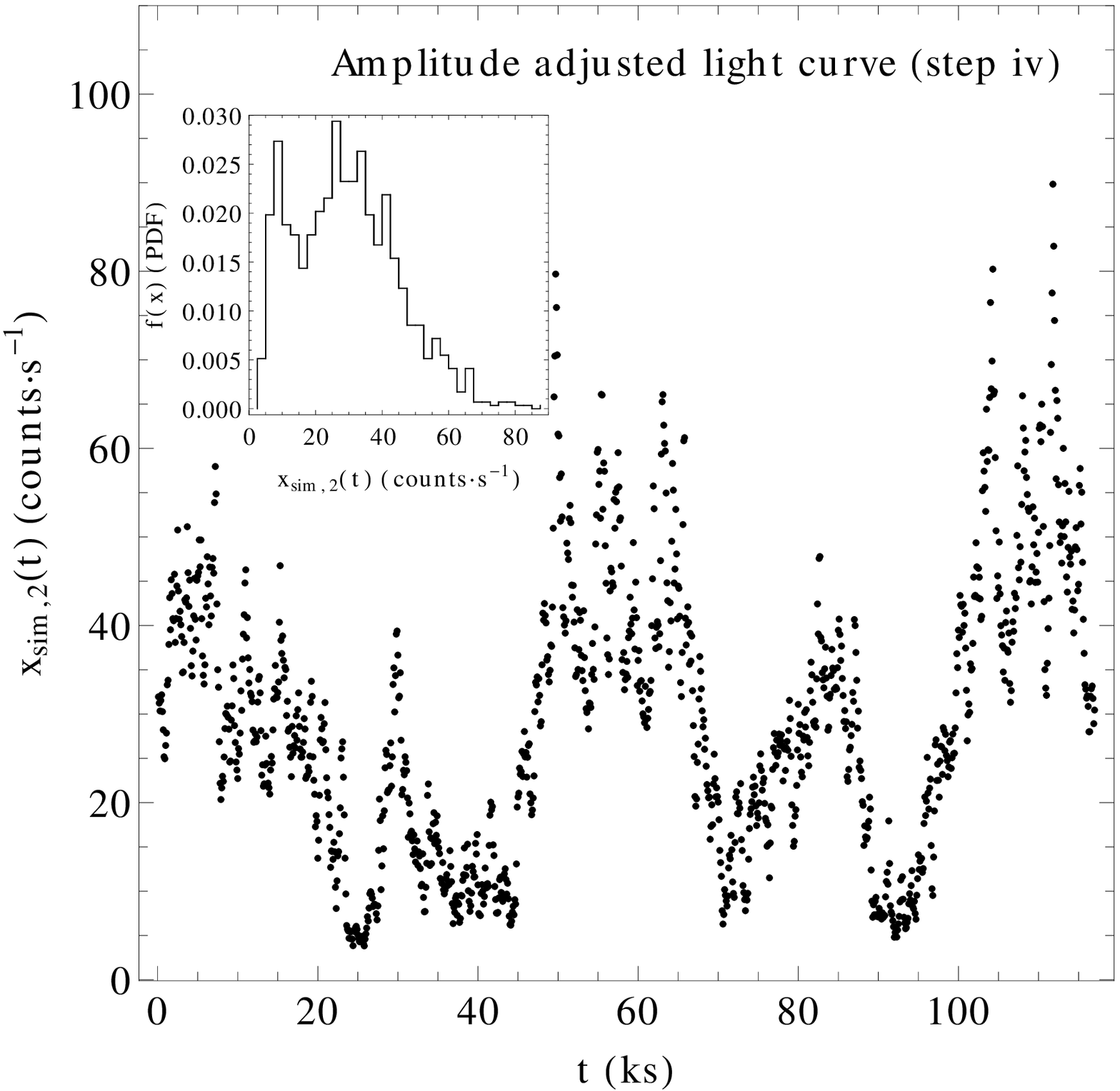}\vspace{-27.8em}}
\parbox{1\linewidth}{\hspace{29em}\parbox{1\linewidth}{\includegraphics[width=3.20in]{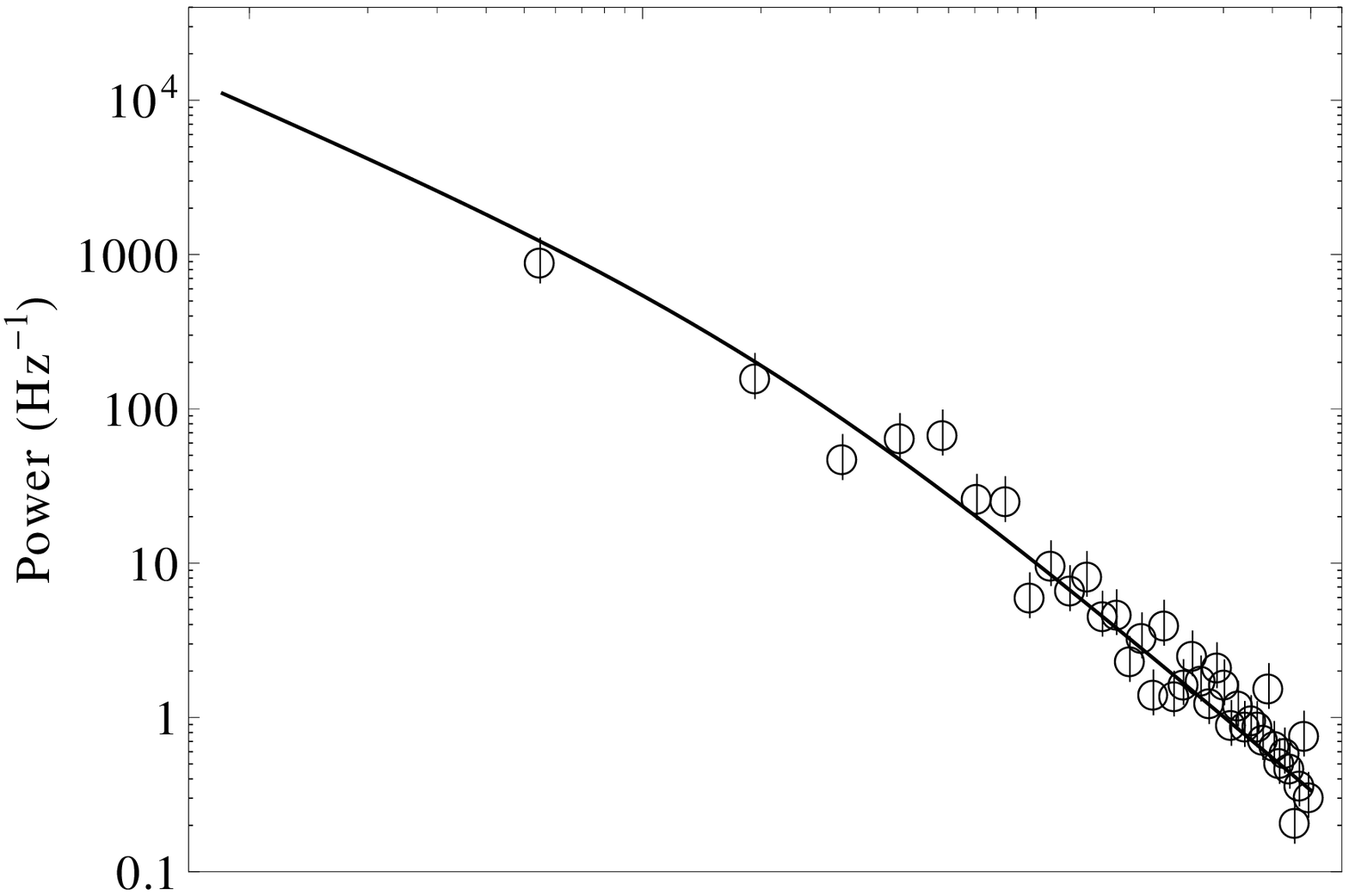}\\[-1.92em]
\hspace*{1.74em}\includegraphics[width=3.04in]{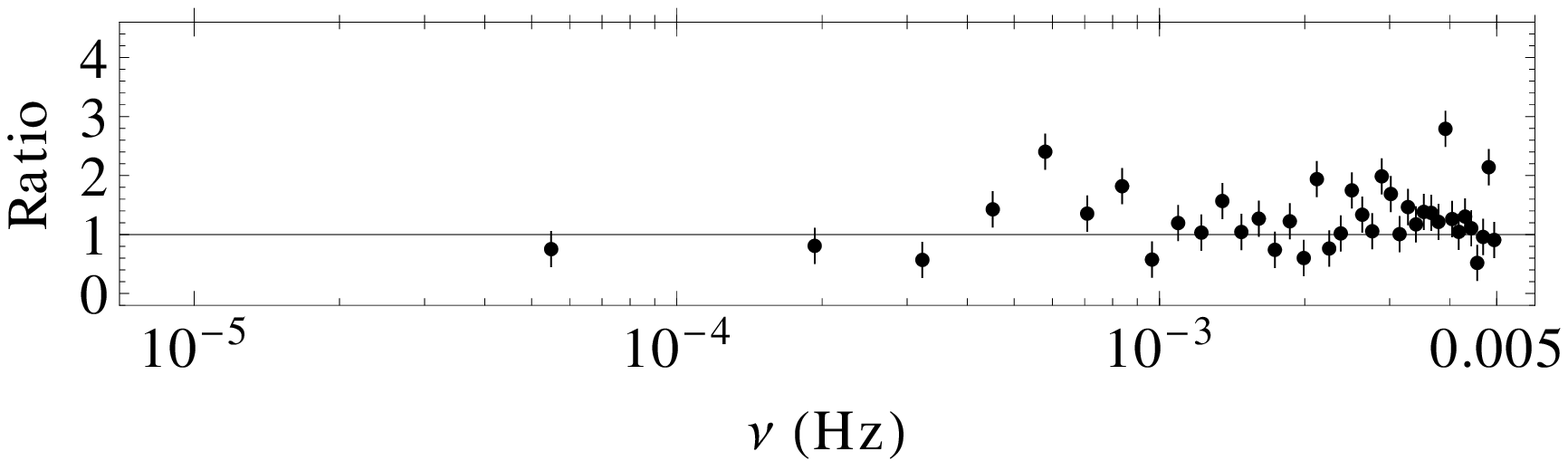}}}\\
\caption{Step iv. [Left panel] The amplitude adjusted light curve together with its PDF histogram (inset). [Right panel] The corresponding binned logarithmic periodogram estimates (open circles) and the underlying target PSD model, $\mathscr{P}(f;\vec{\gamma}_{\rm bf},0)$ (black line) having attached in the bottom the ratio plot, data/model.}
\label{fig:sim1_adjust_lc_prdgr}
\end{figure*}
Having defined the best-fitting PSD and PDF we continue to the actual production of the artificial light curves:
\begin{enumerate}
\item The best-fitting PSD model with $c=0$ is then used to create the normally distributed time series with a periodogram, $P_{\rm norm}(f_j)$. A simulated light curve of this kind together with its periodogram is shown in Fig.~\ref{fig:artif_lc_prdgr} (left and right panel, respectively). By estimating the DFT of $x_{\rm norm}(t)$, for each Fourier frequency, $f_j$, we derive the corresponding amplitudes and phases, $\mathscr{A}_{\rm norm}(j)$ and $\phi_{\rm norm}(j)$ respectively.\par
\item The best-fitting PDF model is used to generate a list of $N$ pseudo-random variates, $x_{\rm sim,1}(t)$, shown in the right panel of Fig.~\ref{fig:ngc4051_simul_hist}. Then, by estimating for each Fourier frequency the DFT of $x_{\rm sim,1}(t)$, $DFT_{\rm sim,1}(j)$, we derive the corresponding amplitude and phases, $\mathscr{A}_{\rm sim,1}(j)$ and $\phi_{\rm sim,1}(j)$, respectively.
\item \underline{Spectral adjustments}: $\mathscr{A}_{\rm sim,1}(j)$ are replaced by $\mathscr{A}_{\rm norm}(j)$, keeping the $\phi_{\rm sim,1}(j)$ unaltered, yielding an adjusted version of $DFT_{\rm sim,1}(j)$, $DFT_{\rm sim.adjust,1}(j)$. By performing an IDFT of we obtain the light curve $x_{\rm sim.adjust,1}(t)$ (Fig.~\ref{fig:sim1_lc_prdgr}, left panel) with an identical periodogram to $P_{\rm norm}(f)$ (Fig.~\ref{fig:sim1_lc_prdgr}, right panel, grey points), but now with measurements which are not longer distributed as $\mathfrak{f}_{\rm mix}(x;\vec{\eta}_{\rm bf})$ (Fig.~\ref{fig:sim1_lc_prdgr}, left panel, inset).  
\item \underline{Amplitude adjustments}: Finally, the values of $x_{\rm sim.adjust,1}(t)$ are replaced by the values of $x_{\rm sim,1}(t)$, based on the ranking of the former. The resulting light curve, $x_{\rm sim,2}(t)$, (Fig.~\ref{fig:sim1_adjust_lc_prdgr}, left panel) has an identical histogram with the sample light curve, but this time the periodogram estimates do not correspond to the target underlying PSD, $\mathscr{P}(f;\vec{\gamma}_{\rm bf},0)$. The right panel of Fig.~\ref{fig:sim1_adjust_lc_prdgr} shows the binned logarithmic periodogram estimates (in bins of 15 consecutive periodogram estimates) of the amplitude adjusted light curve together with $\mathscr{P}(f;\vec{\gamma}_{\rm bf},0)$ and the ratio plot i.e.\ data/model. From the ratio plot it is obvious that particularly the high frequency periodogram estimates, in particular above $10^{-3}$ Hz, are systematically larger than the corresponding $\mathscr{P}(f;\vec{\gamma}_{\rm bf},0)$ values by a factor of 1.5.
\end{enumerate}
All the abovementioned procedure is a single iteration step of the method. Exactly the same process is repeated iteratively from step iii, by replacing the $x_{\rm sim,1}(t)$ with the amplitude adjusted light curve $x_{\rm sim,2}(t)$, the $x_{\rm sim,2}(t)$ with the $x_{\rm sim,3}(t)$ and so on. For this particular case, after the 55\textsuperscript{th} iteration the synthetic light curves remain the same.\par

\subsubsection{Convergence of a single artificial light curve}
\label{sssect:convergence_singleDS}
\begin{figure}
\includegraphics[width=3.1in]{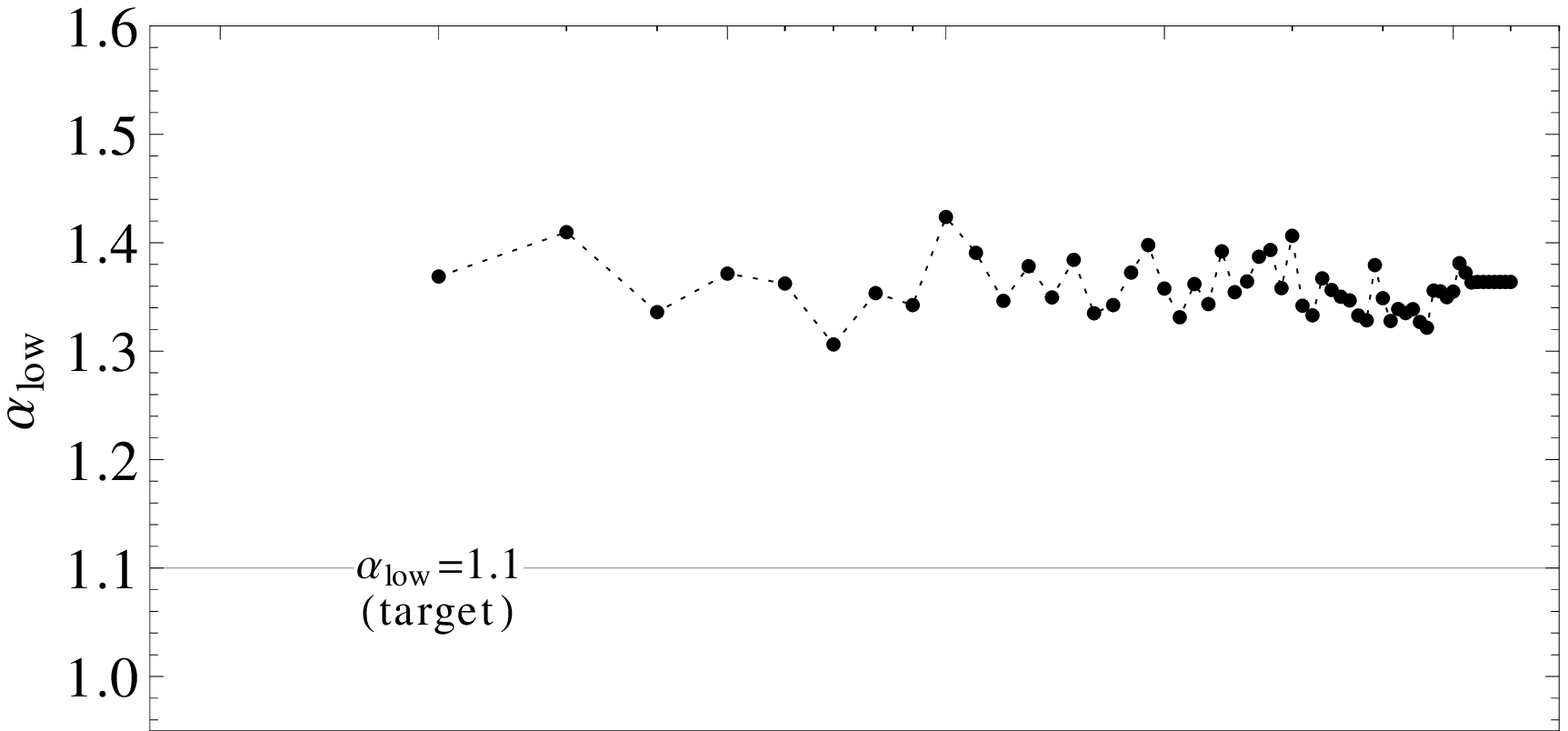}\\[-1.54em]
\hspace*{-0.8em}\includegraphics[width=3.195in]{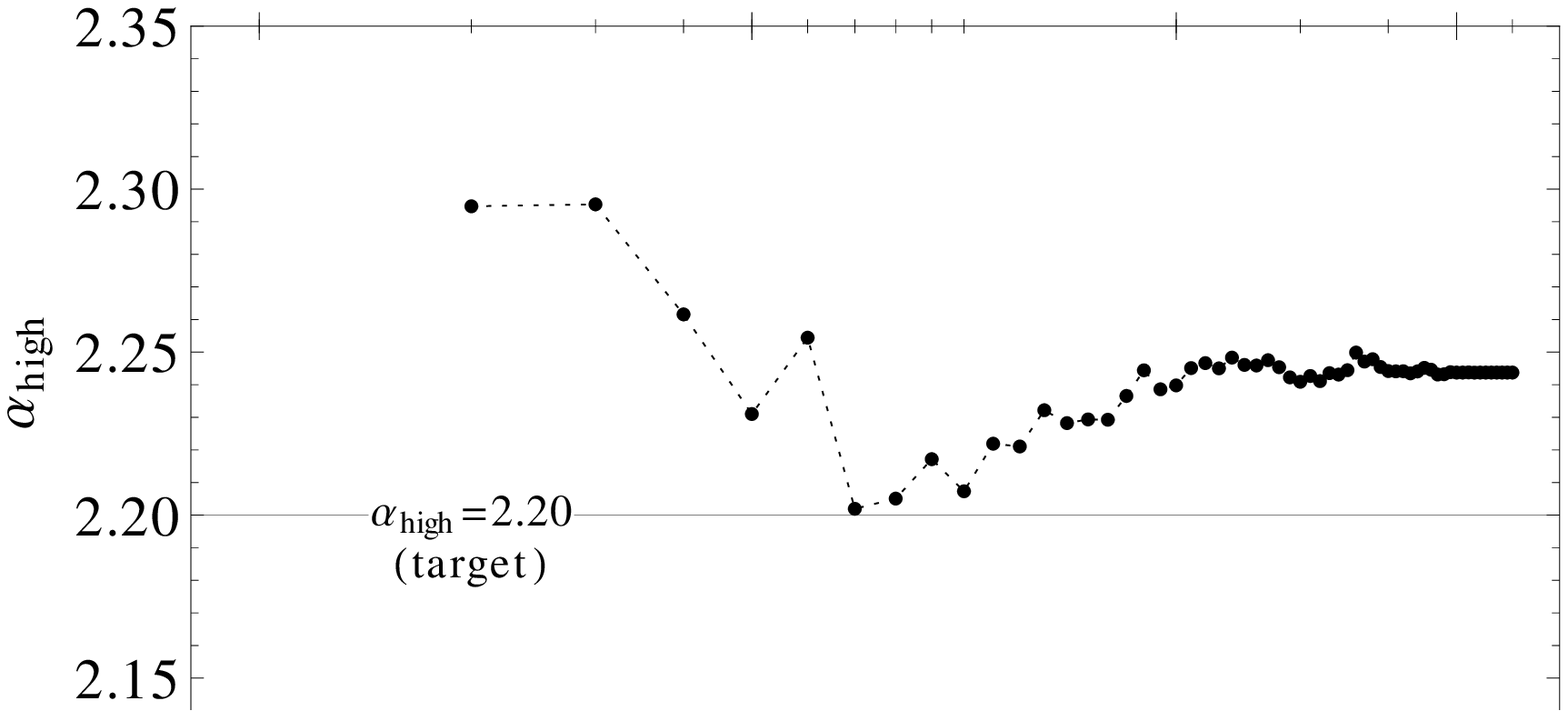}\\[-1.55em]
\hspace*{-0.97em}\includegraphics[width=3.216in]{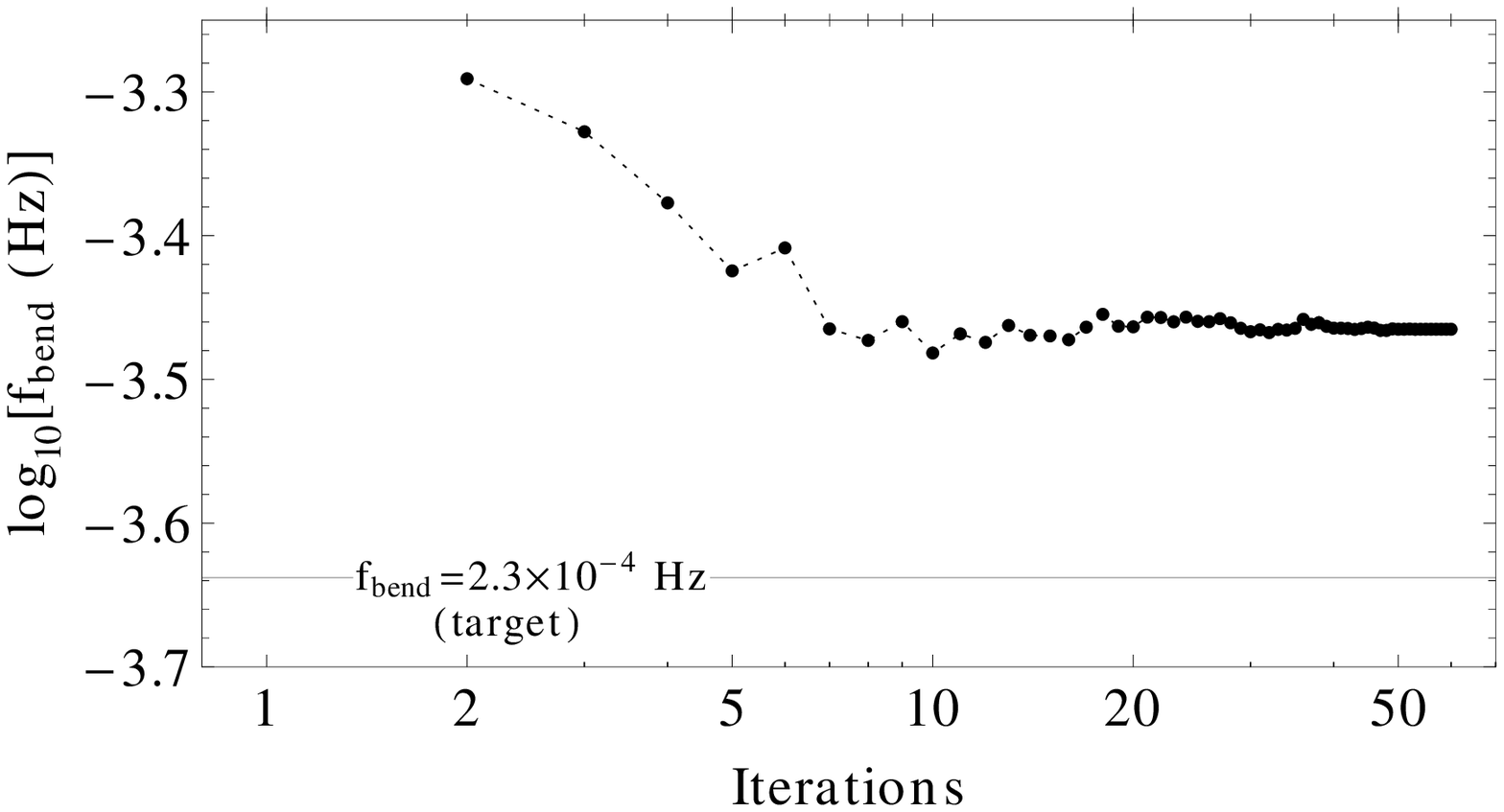}
\caption{Convergence of the iteration process (the first iteration is not shown -see footnote \ref{ftnt:omit1st}-). The horizontal solid grey lines indicate the corresponding target values i.e.\ the best fitting values as derived from the observed data. The dotted line among the various iterations is a linear interpolation intended only to guide the eye. [Top panel] The $\alpha_{\rm low}$ estimates as a function of the iteration number (in logarithmic scale), stabilising at 1.36. [Middle panel] The $\alpha_{\rm high}$ estimates as a function of the iteration number (in logarithmic scale), stabilising at 2.24. [Bottom panel] The logarithm of $f_{\rm bend}$ estimates as a function of the iteration number (in logarithmic scale), stabilising at -3.47.}
\label{fig:convergence}
\end{figure}
In order to check the convergence of the method, we fit the bending power-law model (equation \ref{eqe:psd_bendMod}) to the corresponding periodogram estimates of each iteration step. The resulting values for the $\alpha_{\rm low}$, $\alpha_{\rm high}$ and $f_{\rm bend}$ are shown in Fig.~\ref{fig:convergence} \footnote{\label{ftnt:omit1st}The results of the first iteration are excluded from the panels in order to cover better the variations of the other iterated products. The omitted values are $\alpha_{\rm low}=2.56$, $\alpha_{\rm high}=3.21$ and $\log_{10}[f_{\rm bend}\;{\rm (Hz)}]=-1.90$), respectively. These large deviations result from the fact that the minimisation routine does not localise the minimum (after 500 iteration steps) for the initial periodogram estimates, $P_{\rm sim,1}(f)$, corresponding to a flat PSD (see footnote \ref{ftnt:flatPSD}). Naturally, by increasing the number of iterations we can correct for this artefact but in this context it is unnecessary since for the next steps the 
localisation of the minimum occurs in less than 30 iterations since the degeneracy $\alpha_{\rm high}=\alpha_{\rm low}\simeq0$ does not exist any more.} which we can see form a plateau after the 55\textsuperscript{th} iteration. Fig.~\ref{fig:sim56_adjust_lc_prdgr} shows the corresponding results for the 56\textsuperscript{th} iteration; the synthetic light curve (left panel) follows the exact distribution of the observed data and the binned logarithmic periodogram estimates (in bins of 15 consecutive periodogram estimates) as expected follow, within the 68.3 percent confidence levels depicted by the error bars, the corresponding $\mathscr{P}(f;\vec{\gamma}_{\rm bf},0)$. This means that the 56\textsuperscript{th} synthetic light curve is the final simulated data set, with all the desired statistical and variability properties of the original data set. The best fitting PSD model for the 56\textsuperscript{th} surrogate yields: $\alpha_{\rm low}=1.36^{+0.22}_{-0.38}$, $\alpha_{\rm high}=2.24^{+0.08}_{-0.05}$ 
and 
$f_{\rm bend}=3.4^{+1.1}_{-1.3}\times10^{-4}$ Hz and the resulting histogram is by construction identical to the original one since it is drawn from its best fitting PDF model (Fig.~\ref{fig:ngc4051_simul_hist}, Right panel).\par
\begin{figure}
\includegraphics[width=3.20in]{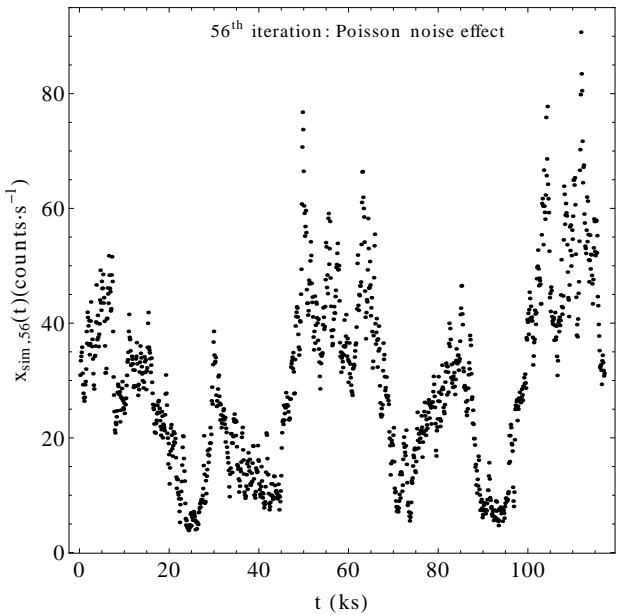}
\caption{The 56\textsuperscript{th} surrogate data set re-sampled from a Poisson distribution as dictated by equation \ref{eq:poisson_variate} ($\Delta t=100$ s).}
\label{fig:poissEffect}
\end{figure}

Finally, in order to take into account the Poisson statistics (Sect.\ref{ssect:poisson}) we re-sample the 56\textsuperscript{th} surrogate data set according to equation \ref{eq:poisson_variate}. The resulting artificial light curve is shown in Fig.~\ref{fig:poissEffect}. This single random synthetic data set encloses all the information of our initial data set and thus can be used in any sort of statistical study.

\subsection{Overall procedure for an ensemble of realisations}
\label{ssect:overall_ngc4051}
\subsubsection{Proposed methodology}
In this section we repeat the abovementioned procedure for an ensemble of 1000 realisations and we compare the statistical properties of the final products to those of the original data set of \n4 i.e.\ the light curve and underlying PSD (Fig.~\ref{fig:ngc4051_lc_psd}). Initially, we perform a goodness-of-fit Kolmogorov-Smirnov hypothesis test \citep{press92a} for the distribution of each artificial light curve, with $H_0$ that the surrogate data set is drawn from the best fit model distribution of \n4 (Fig.~\ref{fig:ngc4051_simul_hist}, left panel) and $H_{\rm a}$ that it was not drawn from that distribution. The mean Kolmogorov-Smirnov statistic derived from the ensemble of light curves is $D_n=0.025^{+0.008}_{-0.006}$ and the mean $H_0$ probability derived is $0.51^{+0.28}_{-0.22}$, depicting the high degree of accordance between the distribution of the artificial data sets and that of the original data set (the error estimates correspond to the 68 per cent confidence intervals). Thus, this method assures 
that the resulting simulated data sets have the same statistical moments as the observed light curve of \n4.\par
\begin{figure*}
\includegraphics[width=3in]{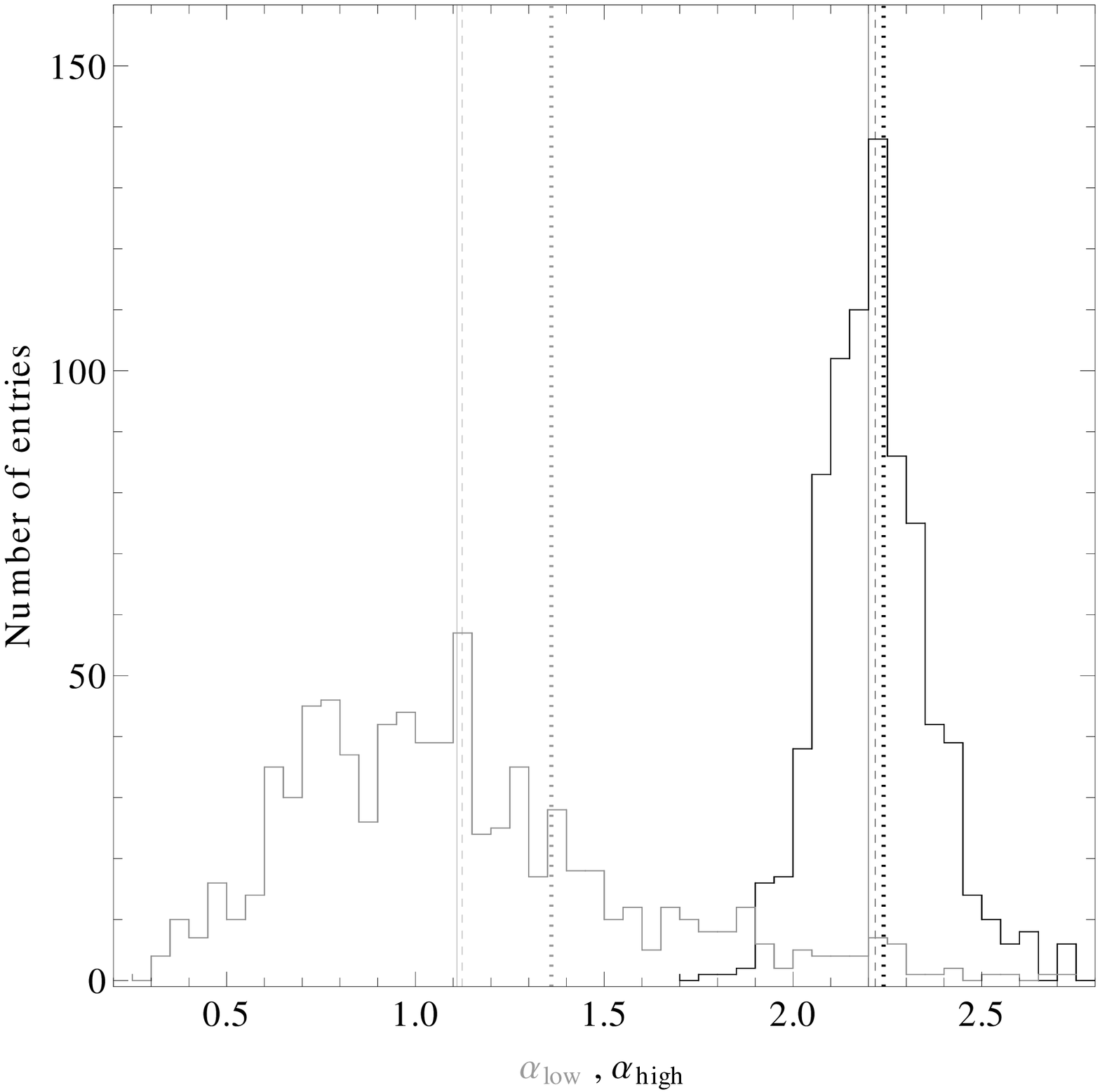}\hspace*{4em}
\includegraphics[width=3.135in]{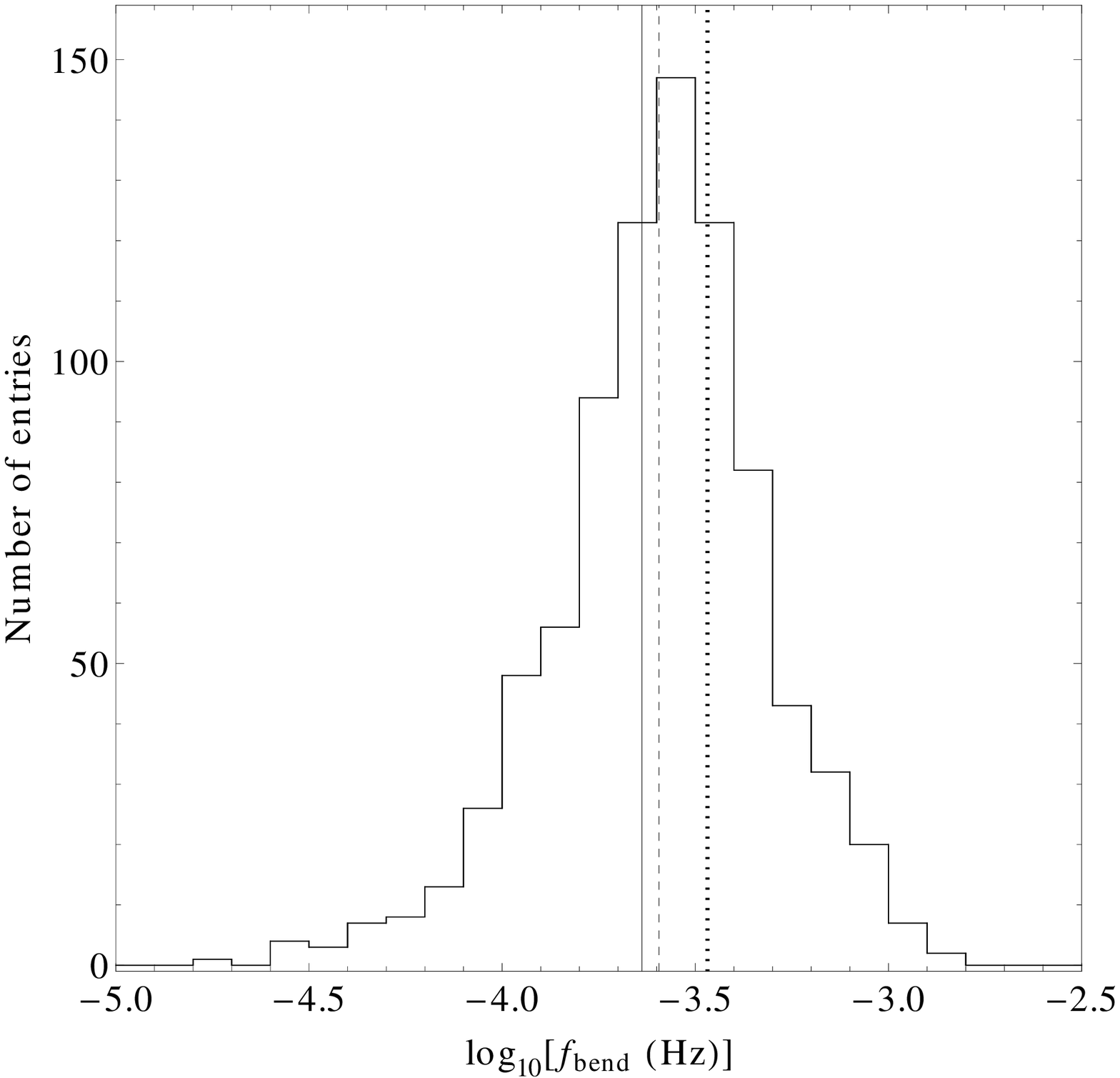}
\caption{Overall simulation results for 1000 artificial light curves. [Left panel] The histogram of the best fitting $\alpha_{\rm low}$ (grey lines) and $\alpha_{\rm high}$ (black lines). The solid lines corresponds to the target values coming directly from the observed data 1.1 and 2.2 respectively, the dashed line corresponds to the mean estimate of the distribution, 1.12 and 2.21 respectively, and the dotted line corresponds to best fitting value as derived from the 56\textsuperscript{th} surrogate of the single realisation, 1.36 and 2.24 respectively (Fig.~\ref{fig:convergence}, top and middle panel, respectively). [Right panel] The histogram of the best fitting logarithms of $f_{\rm bend}$. The solid line corresponds to the target value coming directly from the observed data  $f_{\rm bend}=2.3\times10^{-4}$ Hz (or -3.64 in $\log$ scale with base 10), the dashed line corresponds to the mean estimate of the distribution, $2.4\times10^{-4}$ Hz (or -3.62 in $\log$ scale with base 10), and the dotted line 
corresponds to best fitting value as derived from the 56\textsuperscript{th} surrogate of the single realisation, $3.4\times10^{-4}$ Hz (or -3.47 in $\log$ scale with base 10, Fig.~\ref{fig:convergence}, bottom panel).}
\label{fig:sims_PSD_results}
\end{figure*}

We then fit the PSD model of equation \ref{eqe:psd_bendMod} to the periodogram estimates of each artificial light curve. The distributions of both the low and the high frequency PSD slopes, as well as the the bending frequency, are shown in the left and right panels of Fig.~\ref{fig:sims_PSD_results}, respectively. The sample mean values, together with their 68.3 per cent confidence limits (i.e.\ standard deviation of the sample mean) and the 68.3 per cent confidence intervals of the distributions for $\alpha_{\rm low}$, $\alpha_{\rm high}$ and $f_{\rm bend}$, are given in Table~\ref{tab:sim_res}. The simulation results which come from the proposed method, are entirely consistent with those derived from the original data set of \n4, indicating that there are no biases towards the PSD model parameters which could cause systematic deviations from the targeted values. Thus, the artificial light curves produced as an ensemble with this algorithm have the same variability power, as a function of Fourier frequency,
 as that of \n4.

\begin{figure*}
\parbox{1\linewidth}{\includegraphics[width=2.96in]{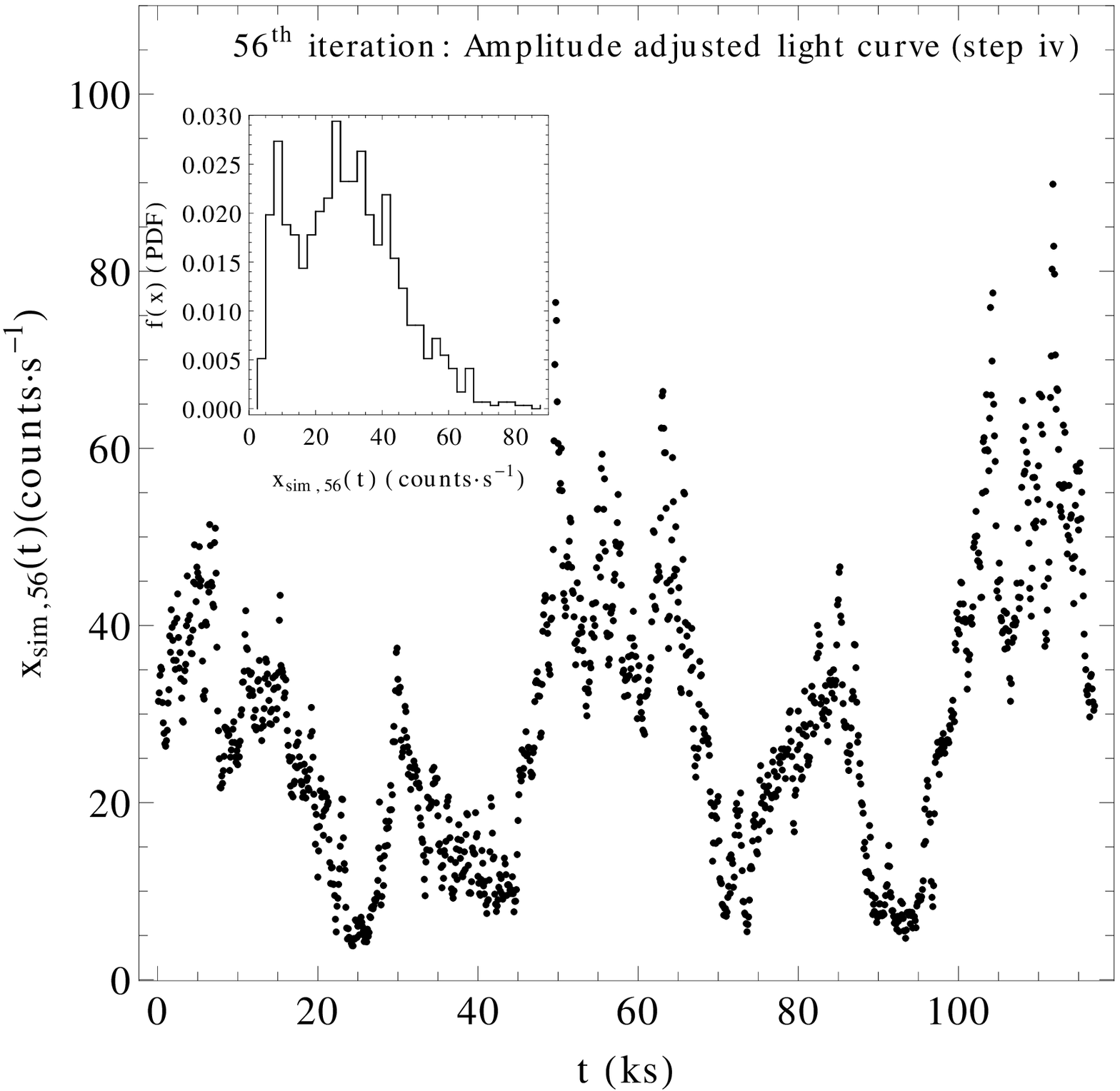}\vspace{-27.8em}}
\parbox{1\linewidth}{\hspace{29em}\parbox{1\linewidth}{\includegraphics[width=3.20in]{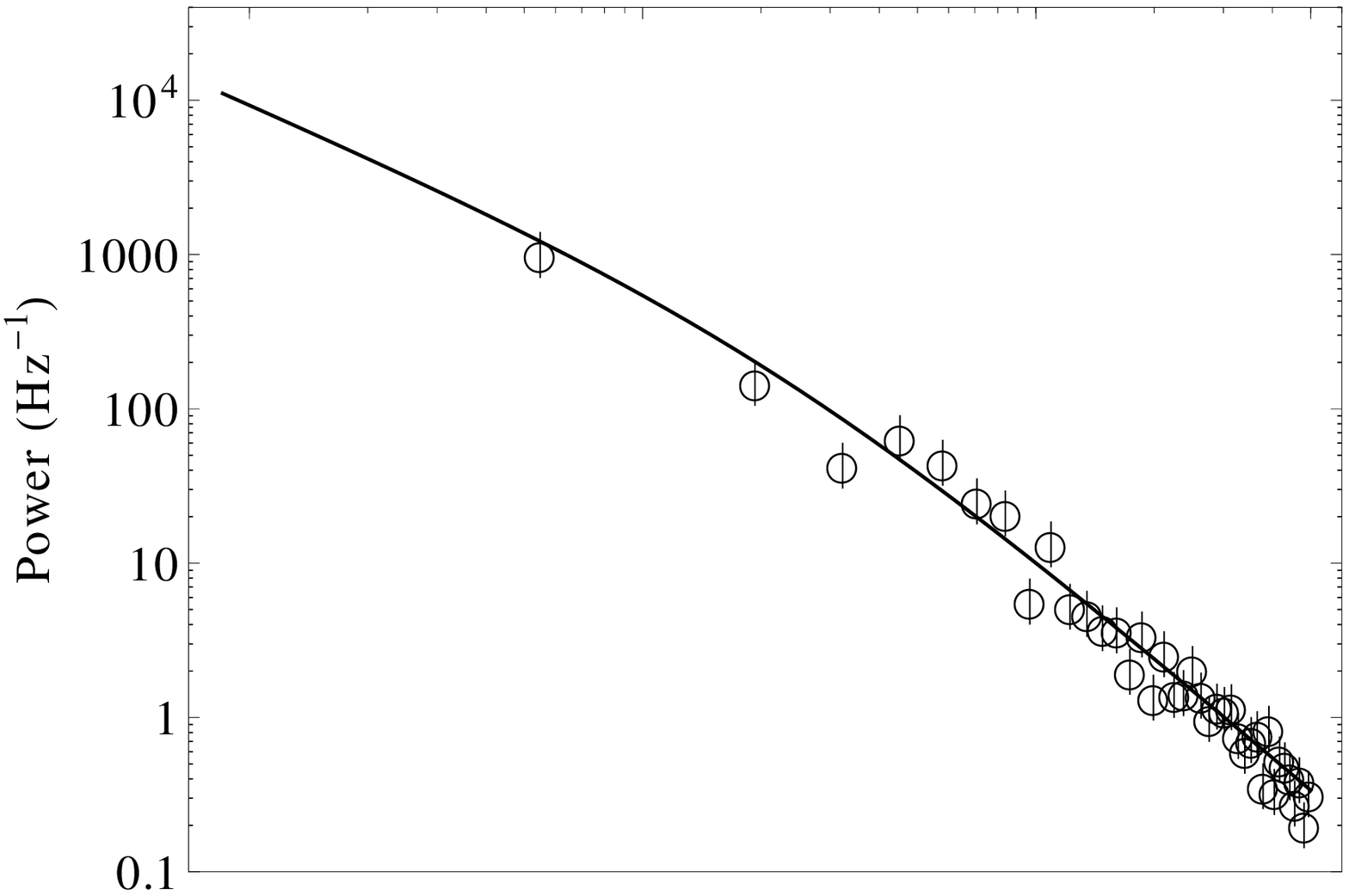}\\[-1.92em]
\hspace*{1.74em}\includegraphics[width=3.04in]{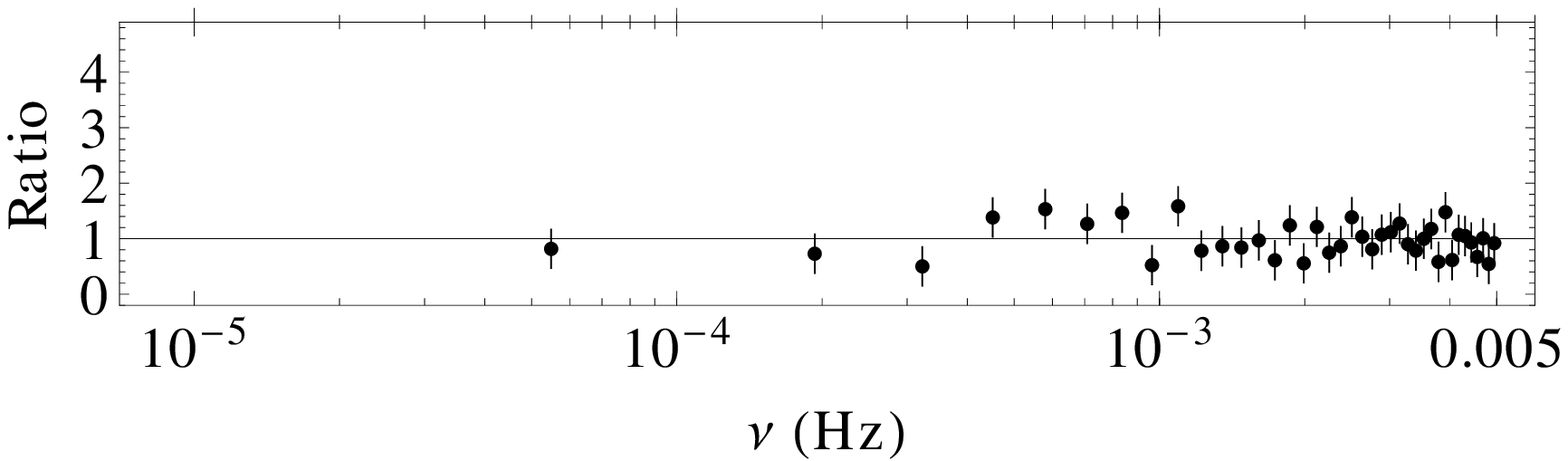}}}\\
\caption{Step iv for the 56\textsuperscript{th} iteration. [Left panel] The amplitude adjusted light curve together with its PDF histogram (inset). [Right panel] The corresponding binned logarithmic periodogram estimates (open circles) and the underlying target PSD model, $\mathscr{P}(f;\vec{\gamma}_{\rm bf},0)$ (black line) having attached in the bottom the corresponding ratio plot, data/model.}
\label{fig:sim56_adjust_lc_prdgr}
\end{figure*}

\subsubsection{Convergence of the ensemble of artificial light curves}
\label{sssect:convergence_ensembleDS}
\begin{figure*}
\hspace{-2em}\includegraphics[width=3.1in]{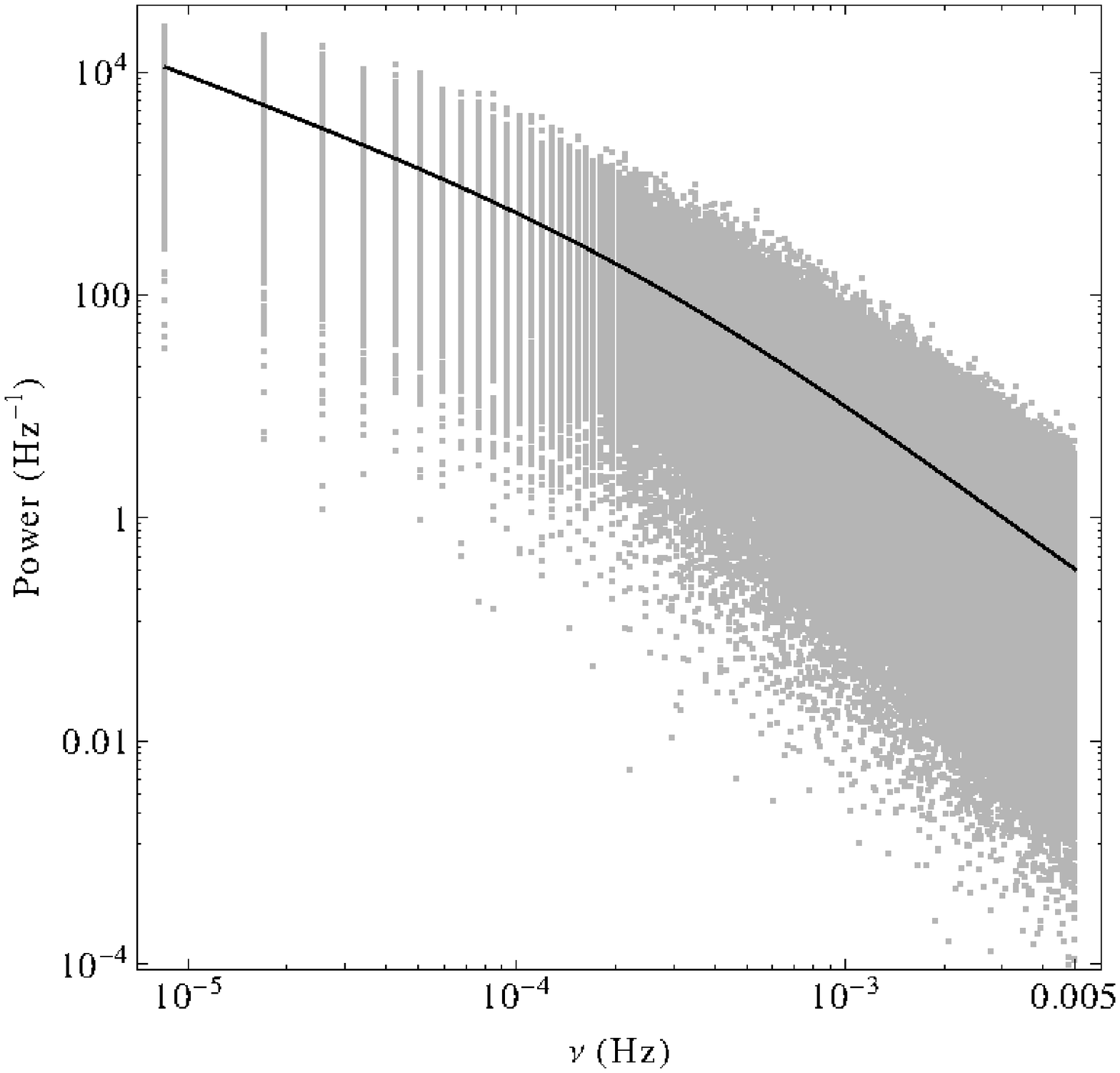}\hspace{4em}
\includegraphics[width=3.01in]{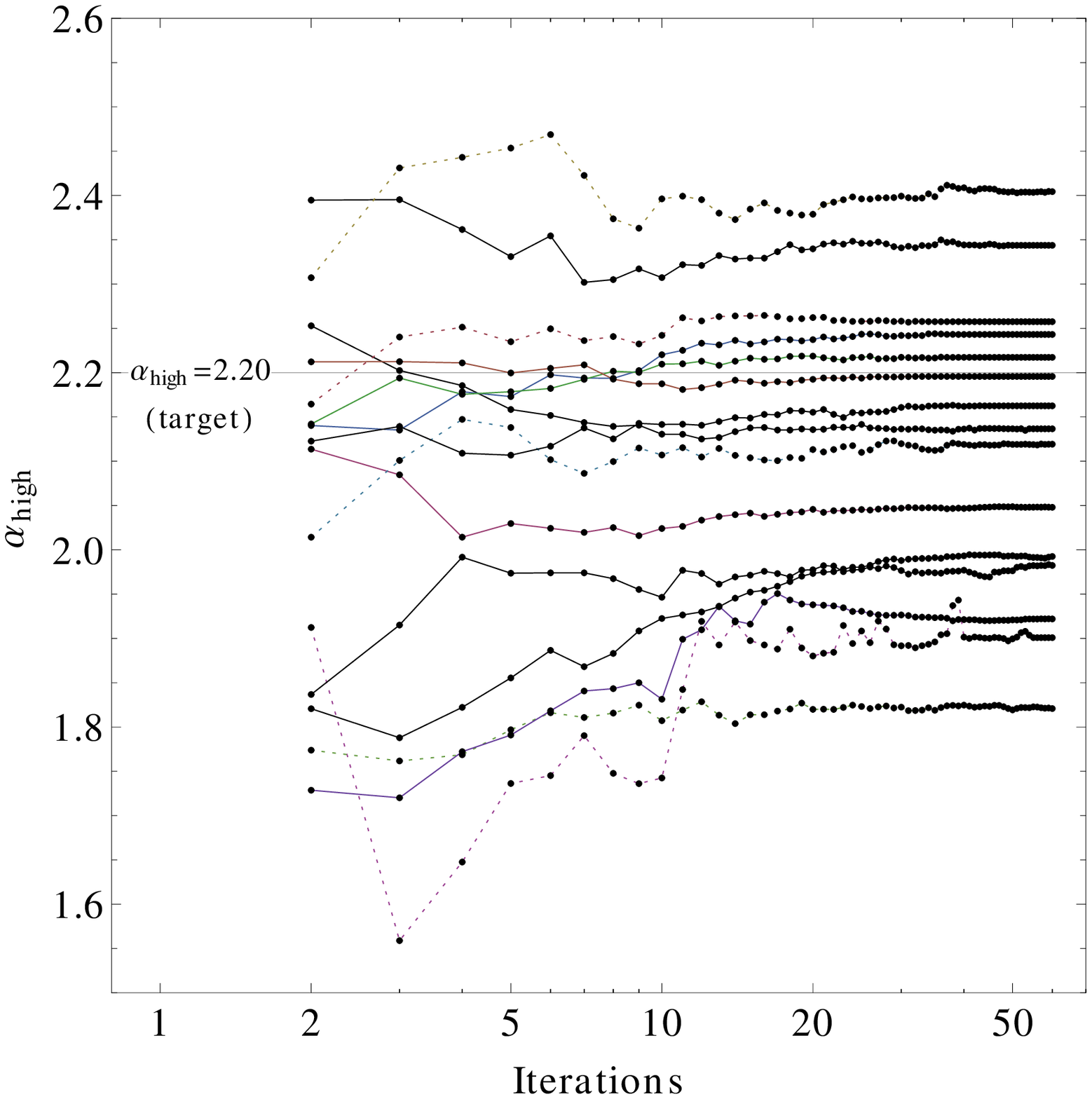}
\caption{Distribution and convergence of the ensemble periodogram estimates. [Left panel] The distribution of the 1000 periodograms (grey points), originating from the 1000 synthetic light curves, around the underlying PSD, $\mathscr{P}(f;\vec{\gamma}_{\rm bf},c_{\rm bf})$, (black, solid line). Note that the synthetic light curves contain Poisson noise. [Right panel] Convergence of the PSD parameter $\alpha_{\rm high}$ as estimated from fitting the periodogram estimates of the 15 synthetic data sets for all the iteration steps.}
\label{fig:simsDist_conv}
\end{figure*}

The scattering in the various estimated PSD model parameters, coming from the 1000 simulated light curves, originates from the asymptotic distribution of the various periodogram estimates, $P(f_j)$, around the input input PSD, $\mathscr{P}(f;\vec{\gamma}_{\rm bf},0)$. As we discussed in Sect.~\ref{ssect:diff_advant} (see for details Appendix~\ref{ssect:psd}) at given Fourier frequency $f_j$, $P(f_j)$ is distributed asymptotically around $\mathscr{P}(f;\vec{\gamma}_{\rm bf},0)$ as a gamma distribution, $\Gamma\left[\nu/2, \mathscr{P}(f;\vec{\gamma}_{\rm bf},0)\right]$ with $\nu$ d.o.f. This behaviour is depicted in the left panel of Fig.~\ref{fig:simsDist_conv} which shows the distribution of the 1000 periodograms around the target PSD. As a sanity check for our simulations, we test whether for a given Fourier frequency, $f_j$, the 
distribution of their various periodogram estimates is indeed in accordance with equation \ref{eqe:prdgrProb}. Thus, we derive for each $f_j$ the distribution of points and we perform an Anderson-Darling test goodness-of-fit test with $H_0$: the periodogram estimates at a given $f_j$ are drawn from the $\Gamma\left[\nu/2, \mathscr{P}(f;\vec{\gamma}_{\rm bf},0)\right]$ distribution and $H_{\rm a}$ that they are not drawn from this distribution. The mean value of the statistic is $2.92^{+0.24}_{-0.13}$ yielding a mean $H_0$ probability $0.18^{+0.09}_{-0.06}$ depicting the high degree of accordance between the estimated distribution of the simulated products and the expected ones (the error estimates correspond to the 68 per cent confidence intervals).\par
Finally, depending on the type of the statistical study, it is not necessary always for each surrogate data set to carry out the iteration process up to the convergence point (as shown in Fig.~\ref{fig:convergence}). Stopping the process in an intermediate step e.g.\ at the 5\textsuperscript{th} iteration step, will yield surrogate data sets which will still have accurate PSD parameters (i.e.\ they will be distributed correctly around the target values without systematic trends), but the various estimates will be less precise than those derived from the final converged products (i.e.\ they will exhibit larger scatter around the target values). Nevertheless, the differences are very small and for this particular example (5\textsuperscript{th} iteration step) are on average of the order of 5 per cent. In the right panel of Fig.~\ref{fig:simsDist_conv} we show this effect by plotting the convergence in the PSD parameter $\alpha_{\rm high}$ for 15 synthetic data sets (as we did in the top panel
of Fig.\ref{fig:convergence}). The publicly available {\sc mathematica} notebook (Sect.~\ref{ssect:avail_code}) contains an animation showing these small differences between all the iteration steps for a single surrogate. 

\begin{table*}
\begin{minipage}{190mm}
\caption{Global simulation results for the PSD parameters.}
\label{tab:sim_res}
\begin{tabular}{@{}cccc}
\hline
Model-parameter & Target values$^\dagger$ & Proposed method\textsuperscript{\textasteriskcentered} & Exponential function method\textsuperscript{\textasteriskcentered}\\
\hline
$\alpha_{\rm low}$ & 1.1 (fixed) & $1.123^{+0.002}_{-0.007}$, $\left[0.87,1.20\right]$ & $0.973^{+0.004}_{-0.006}$, $\left[0.86,1.03\right]$ \\
$\alpha_{\rm high}$ & $2.20^{+0.07}_{-0.04}$ & $2.213^{+0.002}_{-0.001}$, $\left[2.15,2.26\right]$ & $2.063\pm0.002$, $\left[2.00,2.10\right]$\\
$f_{\rm bend}$ ($\times10^{-4}$ Hz) & $2.3^{+1.2}_{-0.9}$ & $2.4\pm0.1$, $\left[2.1,3.3\right]$ & $3.9\pm0.1$, $\left[3.2,5.0\right]$\\
\hline
\end{tabular}
\medskip\\
$^\dagger$ These are the values of \n4 derived in Section\,\ref{ssect:single_ngc4051}.\\
\textsuperscript{\textasteriskcentered} The first value is the sample mean together with its 68.3 per cent confidence limits and the second value in the square brackets\newline corresponds to the 68.3 per cent confidence intervals of the distribution around the mean.
\end{minipage}
\end{table*}

\subsubsection{Exponential light curves}
\label{ssect:exp_lc}
In this section we follow the recipe of \citet{uttley05a} and exponentiate (in base $e$) the TK95 products which are produced by the renormalised PSD model \citep[using equations 13 and 14 in][]{uttley05a}. In this case, the artificial data sets always follow by construction a log-normal distribution which differs intrinsically from the observed statistical properties (described by equation~\ref{eq:pdf}) which we are interested in reproducing for the given illustrative purposes. Thus we do not need to perform a goodness-of-fit hypothesis test for the distributions, since we know {\it a priori} that they are by construction different.\par
\begin{figure*}
\hspace{-1.2em}\includegraphics[width=3in]{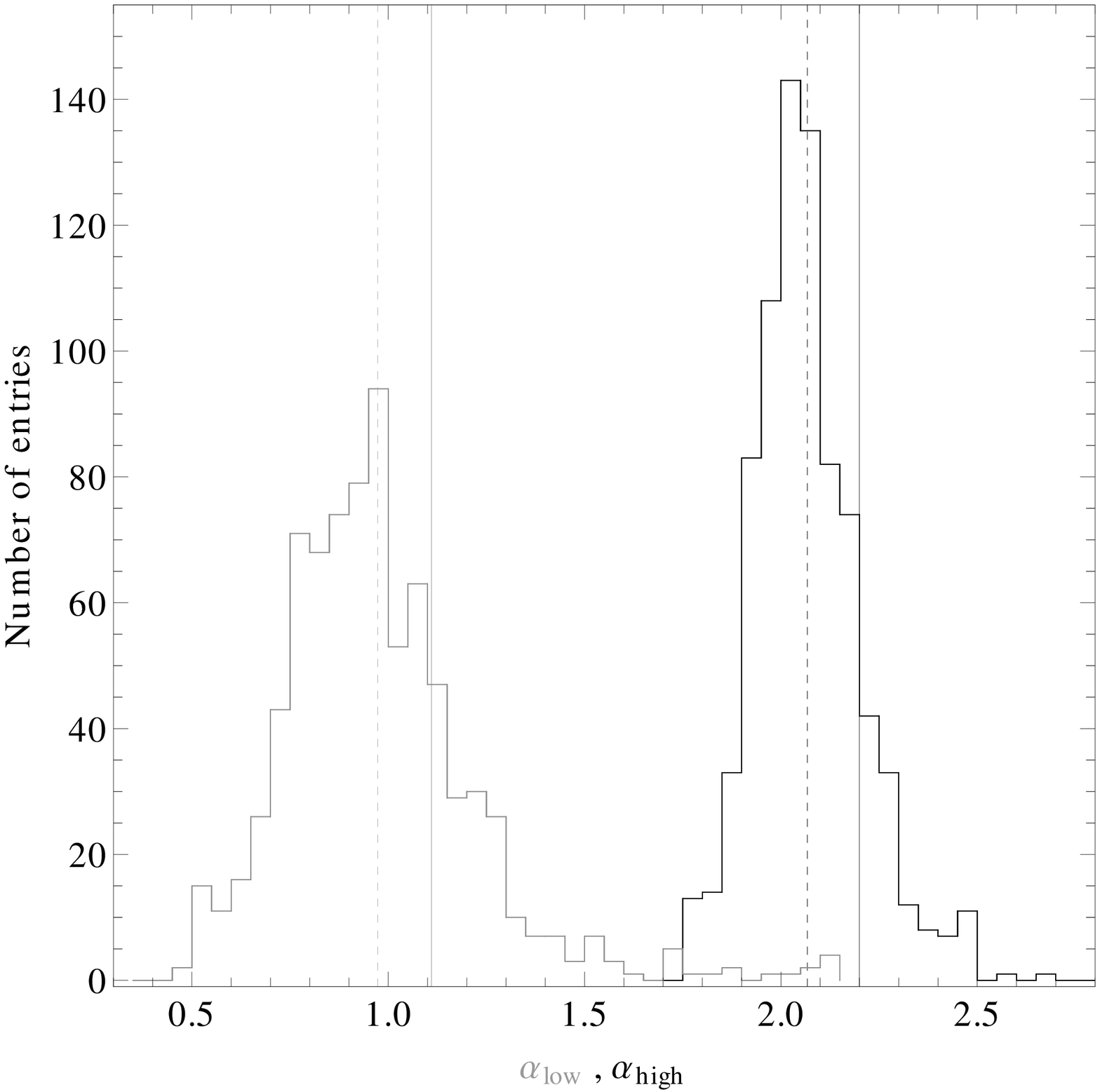}\hspace*{4.2em}
\includegraphics[width=3in]{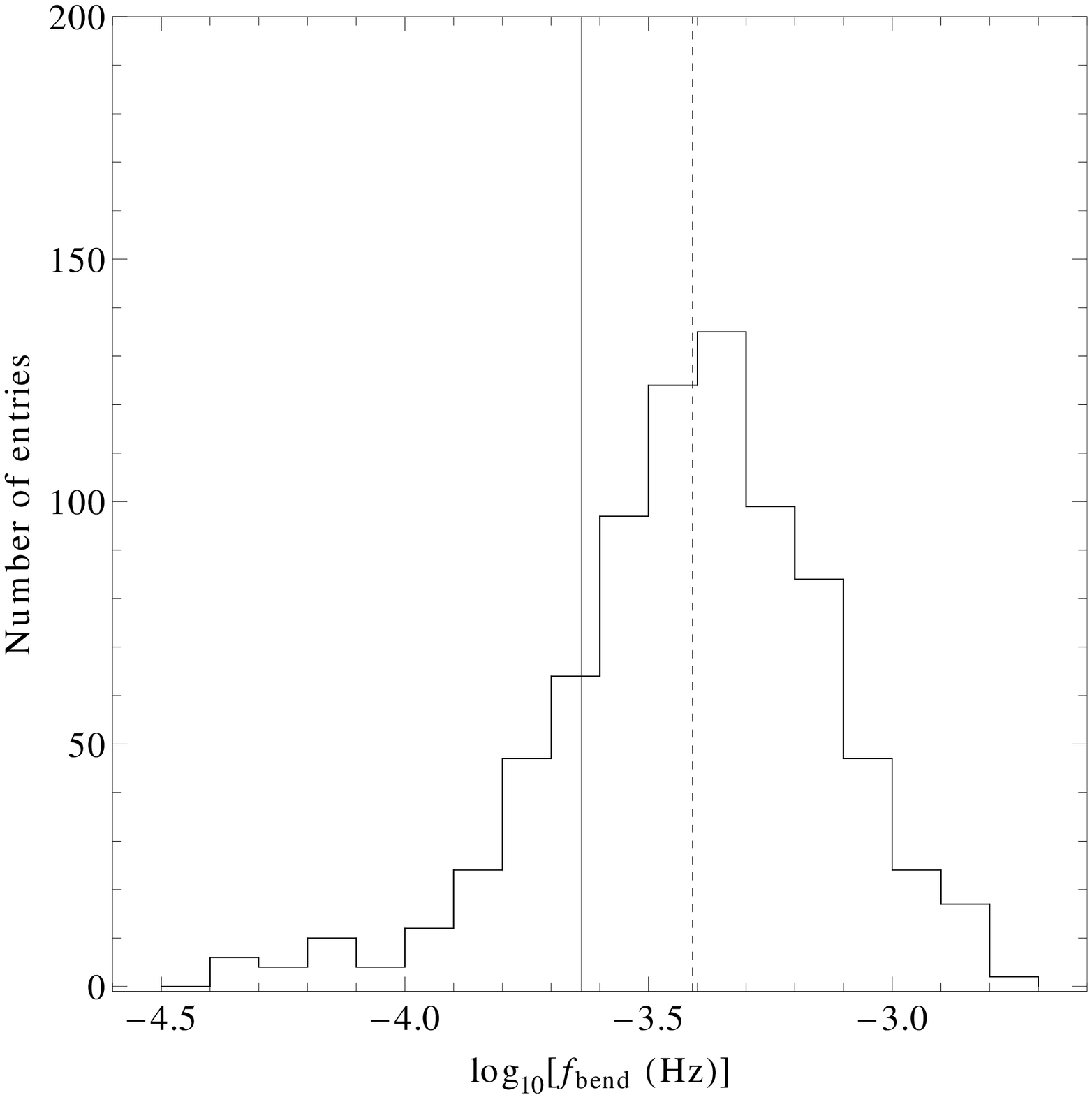}
\caption{Overall results for 1000 exponential light curves. [Left panel]  The histogram of the best fitting $\alpha_{\rm low}$ (grey lines) and $\alpha_{\rm high}$ (black lines). The solid lines corresponds to the target values coming directly from the observed data 1.1 and 2.2 respectively, the dashed line corresponds to the mean estimate of the distribution, 0.97 and 2.06 respectively. [Right panel] The histogram of the best fitting logarithms of $f_{\rm bend}$. The solid line corresponds to the target value coming directly from the observed data  $f_{\rm bend}=2.3\times10^{-4}$ Hz (or -3.64 in $\log$ scale with base 10), the dashed line corresponds to the mean estimate of the distribution, $3.9\times10^{-4}$ Hz (or -3.41 in $\log$ scale with base 10).}
\label{fig:exp_PSD_results}
\end{figure*}

In the next step, we repeat the PSD model fitting procedure for the periodogram estimates of the exponential light curves. The results for the distribution of the best fitting parameters of $\alpha_{\rm low}$, $\alpha_{\rm high}$ and $f_{\rm bend}$ are shown in Fig.~\ref{fig:exp_PSD_results}. Finally, as above, the sample mean values together with their 68.3 per cent confidence limits (i.e.\ the standard deviation of the sample mean) and the 68.3 per cent confidence intervals of the distributions for $\alpha_{\rm low}$, $\alpha_{\rm high}$ and $f_{\rm bend}$ are given in Table~\ref{tab:sim_res}. We can see that the PSD becomes systematically softer by around 8 per cent something which is also shown in fig.~B1 in \citet{uttley05a}. Most importantly for this particular case the most noticeable distortion appears in the bend frequency which systematically shifts towards higher frequencies, deviating in this way by 70 per cent from the target value.\par
Using these simulated data sets for the recovery of the bend frequency of an irregularly sampled light curve \citep[using the procedure of][]{uttley02} will yield systematic deviations from the true underlying value. Note that the degree of the various PSD distortions of the exponential light curves depend on the particular variability properties of the light curves, as well as the actual values of the underlying PSD model.\par
A potential solution to these spectral alterations could be the following: To consider the logarithm of the observed data set (which is Gaussian distributed for the case of a log-normal distribution) and estimate its PSD which is then going to be used as the input PSD for the TK95 simulation. The exponentially transformed TK95 products should follow the original PSD of the observed data set which is log-normally distributed. Before following this recipe, further investigation of this approach should be carried out, something which is out of the scope of this manuscript.

\section{COMPLIMENTARY APPLICATIONS: \textit{FERMI} AND \textit{RXTE} DATA SETS}
\label{sect:compl_applic}
In order to show the wide applicability of our newly proposed method, we further apply it on two radically different looking data sets: a $\gamma$-ray \textit{Fermi-LAT} data set for the blazar 3C\,454.3, an X-ray \textit{RXTE} data set for the XRB Cyg\,X-1.\par\noindent
\underline{The $\gamma$-ray blazar 3C\,454.3}\par
\begin{figure*}
\includegraphics[width=2.95in]{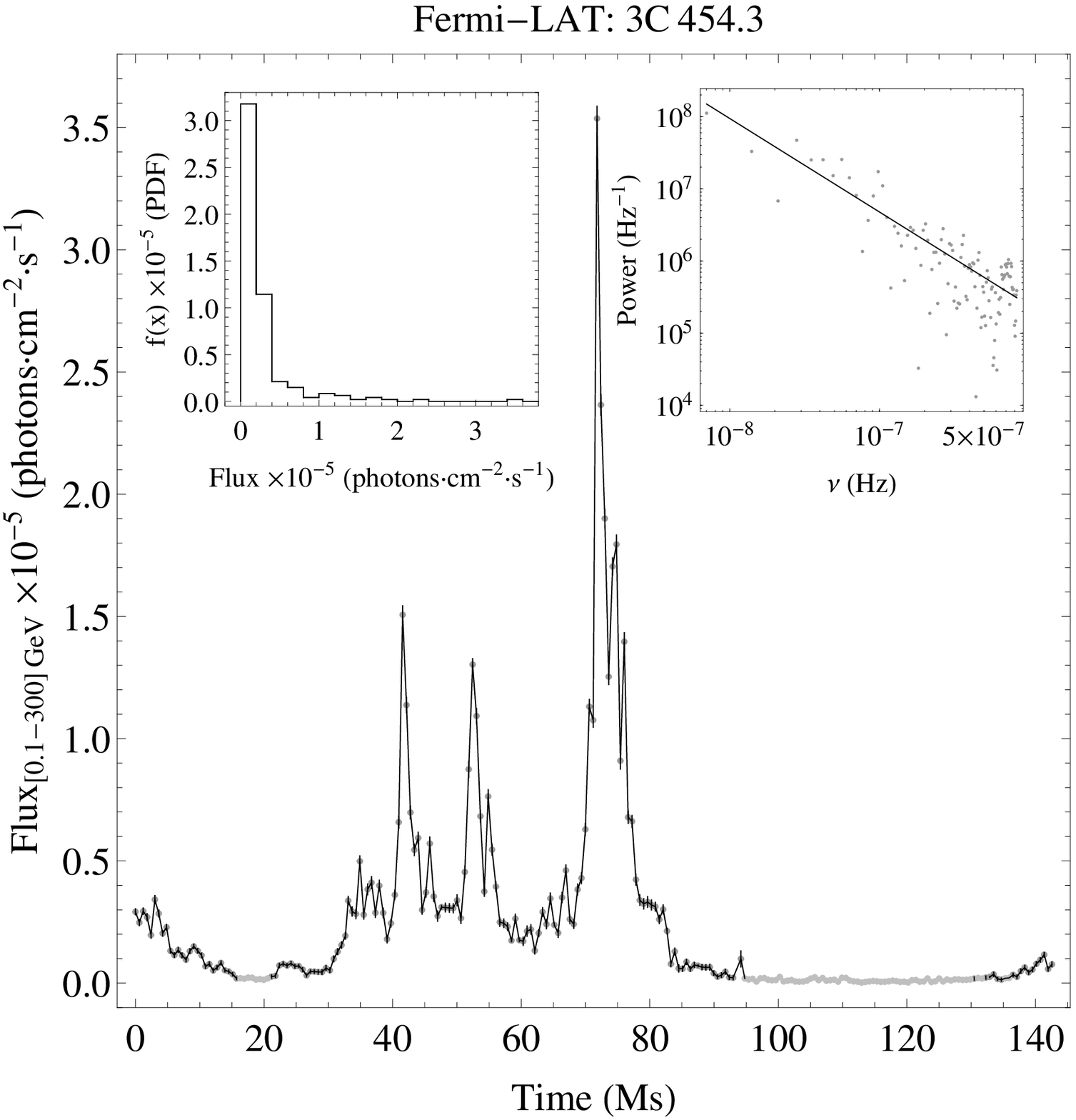}\hspace{4em}
\includegraphics[width=2.95in]{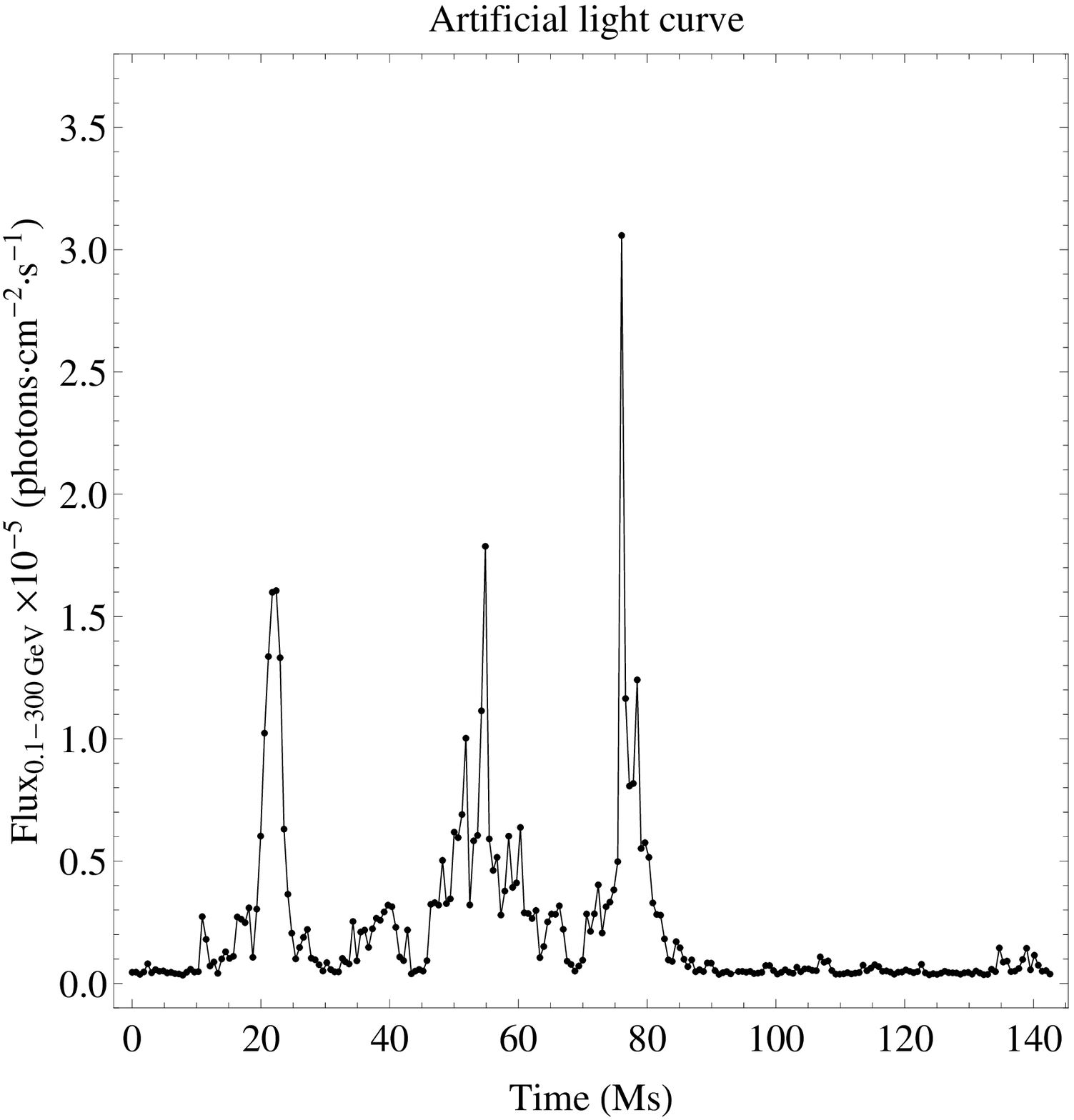}
\caption{The \textit{Fermi-LAT} data set of the blazar 3C\,454.3. [Left panel] The weekly averaged $\gamma$-ray light curve in the energy range of 0.1--300 GeV with black points corresponding to actual flux measurements and grey points to the 90 per cent confidence upper limits. The left and right insets show the PDF histogram and the periodogram estimates (grey points) together with the best fit power-law model (black line), respectively. [Right panel] A single artificial light curve coming after 22 iterations (convergence).}
\label{fig:aplic_fermi}
\end{figure*}

\noindent
We use the weekly \textit{Fermi-LAT} light curve of 3C\,454.3 consisting of 236 points between 54684 and 56334 MJD in the 0.1--300 GeV energy range\footnote{The \textit{Fermi-LAT} data has been retrieved from: \url{http://fermi.gsfc.nasa.gov/ssc/data/access/lat/msl_lc/}.}. The light curve is shown in the left panel of Fig.~\ref{fig:aplic_fermi} with the black points corresponding to the actual flux measurements and the grey points (around 20 Ms and 90--130 Ms) to the 90 per cent confidence upper limits. For the purposes of this study we have simply used the upper limits as actual flux measurements but more precise treatment using survival analysis techniques will be presented in a future work.\par
To remind the readers, the two basic components for the method are the distribution of the data as well as the corresponding PSD. The PDF histogram of the data is shown in the left panel of Fig.~\ref{fig:aplic_fermi} (left inset) and as we can see it is characterised by a long right-tail which becomes zero at much higher flux values from those expected by a simple exponential distribution. Note that if we were about to fit an exponential distribution PDF model to this data set, it would yield a best-fitting inverse scale of $4.25\pm0.06$ having a very poor fit quality, with the Anderson-Darling test statistic value of 30.21 and an $H_0$ probability (i.e.\ the data set is drawn from a population with the fitted distribution) of 0. The right inset in the same plot shows the periodogram estimates of the light curve together with the best fit bending power-law model (eq.~\ref{eqe:psd_bendMod}), $\alpha_{\rm low}=\alpha_{\rm high}=1.3^{+0.4}_{-0.1}$ and $c=2.4\pm0.3$ Hz$^{-1}$ ($f_{\rm bend}=7.8^{+2.1}_{-1.9}\
\times10^{-8}$ Hz), implying that a simple power-law model is enough to describe the data. Using these two data components, we apply our method and we produce an artificial light curve (convergence occurs after 22 iterations) which conserves all the statistical and variability properties of the original data set (Fig.~\ref{fig:aplic_fermi}, right panel). Thus, ensembles of such artificial light curves can be used in any sort of statistical analysis that requires establishment of confidence intervals e.g.\ in CCF analysis during multiwavelength campaigns.\par\noindent
\underline{The XRB Cyg\,X-1}\par
\begin{figure*}
\includegraphics[width=2.95in]{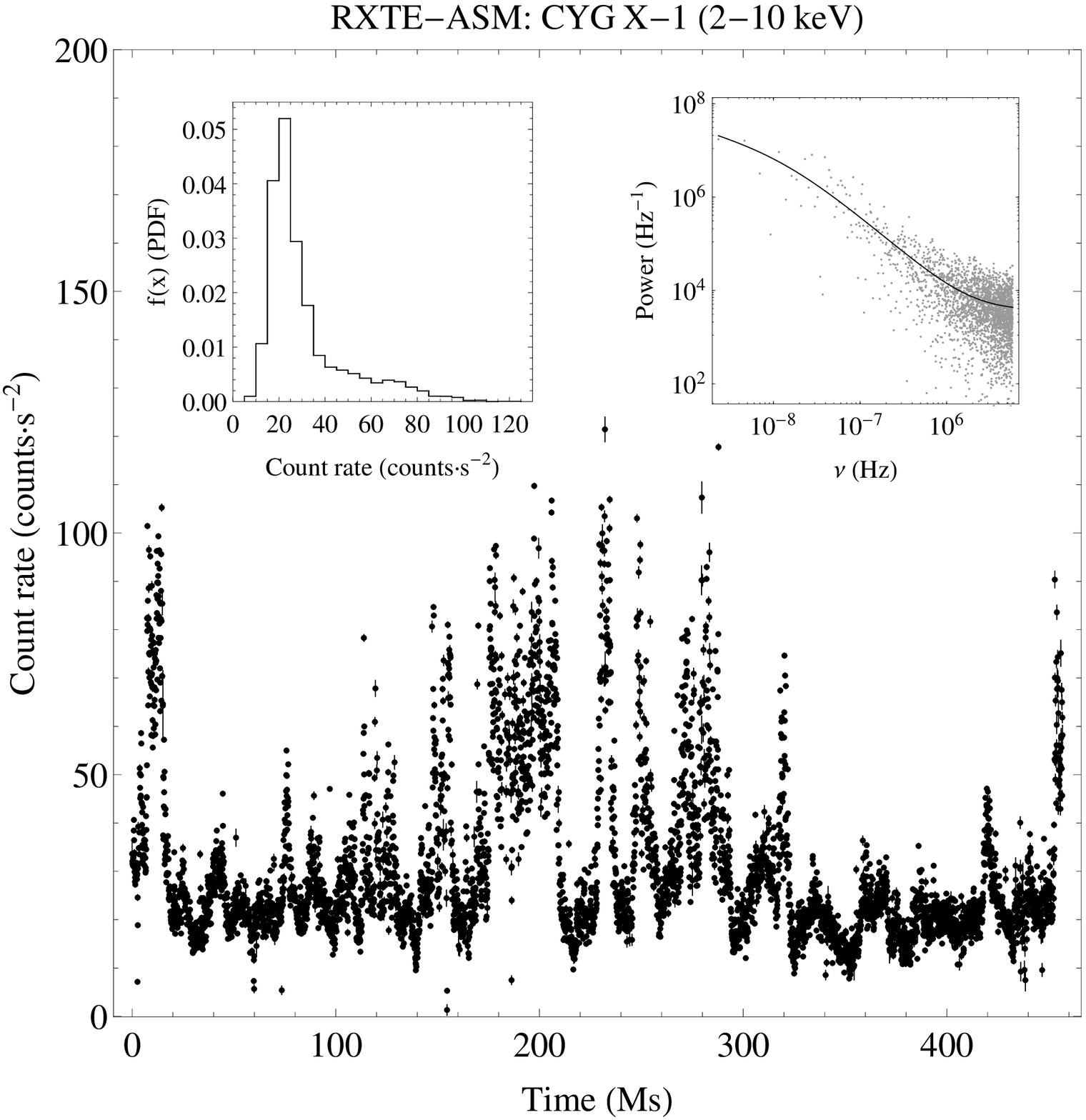}\hspace{4em}
\includegraphics[width=2.95in]{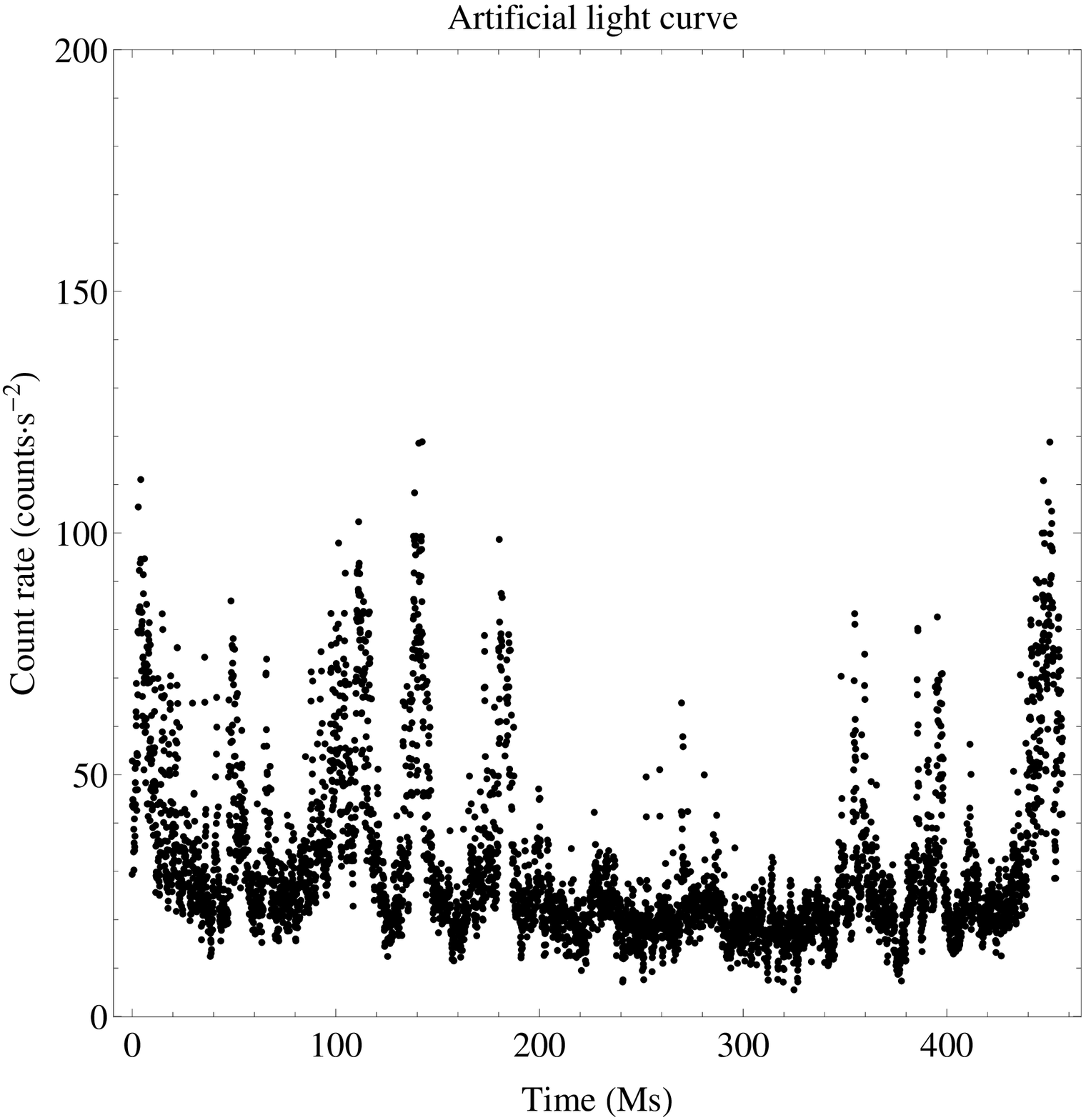}
\caption{The \textit{RXTE-ASM} data set of the XRB Cyg\,X-1. [Left panel] The daily averaged X-ray light curve in the energy range of 2--10 keV. The left and right insets show the PDF histogram and the periodogram estimates (grey points) together with the best fit bending power-law model (black line), respectively. [Right panel] A single artificial light curve coming after 267 iterations (convergence).}
\label{fig:aplic_rxte}
\end{figure*}

\noindent
We use the daily-averaged \textit{RXTE-ASM} light curve of Cyg\,X-1 consisting of 5000 points between 50135.4 and 55422.6 MJD in the 2--10 keV energy range\footnote{The \textit{RXTE-ASM} data has been retrieved from: \url{http://xte.mit.edu/ASM_lc.html}.}. The light curve is shown in the left panel of Fig.~\ref{fig:aplic_rxte} and contains a small number of gaps (289 in total), which we have been filled up with linearly interpolated values. Appropriate treatment, using bootstrapping, should be performed in order to check the effects of the gaps during the following PSD estimation, but for the purposes of this study we ignore this step\footnote{The frequency domain bootstrap methodologies are used to estimate the distribution of the re-sampled periodogram estimates. The main difficulty is to select appropriate statistical estimators whose variance fits that of the re-sampled periodogram estimates (at a given frequency). Useful analyses on this topic have been performed by several authors \citep[e.g.][]
{franke92,dahlhaus96,kreiss03}.}.\par
The PDF histogram of the data is shown in the left panel of Fig.~\ref{fig:aplic_rxte} (left inset) having a characteristic bimodal shape, depicting the high and the low flux states of the source. Assuming that no artefacts are induced to the periodogram estimates, due to the interpolation, the best fit PSD model yields $\alpha_{\rm low}=0.49^{+0.12}_{-0.21}$, $\alpha_{\rm high}=1.58^{+0.14}_{-0.16}$, $f_{\rm bend}=1.32^{+2.6}_{-0.43}\times10^{-8}$ Hz and $c=3692^{+18}_{-12}$ Hz$^{-1}$ (Fig.~\ref{fig:aplic_rxte}, left panel, right inset). After applying our method (convergence occurs after 267 iterations), the resulting artificial light curve (Fig.~\ref{fig:aplic_rxte}, right panel) resembles remarkably to the original data set of Cyg\,X-1 which was chosen as an example of extreme `bursticity'.\par

\section{APPLICATION TO CCF ANALYSIS}
\label{sect:ccf_analysis}
CCF analysis is one the most common method used for analysing multiwavelength light curves obtained in a simultaneous fashion. There are several different flavours and implementations of the CCF e.g.\ discrete correlation function (DCF) \citep{edelson88}, interpolated CCF \citep{gaskell86}, modified CCF \citep{li04}, {\it z}-transform discrete correlation function \citep{alexander97} that are used within the astronomical community. Particularly for the case of irregularly sampled light curves, estimation of the confidence levels, in both the CCF values and/or time delays, is usually done by performing Monte Carlo simulations. Application of a given CCF method to an ensemble of paired random artificial light curves, which have the same PSD as the observed data sets, yields the probability of getting a given CCF estimate purely by chance coincidence. At present the simulated light curves are produced using the TK95 formalism.\par
\begin{figure*}
\hspace*{2em}\parbox{1\linewidth}{\includegraphics[width=2.88in]{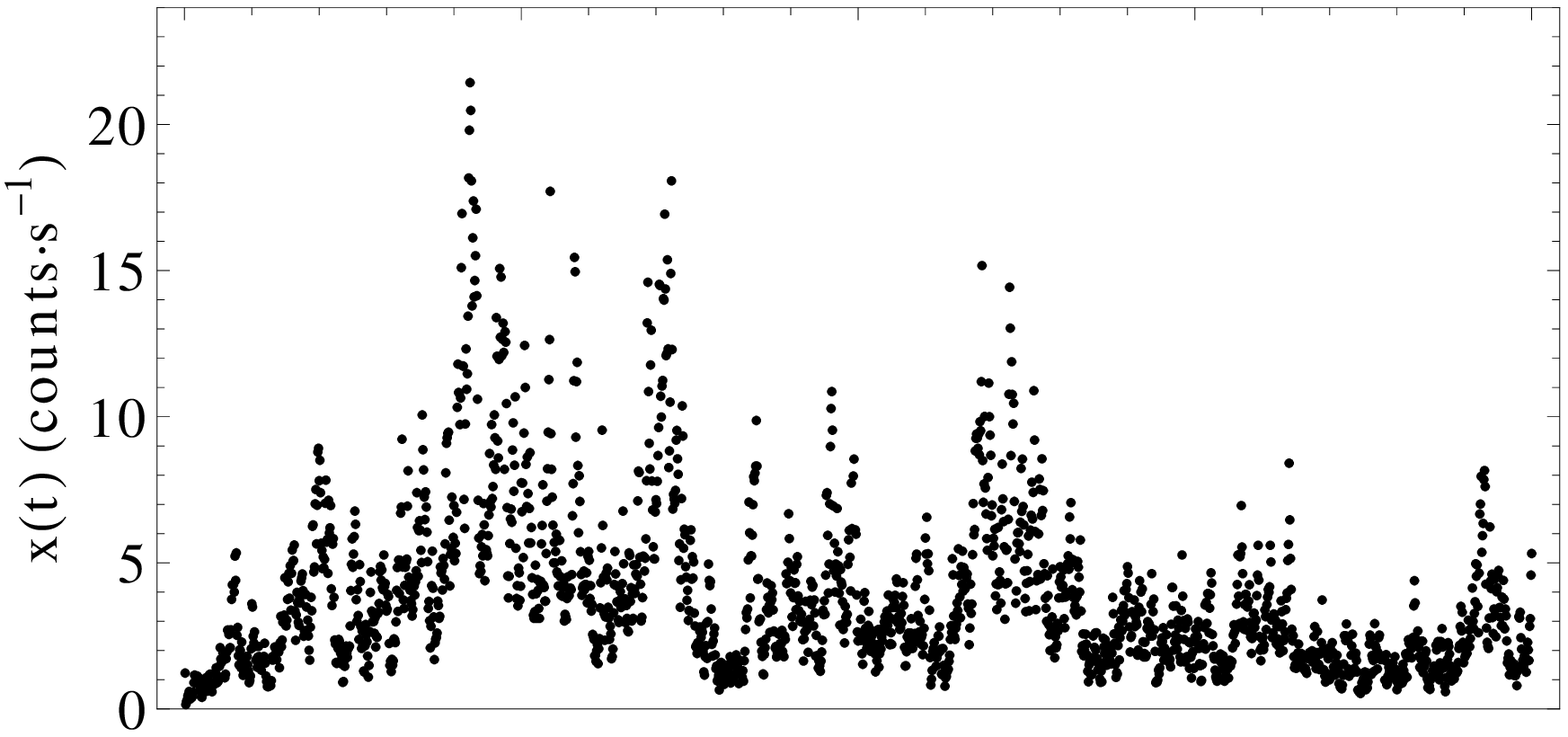}\\[-1.91em]
\hspace*{-0.17em}\includegraphics[width=2.932in]{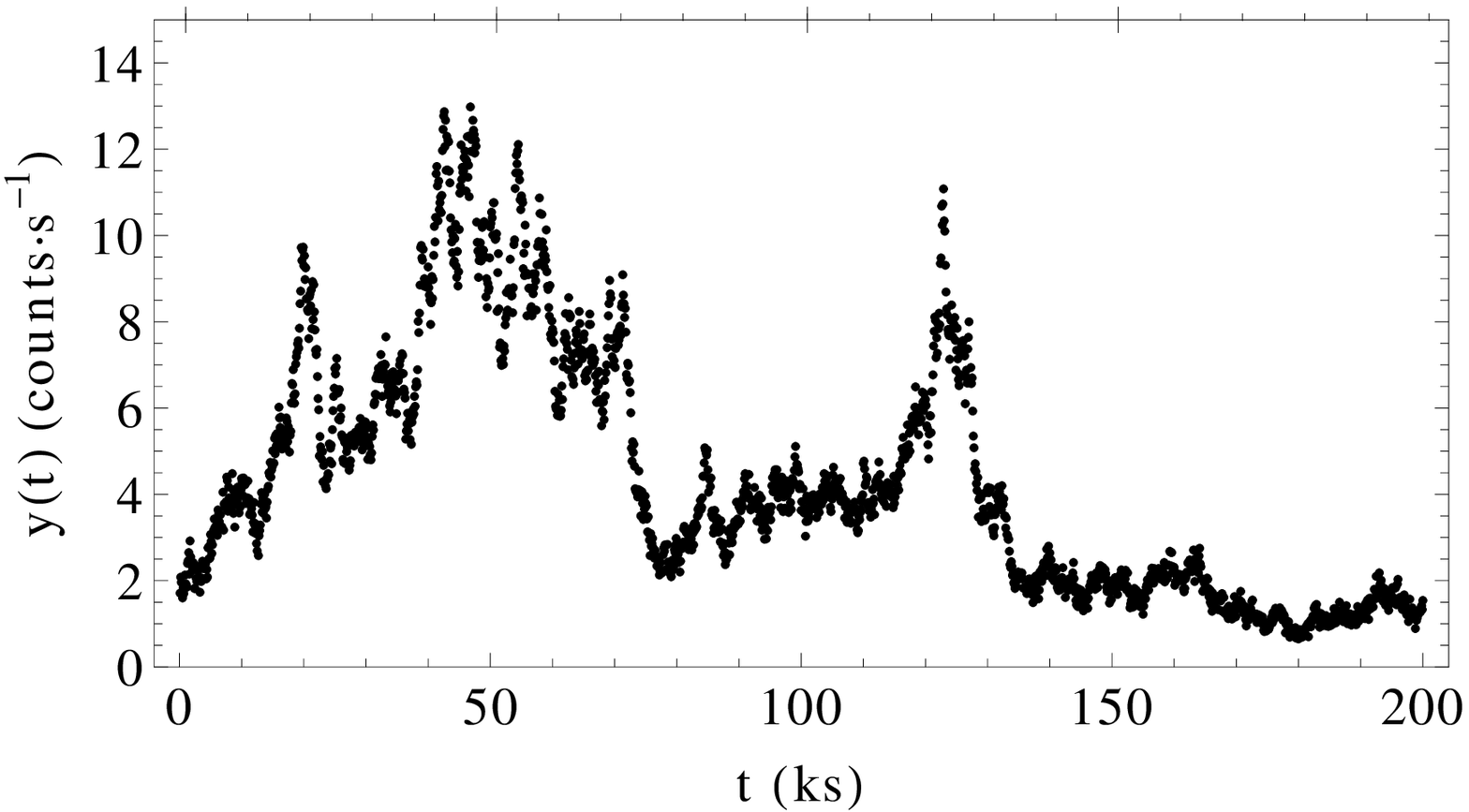}}\\[-24.8em]
\hspace*{29em}\includegraphics[width=2.98in]{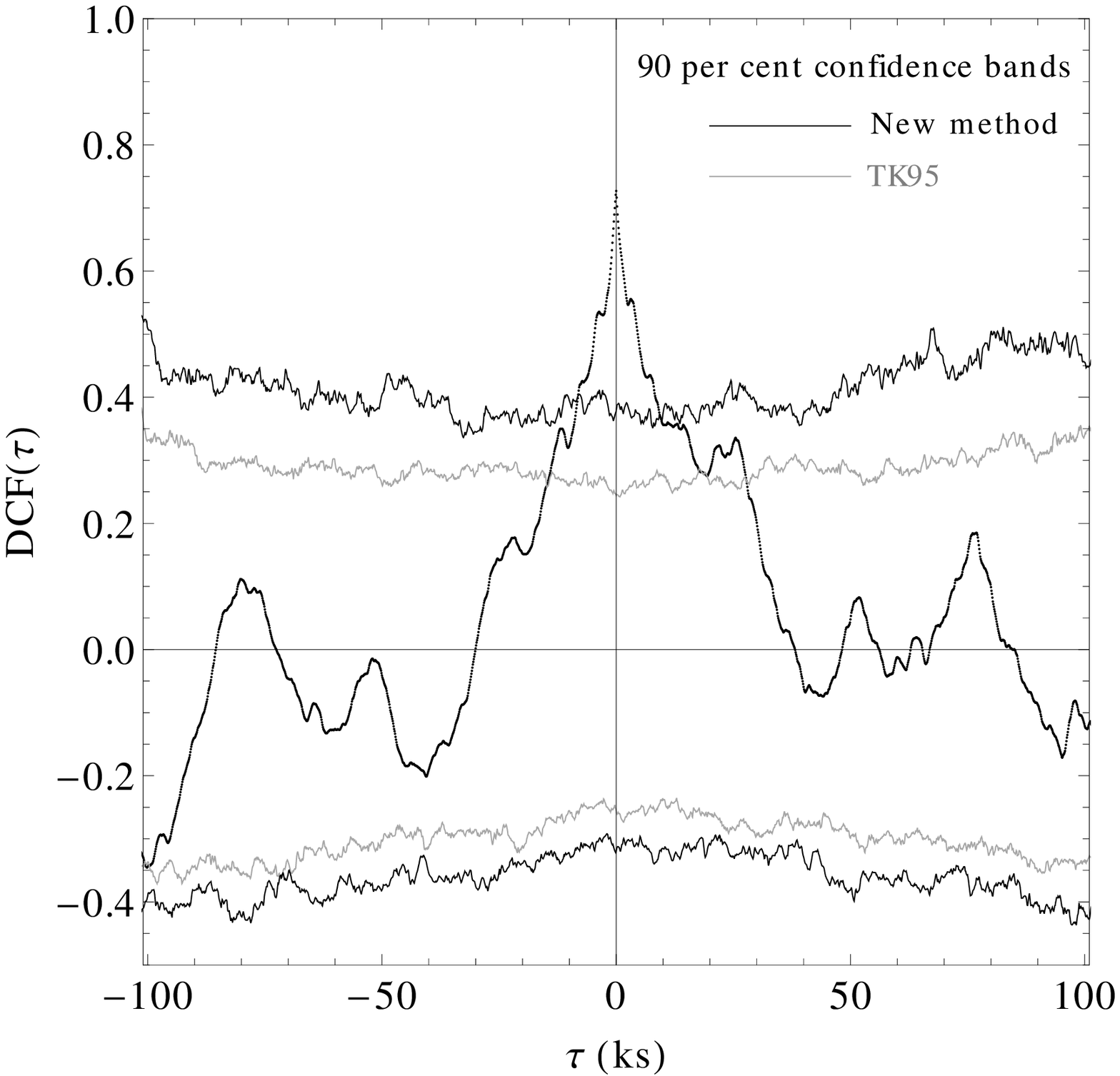}
\caption{Establishment of the statistical significance on the DCF estimates. [Left panel] The two artificially produced `bursty' light curves, $x(t)$, $y(t)$ having the same random seed. [Right panel] The black points correspond to the DCF estimates of of $x(t)$ and $y(t)$, separated by 0.1 ks (100 s). The grey and black horizontal lines corresponds to the 90 per cent confidence bands as derived from the synthetic data sets of TK95 and the newly proposed method, respectively.}
\label{fig:ccf_lcs}
\end{figure*}

Since the TK95 synthetic data sets are distributed normally by construction, the method is appropriate to yield CCF confidence levels only for Gaussian light curves. In order to show that deviations in the CCF levels can occur for the case of non-Gaussian light curves, for illustrative purposes, we create two `bursty' light curves in the following way: Using the TK95 procedure, we produce two artificial light curves having different bending PSDs, $x_{\rm G}(t)$ and $y_{\rm G}(t)$, 200 ks long with a bin size $\Delta t=100$ s, using the same random seed (in order to be correlated). During the production of $y_{\rm G}(t)$, we multiply the imaginary part with a random number between $\left[-0.15,0\right)$. This will slightly modify the final flare profiles i.e.\ amplitudes and phases (equations \ref{eq:amplitude} and \ref{eq:phase}) yielding an asymmetric CCF profile around the zero time delay. Finally, the resulting normally distributed numbers are used as exponents for the bases 2 and 1.5 respectively in 
order to produce two `bursty-like' light curves, $x(t)$ and $y(t)$, (Fig.~\ref{fig:ccf_lcs}, left panel) having of course different underlying PSD parameters from the initial ones. The initial PSD parameters, used for the TK95 methodology, together with the final PSD parameters, estimated after fitting the periodogram estimates of the exponentiated light curves (in base 2 and 1.5, respectively), are given in Table~\ref{tab:sim_ccf}. Note that in both light curves we have added Poisson noise following the recipe described in Sect.\,\ref{ssect:poisson}. For the purposes of this study these two `bursty' light curves will be used as two simultaneously obtained observations of the same object, in different energy bands, for which we will perform CCF analysis.\par
Initially, we estimate the DCF for the two `bursty' light curves (Fig.~\ref{fig:ccf_lcs}, right panel, black points). Then, in order to assess the confidence level of the correlation we produce two ensembles of 1000 pairs of artificial light curves; one following the classical procedure of TK95 and another one using the proposed methodology described in Sect.\,\ref{sect:methodology}. Then, for each method we estimate the DCFs between all the pairs and for each time delay, $\tau$, we estimate the 0.025 and 0.975 quantiles corresponding to the upper and lower limits of the 90 per cent confidence bands.\par
As we can see from the right panel of Fig.~\ref{fig:ccf_lcs} (TK95: grey lines, new method: black lines) realistic representation of the non-Gaussian light curves yields in generally an increase in the confidence level range of the order of 25 per cent which reduces the detection significance of the DCF peak. The reason is that Gaussian light curves have on average the same number of flares above (positive direction) and below (negative direction) the mean, in contrast to the `bursty' light curves which exhibit flares only in the positive direction. That means that between two `bursty' light curves it is much more likely to get a fake correlation by chance coincidence since there is only one possible flare direction.\par
\begin{figure}
\includegraphics[width=2.9in]{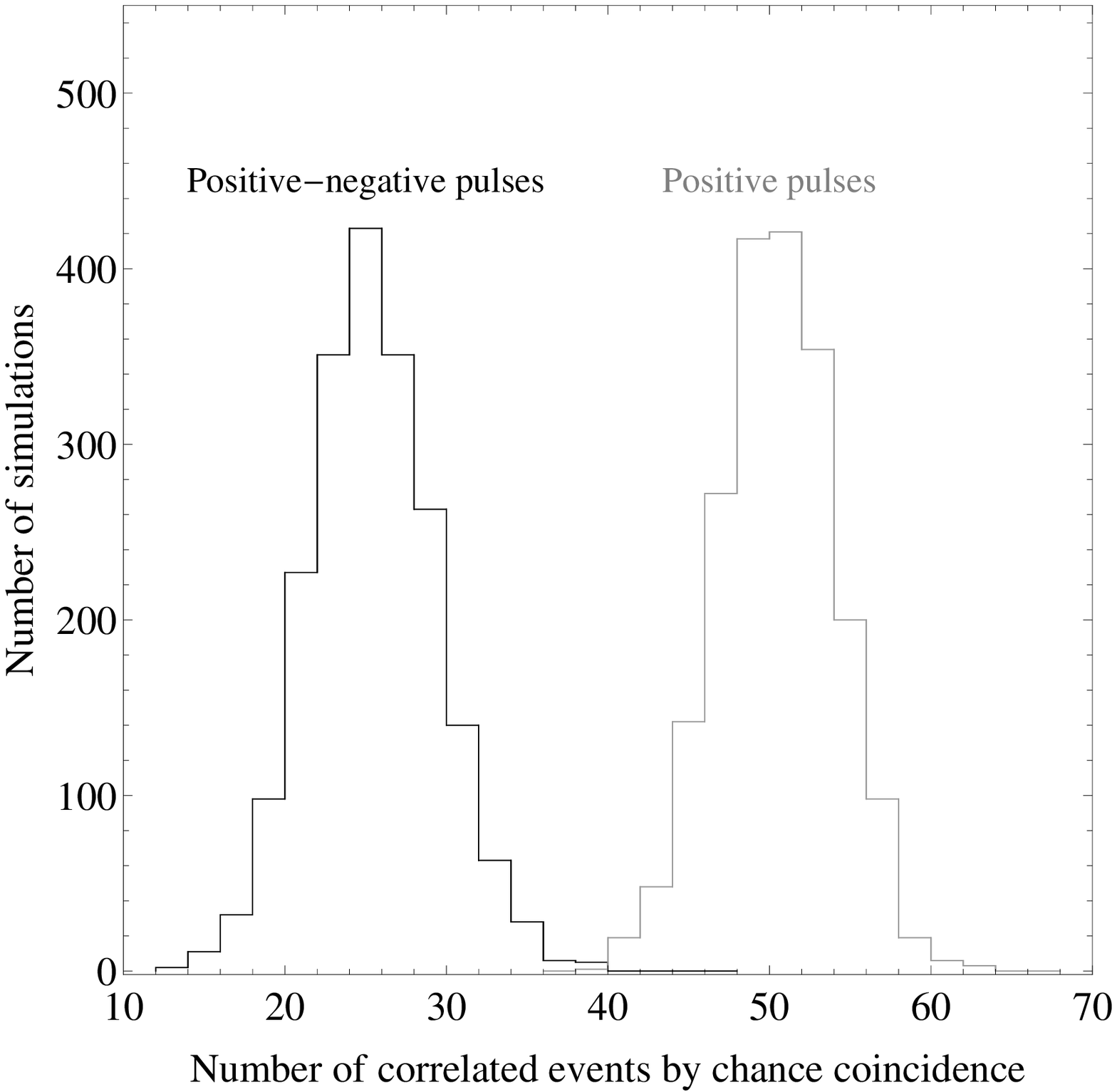}
\caption{Correlated events for the toy-simulation. The black line corresponds to the distribution of the correlated events for the case of positive-negative triangular unit pulses. The grey line corresponds to the same distribution for the case of positive triangular unit pulses.}
\label{fig:ccf_simToy}
\end{figure}

This can be very well understood with the following toy-simulation. We produce a series of 2000 pairs of time series each one consisting of 100 positive and negative triangular positive pulses (the simplified analogue of a Gaussian light curve) occurring at uniformly random non-repetitive integer numbers and we measure the number of simultaneous pulse occurrences by chance coincidence between all the pairs. Then, we repeat the simulation but now the time series consist only of positive triangular unit pulses (the simplified analogue of a `bursty' light curves). As we can see from Fig.~\ref{fig:ccf_simToy} the mean occurrence of chance coincidence correlated events is almost doubled for the case of the positive triangular pulses due to the flare directionality property. This shows that for non-Gaussian light curves TK95 underestimates the chance coincidence occurrences of correlated events and thus yields erroneous smaller estimates for the confidence intervals i.e\ yielding an overestimation of the CCF's 
peak significance.\par

\begin{table}
\begin{minipage}{190mm}
\caption{PSD model-parameters for the CCF simulations.}
\label{tab:sim_ccf}
\begin{tabular}{@{}ccc}
\hline
\multirow{2}{*}{Model-parameter} & \multirow{2}{*}{Initial PSD} & \multirow{2}{*}{Final PSD$^\dagger$}\\[0.4em]
 & $x_{\rm G}(t)$, $y_{\rm G}(t)$ & $x(t)$,$y(t)$ \\
\hline
$\alpha_{\rm low}$ & 0.9,1.8 & $0.91^{+0.09}_{-0.08}$,$0.96^{+0.08}_{-0.07}$\\[0.4em]
$\alpha_{\rm high}$ & 2.3,2.8 & $2.58^{+0.08}_{-0.05}$,$2.67\pm0.06$\\[0.4em]
$f_{\rm bend}$ ($\times10^{-4}$ Hz) & 2.6,10 & $14^{+1}_{-2}$,$18^{+3}_{-2}$\\[0.4em]
\hline
\end{tabular}
\medskip\\
$^\dagger$ These are the values used in the simulations.
\end{minipage}
\end{table}

\section{THE RMS-FLUX RELATION}
\label{sect:rms_flux}
In the special case of a parent log-normal distribution, the rms-flux relation (see Sect.~\ref{sect:intro}) may be sometimes of vital importance for the needs of a statistical study or a theoretical model. The surrogate data sets, following a parent log-normal distribution, have embedded this property in a natural way without the need for further adjustments or tuning.\par
\begin{figure*}
\includegraphics[width=2.85in]{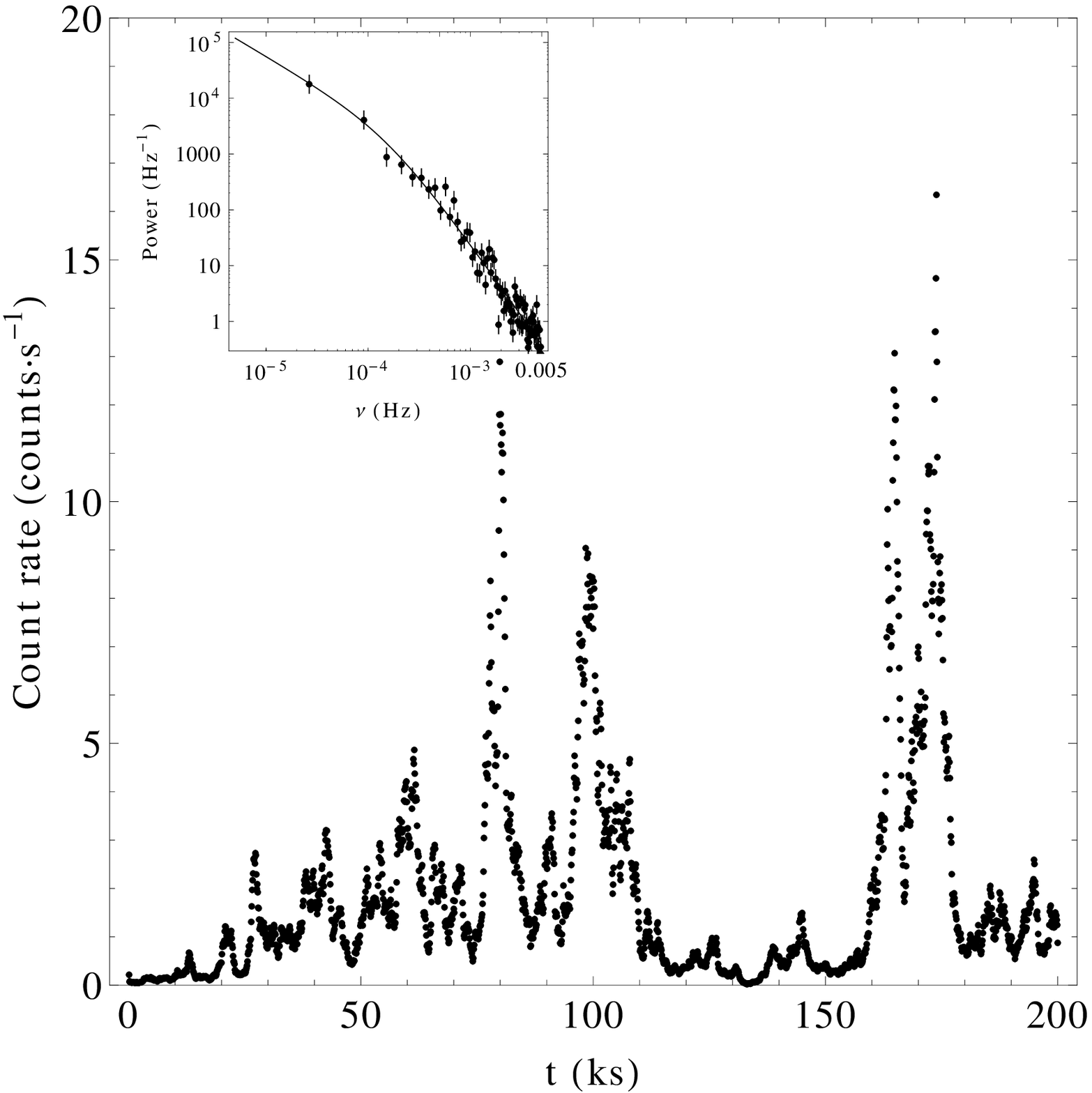}\hspace{4.9em}
\includegraphics[width=2.85in]{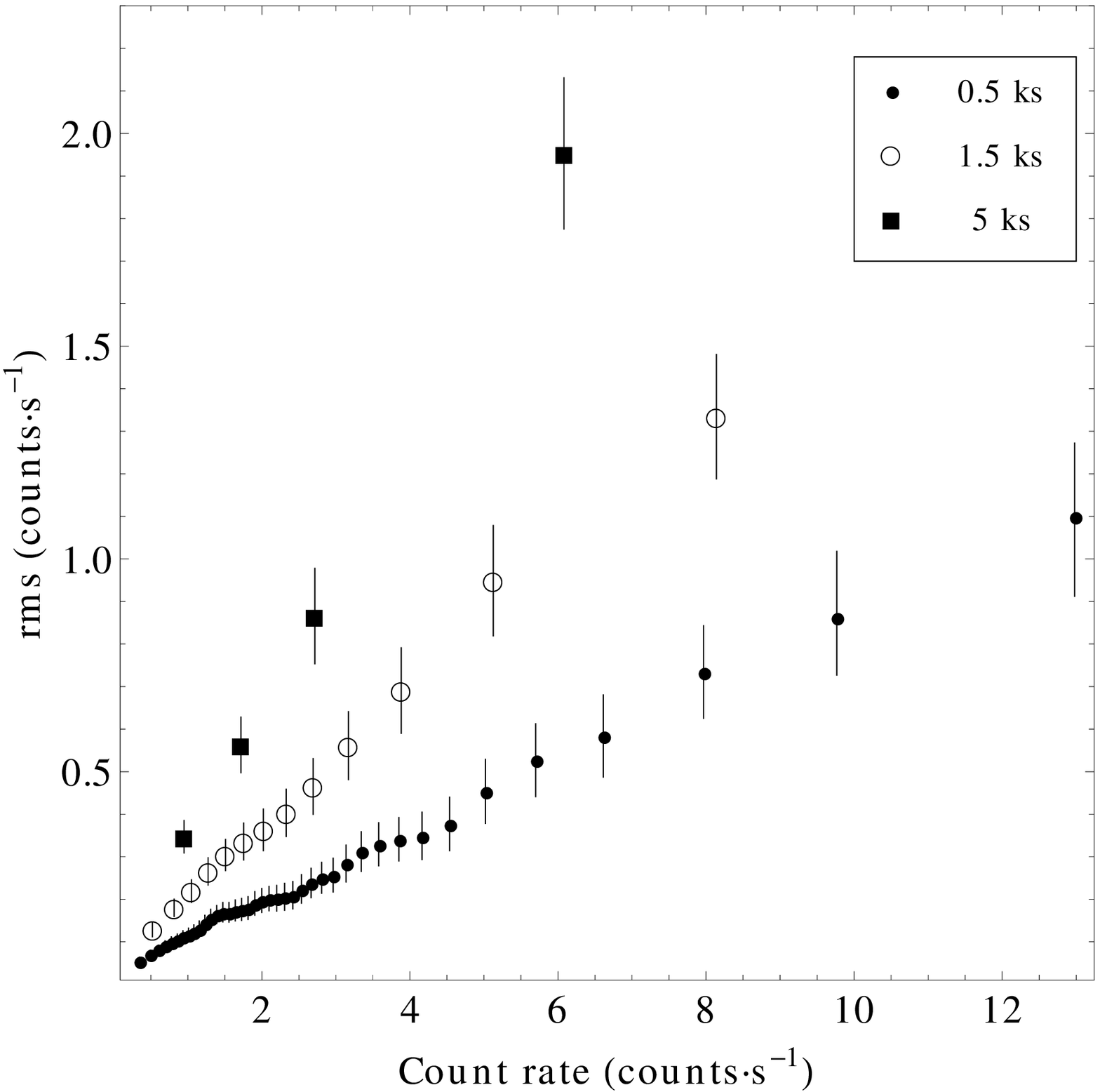}
\caption{The rms-flux relation. [Left panel] The exponentiated light curve together with its binnned logarithmic periodogram and final PSD model (inset). [Right panel] The average rms-flux estimates coming from the 1000 simulated light curves using three different binning schemes i.e.\ 0.5, 1.5 and 5 ks, respectively.}
\label{fig:flux_rms}
\end{figure*}

To show that our method automatically produces the rms-flux relationship, we first create a sample light curve which inherently has the rms-flux relation by following a parent log-normal distribution. Using the TK95 procedure (with initial PSD parameters: $\alpha_{\rm low}=1.5$, $\alpha_{\rm high}=2.8$ and $f_{\rm bend}=1\times10^{-3}$ Hz) we produce a synthetic data set, being 200 ks long in bins of 100 s, and then we exponentiate the resultant data set (having final PSD parameters: $\alpha_{\rm low}=1.11^{+0.05}_{-0.04}$, $\alpha_{\rm high}=2.48^{+0.09}_{-0.06}$ and $f_{\rm bend}=(2.2\pm0.6)\times10^{-4}$ Hz (Fig.~\ref{fig:flux_rms}, lefy panel). This light curve will be used as the observed light curve that we want to simulate.\par
We produce 1000 artificial light curves using our proposed method and then for each one of them we estimate the rms-flux relation using the prescription of \citet{uttley05a}. We select three different length segments of 0.5, 1.5 and 5 ks consisting of 5,15 and 50 bins, respectively. Under a given binning scheme, for each flux value we estimate an average rms and its standard deviation coming from the 1000 surrogate data sets. The results are shown in the right panel of Fig.~\ref{fig:flux_rms} and as we can readily see the simulated light curves follow remarkably well the linear rms-flux relation for a variety of time-scales below and above the $f_{\rm bend}$, corresponding approximately to 4.55 ks. This widely observed variability property is embedded in our artificial light curves in a natural way depicting in a vivid way the fact that our artificial light curves are exact replicas of the observed light curves.\par
This flare-directionality, which is actually mapped on the histogram of the `bursty' light curves in the form of their positive skewness, is taken automatically into account during this newly proposed light curve simulation method. This means that for the establishment of confidence intervals it is of great importance to take correctly into consideration the distribution of the measurements since this can affect significantly the level of chance coincidence occurrences. Note, that for the case of Gaussian light curves (i.e.\ with minuscule skewness) the method automatically is in accordance with the confidence intervals derived by TK95.

\section{INVARIANT QUANTITIES AND STATISTICAL DEPENDENCIES}
\label{sect:invariants}
During any simulation process, it is very import understand which light curve properties are preserved and which are not. The TK95 procedure preserves only the underlying PSD of the observed data set, assuming a Gaussian distribution of measurements for all the cases. Our method preserves both the underlying PSD and the PDF (observed or parent).\par
The preservation of the PDF means that all the statistical moments of a given data set i.e.\ mean ($\mu$), variance ($\sigma^2$), skewness ($\gamma_1$), kurtosis ($\gamma_2$), and so on, are identical between the observed and the surrogate data sets. Since, all these quantities are included in the input PDF (they describe its shape e.g.\ asymmetry, peakedness), they are conserved by construction; during the last iteration step (ranking, step iv) all the measurements are redistributed based on the input PDF (the insets in the right panel of Fig~\ref{fig:ngc4051_simul_hist} and the left panel of Fig.~\ref{fig:sim56_adjust_lc_prdgr} are identical). The fact that all the statistical moments are conserved does not mean that all the statistical dependences of the measurements (e.g.\ non-linear interactions) are preserved. These are two completely different statistical quantities.\par
The various statistical dependences of the measurements are characterised only by the Fourier transform of the joint cumulant functions, known as polyspectra, e.g.\ autospectrum, bispectrum, trispectum and so on (Appendix~\ref{app:cumul_statiDependnce}). Our method preserves only the second order joint cumulant (i.e.\ the covariance) of the input data set, and thus its Fourier transform, autospectrum, (and thus its squared amplitude, the PSD) is the only spectral quantity which is preserved in the final converged synthetic light curve. Thus, the only dependencies that are preserved are those corresponding to the covariance --\ all the higher-order dependencies are ignored.\par
A very common source of confusion and misunderstandings is that there is a notion that preservation of higher order statistical moments (e.g.\ $\gamma_1$, $\gamma_2$ and so on) means preservation of the higher order spectra. The statistical moments characterise only the shape of the PDF, whilst existence of potential dependencies between the data points are mapped only on the polyspectra. As we can see in Appendix~\ref{app:cumul_statiDependnce} this confusion originates from the fact that $\sigma^2$ and $\gamma_1$ (depicting the shape of the PDF) and are equal to the zero delayed joint cumulants $(C_2(0),C_3(0,0))$ (i.e.\ for $s_1=s_2=0$), however polyspectra are the Fourier transform of the joint cumulants i.e.\ summed over all $s_i$.\par

\begin{figure*}
\includegraphics[width=2.85in]{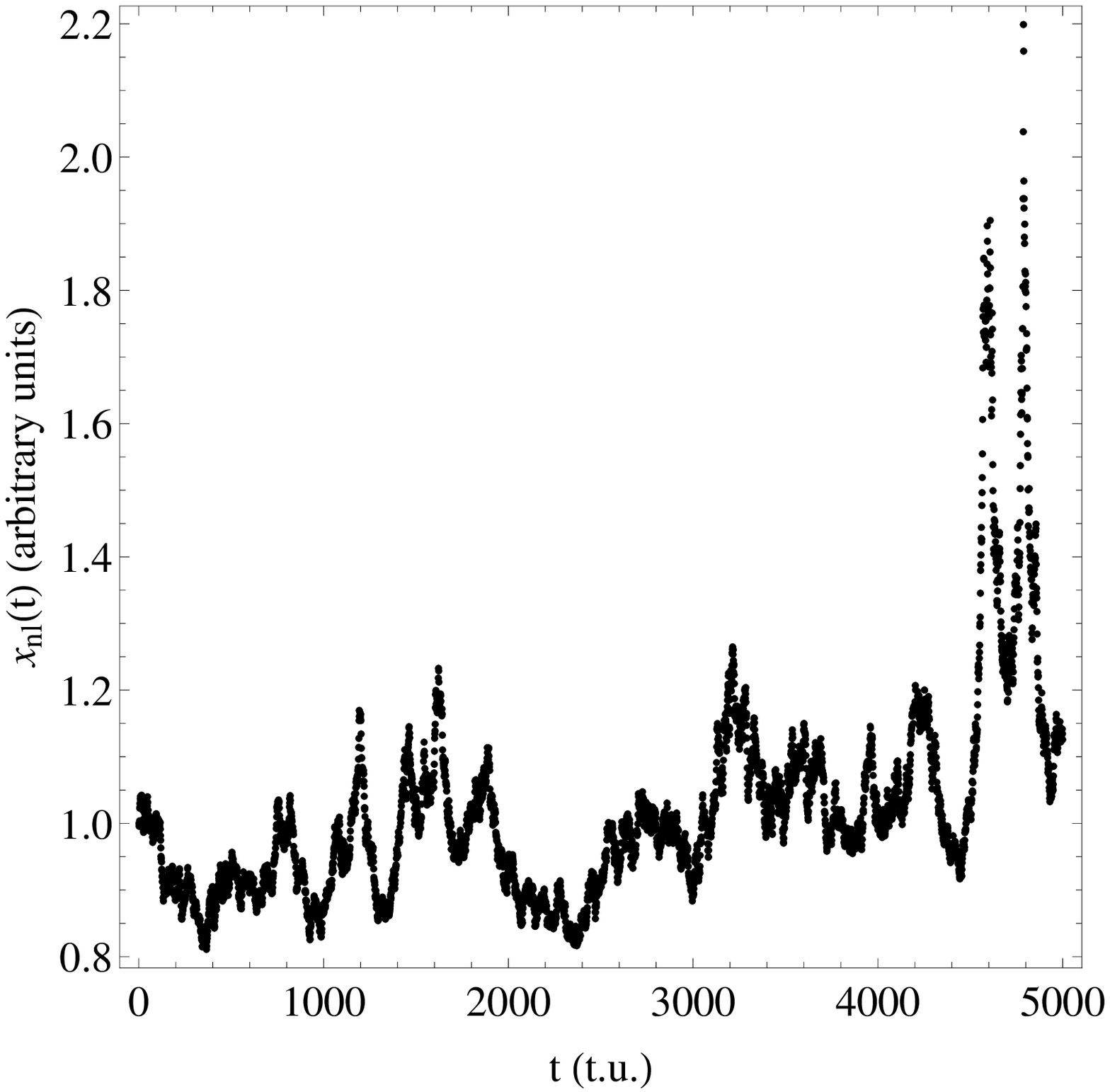}\hspace{4.9em}
\includegraphics[width=2.85in]{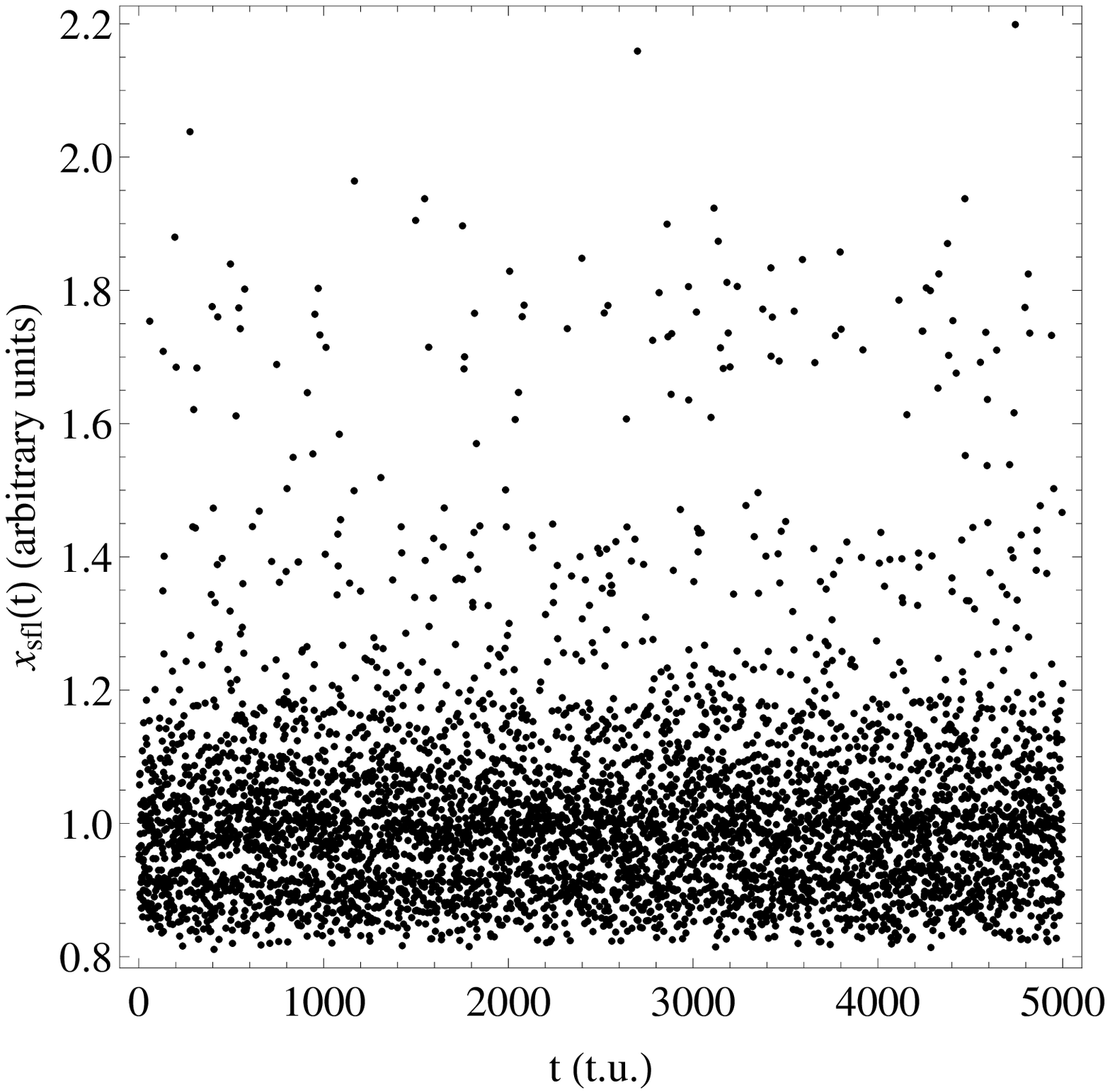}
\caption{A time series process. [Left panel] A realisation of the nonlinear process (equation \ref{eq:nlProcess}), $x_{\rm nl}(t)$, with the times rescaled in the range $t\in[1,5000]$ t.u. and $\Delta t=1$ t.u. [Right panel] A random shuffle of the $x_{\rm nl}(t)$, $x_{\rm sfl}(t)$.}
\label{fig:ts_process}
\end{figure*}

\begin{figure}
\includegraphics[width=2.85in]{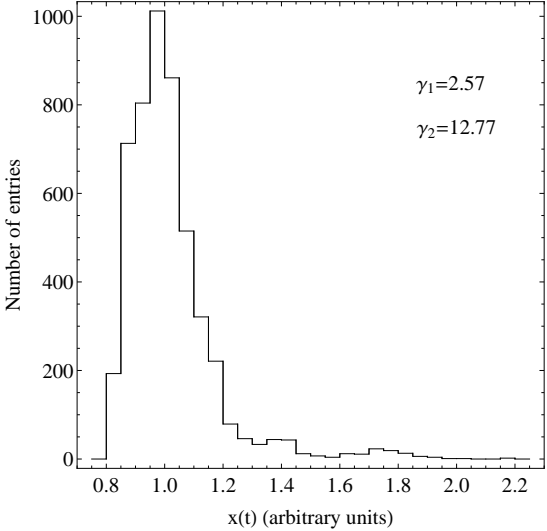}
\caption{The PDF of the time series process shown in Fig.~\ref{fig:ts_process} which has a skewness and a kurtosis of $\gamma_1=2.57$ and $\gamma_2=12.77$, respectively.}
\label{fig:ts_hist}
\end{figure}

A simple example to demonstrate this is the following. We consider a genuine non-linear process similar to the one given in \citet{provenzale92} \citep[also used by][]{vio92}:
\eqb
x^\prime(t)=1-x(t)^{1/5}+x(t)^3wn(t)
\label{eq:nlProcess}
\eqe
in which $wn(t)$ is a Gaussian white noise process with mean value of 0 and standard deviation of 1. We solve this equation for $x(0.01)=1$ in $t\in[0.01,50]$ with a $\Delta t=0.01$ and then we scale the time axes from 1 to 5000 time units (t.u.) in steps of 1 t.u. One realisation of the corresponding process, $x_{\rm nl}(t)$, is shown in the left panel of Fig.~\ref{fig:ts_process}. Then, we shuffle randomly the data of this process (having an equal probability among all the numbers) yielding a white noise process, $x_{\rm sfl}(t)$, and we plot the data in the right of Fig.~\ref{fig:ts_process}. In this way none of the initial dependencies between the data points are preserved, but the two data sets still have identical PDFs, since they consist of exactly the same data points (Fig.~\ref{fig:ts_hist}). This PDF has non-zero higher order statistical moments i.e.\ skewness and kurtosis of $\gamma_1=2.57$ and $\gamma_2=12.77$, respectively.\par

For each data set, we then estimate the normalised square amplitudes of its bispectrum, known as bicoherence, following \citet{kim79}, for the frequencies inside the inner triangle of the principal domain \citep{hinich95}. We have divided each data set into 100 segments, each one consisting of 50 t.u. (i.e.\ 50 consecutive data points), and for the estimation of the bicoherence, he have averaged the corresponding Fourier transforms and biperiodograms.\par
\begin{figure*}
\includegraphics[width=3.3in]{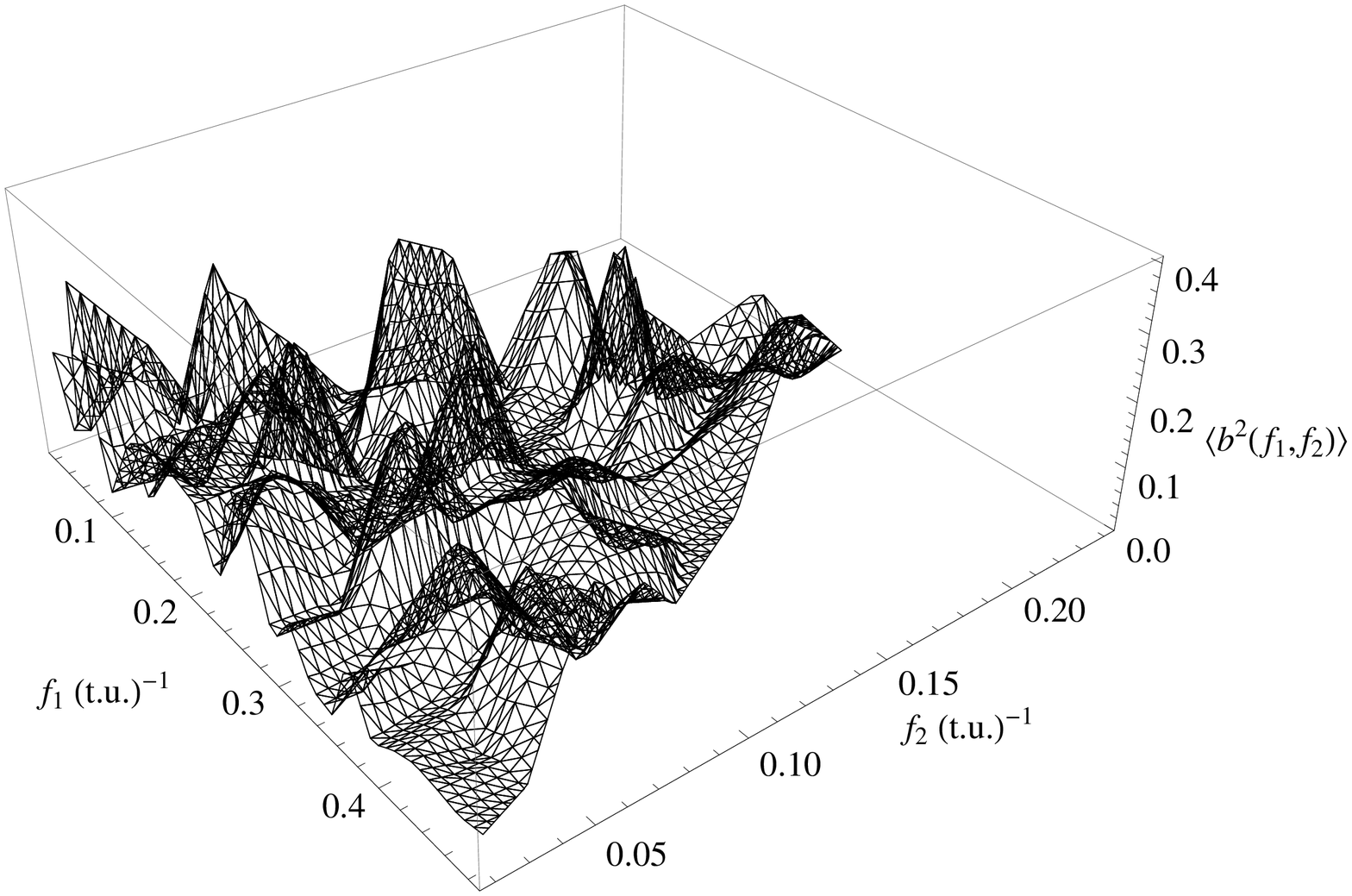}\hspace{1em}
\includegraphics[width=3.3in]{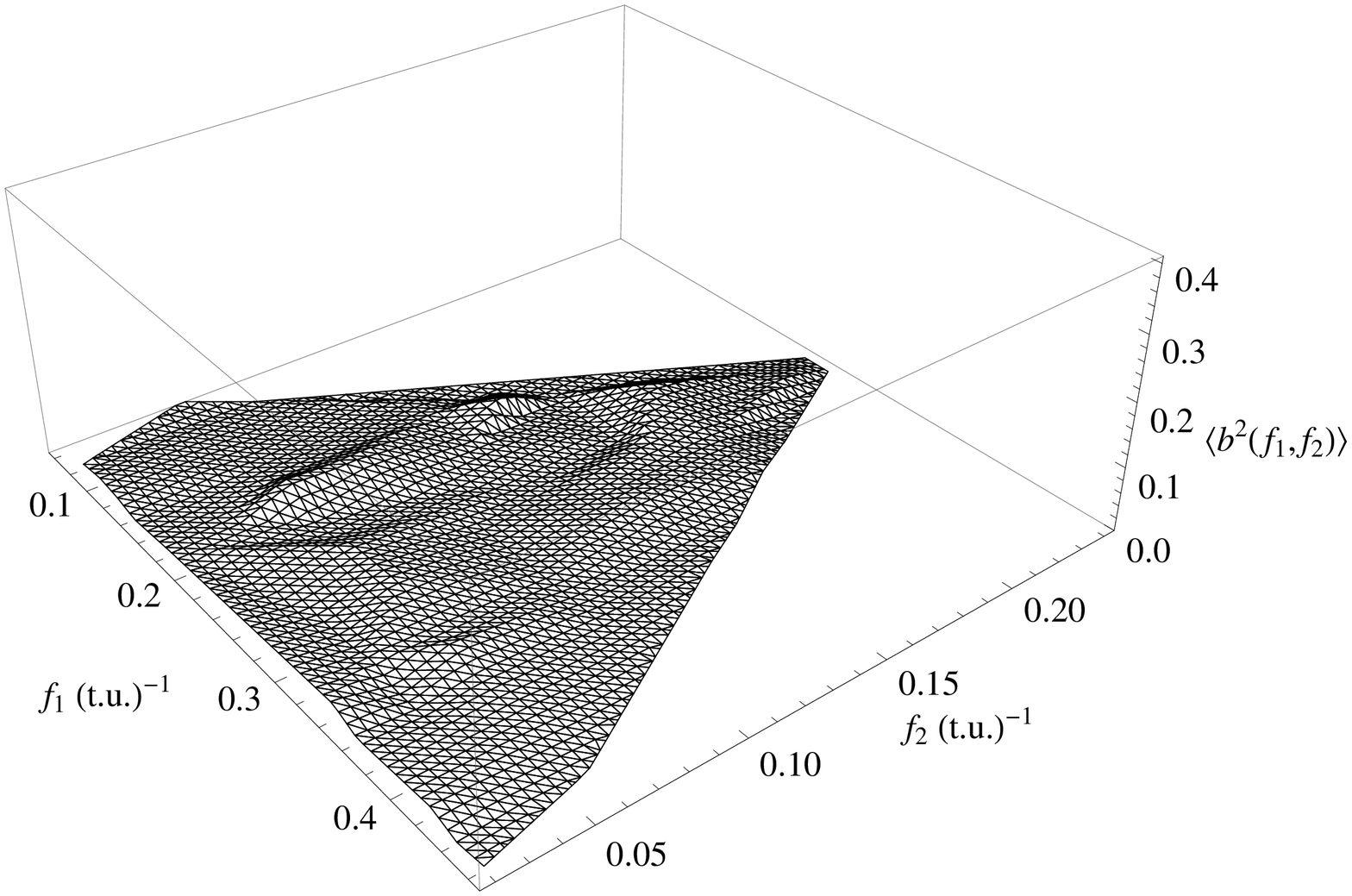}
\caption{Bispectrum analysis. [Left panel] The bicoherence of the nonlinear data set, $x_{\rm nl}(t)$ (Fig.~\ref{fig:ts_process}, left panel). [Right panel] The bicoherence of the shuffled data set, $x_{\rm sfl}(t)$ (Fig.~\ref{fig:ts_process}, right panel).}
\label{fig:ts_bispec}
\end{figure*}

As we can see from Fig.~\ref{fig:ts_bispec}, the two data sets have genuinely different bicoherences i.e.\ genuinely different bispectra. The left panel of Fig.~\ref{fig:ts_bispec} exhibits a great deal of structure for various combinations of $(f_1,f_2)$ depicting the nonlinear dependencies for the data set $x_{\rm nl}(t)$. On the contrary, the right panel of Fig.~\ref{fig:ts_bispec} shows (as expected) a rather quiescent behaviour for the shuffled data set, $x_{\rm sfl}(t)$, despite the fact that it shares exactly the same PDF with the previous data set.\par
This simple example shows vividly that despite the fact that the two data sets share exactly the same PDF, i.e.\ have the same high order statistical moments, they do not share the same bispectra i.e.\ the various dependences between the measurements are genuinely different.

\section{SUMMARY AND DISCUSSION}
\label{sect:summar_discu}
We have presented a new algorithm able to produce artificial light curves which are distributed based on a given PDF (parent or observed) and a given underlying PSD. Our publicly available algorithm combines and enhances the methods of TK95 and SS96. The new method improves significantly on the widely used procedure of TK95 which is able to produce artificial light curves which are only normally distributed. Thus, for any sort timing studies, in which simulated data sets are needed, our algorithm preserves all the genuine variability and statistical source properties yielding ensembles of truly random artificial data sets.\par 
The merits of our method can be summarised in the following lines: 
\begin{itemize}
\item It reproduces the exact variability properties of the observed data, since the synthetic light curves follow the input PSD. The input PSD can originate either from an actual observation or a theoretical model. 
\item It reproduces the exact statistical properties of the observed data/theoretical model since it uses their/its PDF. Thus, the surrogate light curves carry all the statistical moments and depending on the nature of the statistical study, the PDF corresponds either to the observed or the parent PDF.
\item Introduction of higher statistical moments (other than mean value and variance that characterise completely only the Normal distribution) as well as definition of genuinely positively probability distributions allow the construction of realistic `bursty' light curves which can not be created by TK95.
\item For the special case of Gaussian light curves the method yields synthetic data sets are by construction equivalent to those of TK95.
\item For the  special case of a parent log-normal distribution the simulated light curves exhibit the rms-flux relation.
\end{itemize}
Particularly for the case of `bursty' light curves, having by definition non-Gaussian positively defined PDFs which can be even sometimes described by right \textit{heavy-tailed} PDFs, representative for extreme flaring states, this new method is the most appropriate for the correct establishment of confidence intervals of a given method e.g.\ CCF analysis.\par
Due to its generality, the method can be employed to a vast variety of statistical analysis purposes involving light curves obtained across the electromagnetic spectrum for any object. The Monte Carlo simulation studies, which are currently performed using the TK95 products, can now be extended to statistically much more accurate synthetic light curves, thus providing us with robust results with respect to e.g.\ cross correlation analysis, establishment of detection significance for future missions (e.g.\ \textit{LOFT}, \textit{CTA}), detection and characterisation of variability, understudying of the effects of irregular sampling.

\section*{ACKNOWLEDGMENTS}
DE and IMM acknowledge the Science and Technology Facilities Council (STFC) for support under grant ST/G003084/1. This research has made use of NASA's Astrophysics Data System Bibliographic Services. Finally, we are grateful to the anonymous referee for the very useful comments and suggestions that helped improved the quality of the manuscript.

\bibliographystyle{mnauthors}

\appendix

\section[]{DEFINITIONS AND NOMENCLATURE}
\label{app:def_nom}
Below we briefly describe the discrete Fourier transform, the calculation of the periodogram, the PSD estimation and the derivation of the PDF.\par 
\subsection{The periodogram}
\label{ssect:prdgr} 
Consider a light curve $x(t)$ consisting of $N$ equidistant observations: $\{t_k, x(t_k)\}$ for $k=1,2,\ldots,N$ with a sampling period $t_{\rm bin}$, a mean value of $\mu$ and a standard deviation of $\sigma$. The discrete Fourier transform (DFT) of the data set is defined following \citet{press92a}\footnote{In this case the exponential function contains as a running index $k-1$ instead of $k$ since the data start for $k=1$ and not $k=0$.}:
\eqb
DFT(j)=\sum_{k=1}^{N}x(t_k)e^{2\pi i (k-1) j/N}
\label{eq:dft}
\eqe
yielding $N$ estimates for $j=0,\ldots,N-1$, each one corresponding to a Fourier frequency $f_j$ depending on the parity of $N$ (i.e.\ even or odd):\par
At $f_0=0$ ($j=0$) the zero Fourier frequency component, $DFT(0)$, corresponds always to the sum of the light curve estimates.\newline
\underline{For even $N$}\newline
$\bullet$ Positive: $f_j^+$=$j/(N t_{\rm bin})$ for $j$=$1,\ldots,N/2-1$.\newline
$\bullet$ Negative: $f_j^-$=$-(N-j)/(N t_{\rm bin})$ for $j$=$N/2+1,\ldots,N-1$.\newline
$\bullet$ Nyquist: $f_{N/2}=f_{\rm Nyq}=1/(2 t_{\rm bin})$ for $j=N/2$.\par\noindent
Note, that the negative frequencies are mirrored versions of the positive frequencies with opposite signs (around $f_{\rm Nyq}$)  e.g.\ $-f_{N/2-1}^-=f_{N/2+1}^+,\ldots,-f_1^-=f_{N-1}^+$.\newline
\underline{For odd $N$}\newline
$\bullet$ Positive: $f_j^+$=$j/(N t_{\rm bin})$ for $j$=$1,\ldots,(N-1)/2$.\newline
$\bullet$ Negative: $f_j^-$=$-(N-j)/(N t_{\rm bin})$ for $j$=$(N+1)/2,\ldots,N-1$.\newline
$\bullet$ Nyquist: There is no Nyquist frequency estimate.\par\noindent
Note again, that the negative frequencies are mirrored versions of the positive frequencies with opposite signs e.g. $-f_{(N-1)/2}^-=f_{(N+1)/2}^+,...,-f_1^-=f_{N-1}^+$.\par
At a given frequency $f_j$, $DFT(j)$ is a complex number of the form $q+wi$\footnote{For the case of even $N$, the $DFT(N/2)$ (i.e.\ at Nyquist frequency) is a real number since, from equation \ref{eq:dft}, the exponential function for $j=N/2$ is equal to 1 (for odd $k$) or -1 (for even $k$).}and  which carries information about the amplitude and the phase of the corresponding sinusoidal component.  
The amplitude of the sinusoid at a frequency, $f_j$, is given by 
\eqb
\mathscr{A}_j=\frac{1}{N}\sqrt{\operatorname{Re}[DFT(j)]^2+\operatorname{Im}[DFT(j)]^2}
\label{eq:amplitude}
\eqe
and its phase is given by
\eqb
\phi_j=\operatorname{arg}[DFT(j)]=\arctan\left\{\operatorname{Im}[DFT(j)],\operatorname{Re}[DFT(j)]\right\}
\label{eq:phase}
\eqe
taking values in the closed-open interval $(-\pi,\pi]$. For the complex number 0 one may use $\phi=0$ but formally its phase angle is indeterminate.\par
The periodogram of $x(t)$ at a given Fourier frequency $f_j$, $P(f_j)$, is defined as the squared amplitude (equation \ref{eq:amplitude}) of the corresponding sinusoid component
\eqb
&&\hspace*{-2em}P(f_j)=\mathscr{A}_j^2=\frac{1}{N^2}\left\{\operatorname{Re}[DFT(j)]^2+\operatorname{Im}[DFT(j)]^2\right\}\nonumber \\
&&\hspace*{-2em}{\rm for}\;j=0,\ldots,N-1
\label{eq:periodogram}
\eqe
Since the light curve consists only of real measurements, $x(t_k)\in\mathbb{R}$, there is a symmetry between the positive and the negative DFT estimates: $DFT(j^-)=[DFT(j^+)]^{*}$ where $j^-$ and $j^+$ represent the indices for the negative and positive frequencies, respectively, and the asterisk denotes complex conjugation. Thus, the amplitudes of the corresponding positive and negative components are equal and the periodogram is estimated as   
\eqb
&&\hspace*{-2em}P(f_j)=\frac{2}{N^2}\left\{\operatorname{Re}[DFT(j)]^2+\operatorname{Im}[DFT(j)]^2\right\}\nonumber \\
&&\hspace*{-2em}{\rm even\;}N\hspace{-0.3em}:j=0,\ldots,N/2\nonumber \\
&&\hspace*{-2em}{\rm odd}\;N\hspace{-0.3em}:j=0,\ldots,(N-1)/2
\label{eq:doublePrdgr}
\eqe
with $f_j=j/(N t_{\rm tbin})$. There is a plethora of normalisation factors that can be applied to the periodogram \citep[e.g.][]{vaughan03}. In this work we employ the fractional root mean square (rms) normalisation: $N t_{\rm bin}/\mu^2$ and the periodogram (equation \ref{eq:doublePrdgr}) becomes
\eqb
P(f_j)=\frac{2t_{\rm bin}}{\mu^2N}\left\{\operatorname{Re}[DFT(j)]^2+\operatorname{Im}[DFT(j)]^2\right\}
\label{eq:normPrdgr}
\eqe
With this normalisation the square root of the integral of the underlying PSD between two frequencies $f_1$ and $f_2$ yields the contribution to the fractional rms squared variability (i.e.\ $\sigma^2/\mu^2$). Thus, integration between $f_1$, and $f_{\rm Nyq}$ (even) or $f_{(n-1)/2}$ (odd) yields the total rms squared variability.

\subsection{Power spectral density estimation}
\label{ssect:psd}
The `statistical natural' estimator of the underlying power spectral density (PSD), $\mathscr{P}(f)$, is the periodogram, $P(f)$. In the manner of \citet{priestley81}, assume that the light curve, $x_t$, originates from a linear process of the form
\eqb
x_t=\sum_{0}^{\infty}g_u\epsilon_{t-u}
\eqe
where $\epsilon_{t}$ is a purely random Gaussian process and $g_u$ is a given sequence of constants satisfying $\sum_{u=0}^{\infty}g_u^2<\infty$. At a given frequency, $f_j$, $P(f_j)$ is then asymptotically distributed around the $\mathscr{P}(f_j)$ as

\eqb
P(f_j) = \left\{
\begin{array}{lr}
\frac{1}{2}\chi^2_2 \mathscr{P}(f_j) & j=1,\ldots,\substack{N/2-1\;{\rm(even}\;N{\rm)}\\(N-1)/2\;{\rm(odd}\;N{\rm)}} \\[1em]
\frac{1}{2}\chi^2_1 \mathscr{P}(f_{\rm Nyq}) & j=N/2 \;{\rm(even}\;N{\rm)}
\end{array}
\right.
\label{eqe:prdgrChiSquare}
\eqe
where $\chi^2_\nu$ represents the $\chi^2$ distribution with $\nu$ degrees of freedom (d.o.f.). This means that, for a given frequency, the standard deviation of the periodogram estimates is 100 per cent, automatically making the ensemble of periodogram estimates an inconsistent estimator of the underlying PSD.\par
In order to retrieve the $\mathscr{P}(f_j)$, one can use either binning or maximum likelihood methodologies. For the former, the binned logarithmic periodogram has been proposed by \citet{papadakis93} ensuring that the logarithmic periodogram estimates are normally distributed within each geometric mean frequency bin. Thus, PSD models can be fitted to the logarithmic periodogram estimates using a simple least-squares method, requiring Gaussianity within each bin. The latter should include at least 10 periodogram estimates a fact which partially limits the usefulness of the method for small data sets.
Another approach is to fit a PSD model directly to the ensemble of periodogram estimates by performing maximum likelihood estimation which makes direct use of the underlying distribution at a given Fourier frequency (equation \ref{eqe:prdgrChiSquare}). This is the approach used in this work and detailed references can be found in \citep[e.g.][]{anderson90,vaughan05,barret12}.\par
Consider an underlying PSD model, $\mathscr{P}(f_j;\vec{\gamma})$, in which $\vec{\gamma}=\{\gamma_1,\gamma_2,\ldots,\gamma_n\}$ is a vector consisting of the unknown model parameters such as normalisation, break/bend frequency, low/high frequency slopes etc. The probability of obtaining a given single periodogram estimate $P(f_j)$ for the given PSD model $\mathscr{P}(f_j;\vec{\gamma})$ is.
\eqb
&&\hspace*{-2em}\lambda_j\left[P(f_j)|\mathscr{P}(f_j;\vec{\gamma})\right] = \nonumber\\[1em]
&&\hspace*{-2em}\left\{
\begin{array}{lr}
\frac{e^{-P(f_j)/\mathscr{P}(f_j;\vec{\gamma})}}{\mathscr{P}\left(f_j;\vec{\gamma}\right)} & j=1,\ldots,\substack{N/2-1\;{\rm(even}\;N{\rm)}\\(N-1)/2\;{\rm(odd}\;N{\rm)}}\\[1em]
\frac{e^{-P(f_{\rm Nyq})/\mathscr{P}\left(f_{\rm Nyq};\vec{\gamma}\right)}}{\left[\pi P(f_{\rm Nyq})\mathscr{P}\left(f_{\rm Nyq};\vec{\gamma}\right)\right]^{1/2}} & j=N/2 \;{\rm(even}\;N{\rm)}
\end{array}
\right.
\label{eqe:prdgrProb}
\eqe
The constituent functions of the above piecewise expression are usually referred to as `scaled $\chi^2$ distributions' with two and one d.o.f. for the top and lower branch respectively. More precisely, these functions are special forms of the gamma distribution, $\Gamma\left[\nu/2, \mathscr{P}(f_j;\vec{\gamma})\right]$ where $\nu$ corresponds to the d.o.f. i.e.\ $\nu=1$ corresponds only to the the Nyquist frequency, $f_{\rm Nyq}$ ($j=N/2$, even $N$), and $\nu=2$ to all other frequencies (for either even or odd $N$)\footnote{An even more general representation can be obtained through the Pearson's Type III distribution \citep[p.\ 930 in][]{abramowitz72} for $\alpha=0$, $\beta=\mathscr{P}(f_j;\vec{\gamma})$ and $p=\nu/2$.}.

The joint probability of obtaining the ensemble of periodogram estimates for the given PSD model is 
\eqb
\mathscr{L}=\prod_{i=1}^{\substack{N/2\;{\rm(even}\;N{\rm)}\\(N-1)/2\;{\rm(odd}\;N{\rm)}}}\lambda_j\left[P(f_j)|\mathscr{P}\left(f_j;\vec{\gamma}\right)\right]
\label{eqe:prdgrMaxLikeli}
\eqe
since asymptotically (i.e.\ $N\rightarrow\infty$) the various periodogram estimates are strictly independent at the Fourier frequencies $f_j$ \citep{priestley81} (this is the reason why the periodogram is estimated only for these frequencies and not for intermediate values). The maximum likelihood estimate of the model function parameters, $\vec{a}$, is obtained by maximizing the above probability, or equivalently by minimizing the log-likelihood function $\mathcal{C}=-2\ln \mathscr{L}$ which is equal to
\eqb
&&\hspace*{-2em}\mathcal{C} = \nonumber \\[1em]
&&\hspace*{-2em}\left\{
\begin{array}{lr}
\left\{2\sum_{j=1}^{j=N/2-1}\left\{\ln\left[\mathscr{P}(f_j;\vec{\gamma})\right]+\frac{P(f_j)}{\mathscr{P}\left(f_j;\vec{\gamma}\right)}\right\}\right\}+ \\[1em]
\ln[\pi P(f_{\rm Nyq})\mathscr{P}(f_{\rm Nyq};\vec{\gamma})]+2\frac{P(f_{\rm Nyq})}{\mathscr{P}\left(f_{\rm Nyq};\vec{\gamma}\right)}\;\;\;{\rm(even}\;N{\rm)}\\[1em]
2\sum_{j=1}^{j=(N-1)/2}\left\{\ln\left[\mathscr{P}(f_j;\vec{\gamma})\right]+\frac{P(f_j)}{\mathscr{P}\left(f_j;\vec{\gamma}\right)}\right\}\;{\rm(odd}\;N{\rm)}
\end{array}
\right.
\label{eqe:logLikelihood}
\eqe
In this work we have employed two minimisation routines \citep[see for details][]{press92a}: a direct search method, Nelder-Mead \citep{nelder65} and a stochastic function minimizer, simulated annealing \citep{kirkpatrick83}. The PSD models, $\mathscr{P}(f_j;\vec{\gamma})$, which are usually fitted to the data have a power-law form (e.g.\ broken power-law, continuous bending power-law etc.) and, for these type of minimisation problems, both methods identical results.\par
The joint confidence intervals for $q$ model parameters from a total of $n$ components of $\vec{a}$, $\{\alpha_1,\alpha_2,\ldots,\alpha_q,\alpha_{q+1},\alpha_{q+2},\ldots,\alpha_n\}$ can be estimated using the method of \citet{cash79}, based on the theorem of \citet{wilks38}. Initially, a global minimum is found by varying all the $n$ model parameters yielding $(\mathcal{C}_{\rm min})_n$. The $q$ parameters of interest are fixed to their best-fitting values and the rest, $q+1,q+2,\dots,n$ then are varied until a global minimum is reached corresponding to $(\mathcal{C}_{\rm min})_{n-q_{\rm bf}}$. The quantity $\Delta \mathcal{C}=(\mathcal{C}_{\rm min})_{n-q_{\rm bf}}-(\mathcal{C}_{\rm min})_n$ is then distributed as a $\chi^2$ distribution with $q$ d.o.f. Thus, the 68.3 and 90 per cent single confidence intervals for one parameter ($q=1$) correspond to a $\Delta \mathcal{C}$ of 1 and 2.71, respectively. Similarly, the 68.3 and 90 per cent joint confidence intervals for one parameter ($q=2$) 
correspond to $\Delta \mathcal{C}$ of 2.30 and 4.61, respectively. In general, for a given confidence interval $p$, and a given value of $q$ the corresponding value of $\Delta \mathcal{C}$ is given by $2Q^{-1}\left(\nu/2,0,p\right)$, where $Q^{-1}$ corresponds to the inverse of the generalised regularised incomplete gamma function \citep[definitions for $Q^{-1}$ can be found in][]{chaudhry94}.\par

\subsection{Probability density function estimation}
\label{ssect:histogram}
The probability density function (PDF) of the observations should always be represented by a positive real-valued distribution and, depending on the purpose of the statistical study, should correspond either to the parent or the observed distribution. The PDF is used in the proposed method (Sect.~\ref{sect:methodology}) to produce a sample of independent and identically distributed random variates.\par
In general there are three ways to derive the probability density function (parent or observed) from a given data set. Note, that for the case of the parent distribution we need very large data sets to be able to match the overall variability profile of the source under study. One approach is to fit a probability density function model, $\mathfrak{f}(x_i;\vec{\eta})$, to the histogrammed data (where $\vec{\eta}$ is a vector consisting of the unknown distribution's model parameters), using the maximum likelihood method in a similar fashion to that described above in App.~\ref{ssect:psd}, i.e.\ maximising the log-likelihood function $\sum_i{\ln \mathfrak{f}(x_i;\vec{\eta})}$. For pathological cases of histogrammed observations exhibiting e.g.\ highly skewed, non-zero kurtosis in conjunction with extreme long-tailed distributions, one can use appropriate methodologies developed for such purposes, such as the method of generalised moments \citep{wooldridge01}, the method of cumulants \citep{frisken01} and the 
method of factorial moments \citep{bialas86} 
the latter being particularly useful in the presence of low count rates i.e.\ high Poisson noise \citep{deWolf96}.\par
Another approach is to use directly a piecewise-constant representation of the unknown PDF \citep{knuth06}, using directly the data set consisting of $x_i$ observations $(i=1,\ldots,N)$:
\eqb
\mathfrak{h}(x)=\sum_{k=1}^{M}\frac{N_k}{N \upsilon_k}\Pi(\xi_{k-1},x,\xi_k)
\eqe
where $N_k$ is the number of data points in the $k$\textsuperscript{th} bin, $\upsilon_k$ is the width of the $k$\textsuperscript{th} bin, $\xi_{k-1}$ and $\xi_k$ are the edges of the $k$\textsuperscript{th} bin, and $\Pi(\xi_\alpha,x,\xi_\beta)$ is the boxcar function being equal to 1 for $\xi_\alpha\leq x \leq \xi_\beta$ and 0 otherwise.\par
Finally, instead of using the PDF representation of the data set, one can use a cumulative distribution function by employing the empirical distribution function of the data set consisting of $x_i$ observations $(i=1,\ldots,N)$. This can be done by estimating the following quantity \citep[after][]{owen01}:
\eqb
\mathfrak{H}(y)=\frac{1}{N}\sum_{i=1}^{N}\Pi(-\infty,x_i,y)
\eqe
This can then be used directly in order to produce random numbers, which is the primary reason that we need to estimate the distribution of the data points.

\section{Statistical moments, cumulants and polyspectra}
\label{app:cumul_statiDependnce}
In the manner of \citet{priestley81}, let $X$ be a random variable with {\it moment generating function} $M(t)$ then the {\it cumulant generating function}, $K(t)$ is defined as:
\eqb
K(t)=\ln\left[M(t)\right]
\eqe
By expanding the above expression in a power series we get:
\eqb
K(t)=k_1t+k_2\frac{t^2}{2!}+\ldots+k_r\frac{t^r}{r!}
\eqe
The coefficient of $t^r/(r!)$ is called the {\it r\textsuperscript{th} cumulant}. Only, the first three cumulants coincide with the first three statistical moments (mean, variance, skewness) and all the other are given by more complicated polynomial expressions i.e\ $k_1=\mu$, $k_2=\sigma^2$, $k_3=\gamma_1$, $k_4=\gamma_2-3\sigma^4$,etc.\par
Generalising the above to more random variables, let $X_t$ be a process stationary up to order $k$ and let $C(s_1,s_2,\ldots,s_{k-1})$ denote the {\it joint cumulant} of order $k$ of the set of random variables $\{X_t,X_{t+s_1},\ldots,X_{t+s_{k-1}}\}$ is the coefficient of $(z_1,z_2,\ldots,z_k)$ in the expansion of the joint cumulant generating function
\eqb
K(z_1,z_2,\ldots,z_n)=\ln\left[M(z_1,z_2,\ldots,z_n)\right]
\eqe
in which $M(z_1,z_2,\ldots,z_n)$ is the {\it joint moment generating function}.\par
The second order joint cumulant, $C_2(s_1)$, is simply the covariance, $\cov\left(X_t,X_{t+s_1}\right)$ and the third order joint cumulant, $C_3(s_1,s_2)$, is identical to the third order joint moment,$\gamma_{1}(s_1,s_2)$ \citep[sometimes in economics this is called co-skewness and the next joint cumulant co-kurtosis e.g.][]{hwang99}.
\eqb
C_2(s_1)=\left<(X_t-\mu)(X_{t+s_1}-\mu)\right>
\eqe
\eqb
C_3(s_1,s_2)=\left<(X_t-\mu)(X_{t+s_1}-\mu)(X_{t+s_2}-\mu)\right>
\eqe
for $s_1=s_2=0$ these two quantities are directly related to the variance and the skewness, $C_2(0)=\sigma^2$ and $C_3(0)=\gamma_1\sigma^3$ and these are two properties that are mapped on the PDF.\par
The Fourier transforms of the corresponding higher order cumulants are called polyspectra.
\hspace*{-2em}\parbox{1\linewidth}{
\eqb
&&h_k(f_1,f_2,\ldots,f_n)\nonumber=\\[1em]
&&\sum_{s_1=-\infty}^{\infty}\ldots\sum_{s_{k-1}=-\infty}^{\infty}C(s_1,\ldots,s_{k-1})e^{-2\pi(f_1s_1+\ldots+f_{k-1}s_{k-1})}\hspace{2.8em}
\eqe
}\\
The second order polyspectrum is the autospectrum and its squared amplitude is the PSD, $|h_2(f)|^2\equiv\mathscr{P}(f)$. The third and the fourth order polyspectra are known as bispectrum and trispectrum respectively. These are the quantities that characterise the various dependences between the various measurements. The fact that two data sets have e.g.\ the same variance and skewness (i.e.\ $C_2(0)$, $C_3(0,0)$), does not mean that they have the same covariance and third order joint cumulant, $C_2(s_1)$ and $C_3(s_1,s_2)$, respectively. Thus, data sets which have the same statistical moments (i.e.\ same PDFs), does not mean that they have the same polyspectra. 

\bsp
\label{lastpage}
\end{document}